\documentclass[a4paper,11pt]{article}
\usepackage{jheppub} 

\usepackage{orcidlink}
\usepackage[normalem]{ulem}
\usepackage{graphicx}
\usepackage{import}

\usepackage{xcolor, graphicx}
\usepackage[export]{adjustbox}
\usepackage{mathrsfs}
\usepackage[normalem]{ulem}
\usepackage{braket}  
\usepackage{tikz}
\usetikzlibrary{decorations.markings}
\usetikzlibrary{calc} 
\usetikzlibrary{arrows.meta}
\tikzset{>=Stealth} 
\AtBeginDocument{%
  \setcounter{tocdepth}{2}
}

\tikzset{
  midarrow/.style={
    postaction={decorate},
    decoration={markings, mark=at position 0.5 with {\arrow{stealth}}}
  }}

\usepackage{mathtools}
\usepackage{tikz}
\makeatletter
\newcommand\mathcircled[1]{%
  \mathpalette\@mathcircled{#1}%
}
\newcommand\@mathcircled[2]{%
  \tikz[baseline=(math.base)] \node[draw,circle,inner sep=1pt] (math) {$\m@th#1#2$};%
}
\makeatother

\title{\boldmath Non-invertible Nielsen circuits and 3d Ising gravity}

\author{Saskia Demulder\,\orcidlink{0000-0002-4337-4489}}
\affiliation{Department of Quantitative Methods, CUNEF Universidad,\\ Calle Almansa 101,
28040 Madrid, Spain}

 \emailAdd{saskia.demulder@cunef.edu}

\abstract{We extend Nielsen's formulation of quantum circuit complexity to include intrinsically non-invertible operations. Such gates arise from fusion with topological defect operators and remove a basic limitation of symmetry-based circuits: the inability to change superselection sectors, or in two-dimensional CFTs, conformal families. We realise fusion operations as completely positive, trace-preserving quantum channels acting between sectors, with consistency ensured by the fusion and associator data of an underlying unitary modular tensor category. In contrast to standard Nielsen circuits, non-invertible circuits lead to an optimisation problem that is no longer governed by geodesics on a continuous group manifold but instead reduces to a discrete shortest-path problem on the fusion graph of superselection sectors. We illustrate the framework in representative rational conformal field theories. Finally, we interpret fusion-induced transitions as discrete changes in boundary stress-tensor data, corresponding to shock-like defects in AdS$_3$ gravity.

}

\begin{document}
\maketitle
\flushbottom

\section{Introduction}

Symmetry has long provided one of the central organising principles of physics, governing conservation laws, phase structure, and universality across disparate types of systems. In quantum field theory, however, the notion of symmetry has undergone a significant generalisation: physically meaningful symmetries need not be invertible. Such non-invertible symmetries are characterised not by group actions but by fusion rules, whereby the composition of two symmetry operators decomposes into a sum of operators rather than a single outcome. These structures were first identified concretely in two-dimensional conformal field theory through the study of topological defect lines and their fusion and transmission properties \cite{Oshikawa:1996dj,Petkova:2000ip,Fuchs:2002cm,Frohlich:2006ch}. In rational conformal field theories, non-invertible symmetries organise superselection sectors, constrain dynamics, and control operator algebras \cite{Verlinde:1988sn,Moore:1988qv,Fuchs:2023ngi}. Related perspectives emphasising non-invertible symmetries as quantum operations have also been discussed in \cite{Okada:2024qmk}. See recent reviews \cite{McGreevy:2022oyu,Shao:2023gho,Schafer-Nameki:2023jdn,Bhardwaj:2023kri} for an exhaustive overview of the scope and literature.

At the same time, ideas from quantum information theory have led to a geometric understanding of quantum circuits through the work of Nielsen \cite{Nielsen:2005mkt,Nielsen:2006cea}. In this framework, circuit complexity is defined as a geometric optimisation problem on the space of unitary operations generated by a fixed gate set, with a right-invariant metric encoding the relative cost of different directions in the Lie algebra. A defining feature of this construction is its reliance on invertible dynamics: circuits correspond to continuous paths on a group manifold, and composition is reversible by construction. While this approach has been extended in various directions, including the introduction of penalty factors, refinements of state complexity, and symmetry-based circuit constructions, see e.g. \cite{Balasubramanian:2019wgd,Balasubramanian:2021mxo,Chagnet:2021uvi,Craps:2023rur,Baiguera:2023bhm,Baiguera:2025uss} for recent works on those directions and \cite{Baiguera:2025dkc,Chapman:2021jbh,Susskind:2018pmk} for reviews, all such frameworks remain intrinsically tied to unitary, sector-preserving operations.

In quantum field theory, and especially in two-dimensional conformal field theory, it is natural to identify the gate set of Nielsen complexity to be generated by symmetry of the system. A particularly successful realisation of Nielsen's geometric approach in this context is obtained by taking the gate set to be generated by the Virasoro symmetry. In this setting, \cite{Caputa:2018kdj} reformulated circuit complexity for Virasoro symmetry circuits and showed that the corresponding cost functional is given by the geometric action on Virasoro coadjoint orbits, yielding a universal and symmetry-controlled notion of circuit geometry in two-dimensional CFT. Subsequent work further refined the AdS$_3$ interpretation of such Virasoro circuits and clarified an important conceptual distinction: one may view the circuit either as a procedure that stitches together distinct bulk configurations, or as physical time evolution within a fixed coadjoint orbit \cite{Erdmenger:2021wzc,Erdmenger:2022lov}.

Despite their success, all symmetry group circuit frameworks share a basic limitation: they are intrinsically tied to invertible dynamics. The circuit is a path on a group manifold, and the corresponding action necessarily preserves the representation sector of the Hilbert space. In two-dimensional conformal field theory, these sectors are the Virasoro conformal families and play the role of superselection sectors, in the sense that symmetry operations cannot mix them. As a result, symmetry-based circuits remain confined to the coadjoint orbit determined by the initial data and cannot describe processes that change conformal families or branch into multiple outcomes as fusion operation do. 

\paragraph{Motivation and summary of results.} In this work we take the next step by incorporating non-invertible symmetry operations into circuit complexity, motivated by their central r\^ole in organising quantum field theories. An immediate consequence of this perspective is that it directly addresses the limitation discussed above, namely the inability of purely invertible circuits to change superselection sectors.

More precisely, we propose a categorical extension of Nielsen's geometric approach to circuit complexity in which elementary non-invertible operations are realised through the fusion data of a unitary modular tensor category. Concretely, we consider extended circuits of the form
\begin{align}\label{eq:new_circuit}
    D_{a_k}D_{a_{k-1}}D_{g_n}\cdots D_{a_1}D_{g_2}D_{g_1}|\psi\rangle\,,
\end{align}
which we use as a convenient representative for circuits built from invertible and non-invertible gates.
Here $|\psi\rangle$ is an initial state of the system on which both invertible and non-invertible symmetries act.  The gates $D_{g_i}$ denote unitary gates associated with elements $g_i\in G$ of a Lie group and satisfy the usual group composition law,
\begin{align}
    D_{g_i}D_{g_j}=D_h
\end{align}
for some $h\in G$. The gates $D_{a_i}$, by contrast, realise the action of non-invertible symmetries: their composition is governed by fusion rules,
\begin{align}\label{eq:fusiondef}
    D_{a_i}\,D_{a_j}=\bigoplus_k N_{a_i a_j}^{\,a_k}\,D_{a_k}\,,
\end{align}
where the $a_k$ label simple objects of the category and $N_{a_i a_j}^{\,a_k}$ are the corresponding fusion multiplicities. 

Physically, the non-invertible gates introduced in this work may be viewed as local insertions of topological defect junctions at fixed circuit (or boundary) time, acting on the Hilbert space of the theory. In the circuit framework developed here, such defect insertions are promoted to genuine gate operations by realising fusion as completely positive, trace-preserving quantum channels acting between distinct Hilbert-space sectors, with composition controlled by the associator of the underlying category. This yields a well-defined extension of Nielsen circuits in which new gate types can change conformal families, allowing one to describe discrete sector changes within an otherwise continuous symmetry circuit and to assign intrinsic costs to such transitions. For full technical control, we focus on rational conformal field theories, where fusion data are finite and exactly known and are encoded by a unitary modular tensor category.

As an application, we study the implications of non-invertible circuits for $3d$ AdS gravity in a fully quantum regime. In the Chern-Simons formulation of AdS$_3$ gravity, boundary Virasoro data determine bulk configurations through the conjugacy classes of $SL(2,\mathbb R)\times SL(2,\mathbb R)$ holonomies rather than through local metric degrees of freedom. This structure underlies the appearance of Virasoro minimal models as admissible boundary theories of Chern-Simons gravity \cite{Bershadsky:1989mf} and, in particular, the demonstration that in a quantum regime of three-dimensional gravity the bulk partition function can be computed and, under specific conditions, agrees with the torus partition function of the Ising CFT ($c=\tfrac12$), furnishing evidence for a duality between strongly coupled AdS$_3$ gravity and the minimal model \cite{Castro:2011zq}.

In this setting, fusion operations act between distinct conformal families and therefore induce discrete changes in Virasoro highest weights. When translated into the Chern-Simons language, such family-changing operations correspond to localised modifications of bulk holonomy data. We interpret these as shock-like defects in a topological gravity theory: instantaneous changes in boundary stress-tensor data that alter the global holonomy class without introducing propagating gravitational degrees of freedom. Related time-dependent probes of Ising gravity have recently been explored in \cite{Janik:2025zji}, providing further evidence that minimal models offer a controlled laboratory for studying quantum aspects of AdS$_3$ gravity beyond semiclassical limits.

Although our framework makes essential use of categorical fusion and associator data, its conceptual role differs sharply from that in topological quantum computation. There, fusion is treated as state preparation or measurement, while computation is implemented through unitary braiding operations. Here, by contrast, fusion itself is promoted to a non-invertible circuit operation acting directly on the physical Hilbert space. No braid group representations are invoked, and the construction is not aimed at fault-tolerant computation. Rather, categorical consistency ensures that fusion-defined quantum channels compose unambiguously and admit well-defined cost assignments, even though they are intrinsically non-unitary.

\paragraph{The structure of the paper.}  In section \ref{sec:reviewNielsen} we review Nielsen complexity for unitary Virasoro circuits and identify the obstruction to changing conformal families. In section \ref{sec:noninv_gates} we introduce fusion operations as non-invertible gates and construct the associated quantum channels, analysing their composition and associativity. The cost for the new set of gates  and complexity of the enhanced Nielsen circuit are defined in section \ref{sec:costs}.  Section \ref{sec:shocklike_def} develops the bulk interpretation in the Ising subsector of AdS$_3$-gravity, relating fusion-induced transitions to jumps in Ba\~nados geometries and analysing the associated energy conditions. In section \ref{sec:conclusions} we conclude, highlighting a number of future directions.

\section{Review of Nielsen complexity and Virasoro circuits}\label{sec:reviewNielsen}

To set the stage, in this section, we first swiftly review Nielsen complexity in 2d CFTs, mostly following the results of \cite{Caputa:2018kdj,Erdmenger:2021wzc}. Quantum circuit complexity provides a quantitative notion of the minimal cost required to prepare a target quantum state from a reference state using a prescribed set of elementary gates. In the Nielsen approach, this cost is defined geometrically as the length of a path generated by continuous gates acting on Hilbert space, with the choice of gate set and metric determining the resulting notion of complexity. 

In two-dimensional conformal field theory, a natural continuous gate set is provided by the Virasoro algebra, whose generators implement infinitesimal conformal transformations. Circuits built from Virasoro generators define unitary, invertible evolutions acting within a fixed highest-weight representation. The unitary gates associated to the Virasory algebra take on the form
\begin{align}\label{eq:inv_Nielsen_gate}
	U(\tau)=\mathcal{P}\exp \left(\int_0^\tau Q(\tau')\,\mathrm d \tau' \right)\,,
\end{align}
where 
\begin{align}\label{eq:gen_vir_circuit}
	Q(\tau)\equiv \int_0^{2\pi}\exp\left(\frac{\mathrm d\sigma}{2\pi}\epsilon(\tau,\sigma)T(\sigma)\right)=\sum_{n\in \mathbb Z}\epsilon_n(\tau)\left(L_{-n}-\frac{c}{24}\delta_{n,0}\right)\,.
\end{align}
Here $\epsilon_n(\tau)$ play the role of velocities along the path driven by the gates.
In this formula $\epsilon(\tau,\sigma)$ has been expanded in terms of the Virasoro algebra generators
\begin{align}
    [L_m,L_n]=(m-n)L_{m+n}+\frac{c}{12}m(m^2-1)\delta_{m+n,0}\,.
\end{align}
Denoting the path corresponding to the trajectory generated infinitesimally through $\epsilon(\tau,\sigma)$ by $f(\tau,\sigma)$, then the $\mathcal F_1$ cost function is \cite{Caputa:2018kdj} 
\begin{align}\label{eq:Vir_cost_KKSaction}
    \mathcal F_1=|\mathrm{tr}[\rho(\tau)Q(\tau)]|=\frac{c}{24\pi}\int_0^{2\pi}\mathrm d\sigma \,\frac{\dot f}{f'}\left(2r^2+\{f,\sigma\}\right)\,,
\end{align}
where $2 r^2=\frac{1}{2}(1-(24h/c)]$. This expression coincides with the Kirillov geometric action on  coadjoint orbits of the Virasoro group. Each Virasoro coadjoint orbit carries a Kirillov-Kostant symplectic structure and an associated geometric action. For a path $f(t,\sigma)\in\mathrm{Diff}^+(S^1)$, this is the Alekseev-Shatashvili action 
\begin{align}\label{eq:SA} 
	S_{\mathrm{AS}}[f] =\frac{1}{2\pi}\int \mathrm d\tau\mathrm d\sigma\, \frac{\dot f}{f'} \left(b_0 (f)^2-\frac{c}{24}\partial_\sigma \left(\frac{ f''}{f'}\right) \right)\,,
\end{align}
where $ \partial_\sigma f =f'$.
When a right-invariant cost function is chosen on the Virasoro group, this action coincides with the Nielsen complexity functional for Virasoro-driven circuits \cite{Caputa:2018kdj}, see also \cite{Erdmenger:2020sup}. In this formulation, a continuous unitary circuit corresponds to a trajectory on a fixed Virasoro coadjoint orbit, and its complexity is given by the length of this trajectory with respect to the induced metric. Geometric quantisation of a fixed coadjoint orbit yields a highest-weight Virasoro representation with fixed conformal weight. Consequently, unitary Virasoro gates act within a single conformal family and do not change the orbit label $h$ .

At the level of holographic interpretation, an important refinement of the purely group-theoretic viewpoint was proposed in \cite{Erdmenger:2021wzc}. In the original work of \cite{Caputa:2018kdj} the circuit parameter $\tau$ is introduced as an auxiliary coordinate along the path in the Virasoro group, serving to parametrise geodesics on a coadjoint orbit. Later it was recognised that a more natural bulk dual arises when the instantaneous generator $Q(\tau)$ in \eqref{eq:gen_vir_circuit} is identified with the physical CFT Hamiltonian, so that $\tau$ coincides with boundary time. Under this identification, the corresponding bulk dual is not a sequence of disconnected geometries, but instead a single time-dependent Ba\~nados geometry that evolves continuously under the same generator of physical time translations \cite{Erdmenger:2021wzc}. This observation provides a concrete framework in which Virasoro complexity measures admit exact gravity duals with a single underlying bulk spacetime and will be critical in section \ref{sec:shocklike_def}.

The discussion above highlights an intrinsic limitation of Virasoro-based Nielsen circuits. Because Virasoro gates act by diffeomorphisms of the circle, they generate continuous, invertible motions along a single coadjoint orbit. Equivalently, at the quantum level, they act within a fixed highest-weight representation $\mathcal H_h$ and therefore preserve the conformal family.  This obstruction motivates an enlargement of the circuit model to include non-invertible gates capable of changing the representation label.

\section{Non-invertible gates from categorical fusion}\label{sec:noninv_gates}

In this section, we construct a circuit framework in which categorical fusion processes are promoted to elementary gate operations and incorporated into Nielsen's approach to circuit complexity. In two-dimensional conformal field theory, fusion is implemented by the action of topological defect operators on superselection sectors. Such operations change conformal families but do not act unitarily within a fixed Hilbert space, and therefore fall outside the standard Nielsen framework.

Our strategy is to reinterpret fusion not as a unitary gate, but as a quantum operation acting between sector Hilbert spaces. This requires translating the categorical data of fusion (multiplicities, junctions, and associativity) into linear maps that compose consistently in a circuit setting. In rational conformal field theories, this translation is natural: finiteness of fusion rules, canonical Hermitian structures ensured by unitarity, and associativity encoded by $F$-symbols together guarantee that fusion-defined operations admit a well-defined notion of circuit ordering and composition.

Throughout we restrict to rational CFTs, for which the chiral data form a unitary modular tensor category (UMTC), which we denote by $\mathscr C$ (see appendix \ref{app:UMTC}).   Its simple objects will be denoted by labels $a\in\mathcal I$; when needed, we write $X_a$ for the corresponding object, but in what follows we will freely identify objects with their labels. These labels correspond to conformal families or superselection sectors, and the bulk Hilbert space decomposes as
\begin{align}\label{eq:Hilbertspace}
\mathcal H = \bigoplus_{a\in\mathcal I}\mathcal H_a = \bigoplus_{a\in\mathcal I} a \otimes \bar a\,.
\end{align}
Each label $a\in I$ is a superselection label specifying a chiral irreducible representation $\mathcal H_a$ of the chiral algebra.

The tensor product of two chiral sectors decomposes according to the fusion rules of the category.  More precisely, one has the canonical decomposition
\begin{align}\label{eq:fusion}
a \otimes b \;\cong\; \bigoplus_{c\in\mathcal I} c \otimes V_{ab}^{\,c}\,,
\end{align}
where the multiplicity spaces
\begin{align}\label{eq:fusionspace}
V_{ab}^{\,c}= \mathrm{Hom}_{\mathscr C}(X_a \otimes X_b, X_c)\,,\quad\dim V_{ab}^{\,c} = N_{ab}^{\,c}
\end{align}
are the fusion spaces of the theory.  Their elements are junction implementing the elementary fusion process $a \otimes b \to c$ and these will form a central ingredient\footnote{ One might instead consider closed Verlinde lines as candidate gates. However, the action of a closed Verlinde line labelled by $a$ on a chiral sector $b$ is diagonal and given by multiplication by the modular $S$-matrix eigenvalue $S_{ab}/S_{0b}$. As a result, Verlinde lines act within each conformal family and cannot change the sector label. See eq. \eqref{eq:diag_action} in appendix \ref{app:UMTC}.} for the realisation of the non-invertible Nielsen gates.

Concretely, in what follows we construct:
(\textit{i}) a canonical Stinespring isometry whose matrix elements are given by a choice of orthonormal intertwiners in $V_{ab}^{\,c}$;
(\textit{ii}) the associated Kraus operators and the induced completely positive, trace-preserving channel acting between sector Hilbert spaces;
(\textit{iii}) the positive operator-valued measure (POVM) describing post-selection onto a definite fusion channel $c$ when junction degrees of freedom are discarded; and
(\textit{iv}) a consistency check showing that successive fusion operations compose associatively, with Kraus operators related by the associator ($F$-symbols) of the underlying unitary modular tensor category. We will subsequently show that fusion gates verify an intrinsic order dependence with respect to invertible circuit segments.

\subsection{Fusion as a non-invertible gate operation}

Fusion junctions implement sector-changing operations: they take a state in a conformal family $a$ to a state in a (generally different) family $c$ after fusing in a defect label $b$. Unlike the invertible gates of the Nielsen framework, such processes are intrinsically non-unitary: a single fusion step may have several admissible output sectors, and no canonical inverse operation exists. To incorporate fusion into a circuit framework, we must therefore keep track of this multiplicity of outcomes in a controlled way. Rather than representing fusion by a single operator on a fixed Hilbert space, we first describe it as a map into an enlarged space that records the fusion-channel data. This enlarged description will later allow us to define both composition of fusion gates and the selection of definite fusion outcomes in a precise and systematic manner.

To construct conformal-family-changing operations, we use local defect junction fields that mediate the fusion of topological lines. In a rational CFT described by a unitary modular tensor category $\mathscr C$, these endpoints are encoded categorically by fusion junctions.
For simple objects $X_a,X_b,X_c\in\mathcal C$, the vector space $V_{ab}^c$ as defined in \eqref{eq:fusionspace} collects the allowed ways in which defects of type $a$ and $b$ can fuse into a defect of type $c$. An element $t_{ab\to c}^{(\mu)} \in V_{ab}^{\,c}$ is a junction operator (intertwiner), with $\mu=1,\dots,N_{ab}^{\,c}$ labelling a basis of the fusion space. A junction operator may be visualised as a local defect junction: two topological defect lines labelled by $a$ and $b$ approach each other and merge into a single defect of type $c$. Diagrammatically, this process is represented by a trivalent vertex,
\begin{align}\label{eq:intertwiner}
t_{ab\to c}^{(\mu)}:\quad
\raisebox{-20pt}{\includegraphics[scale=0.102]{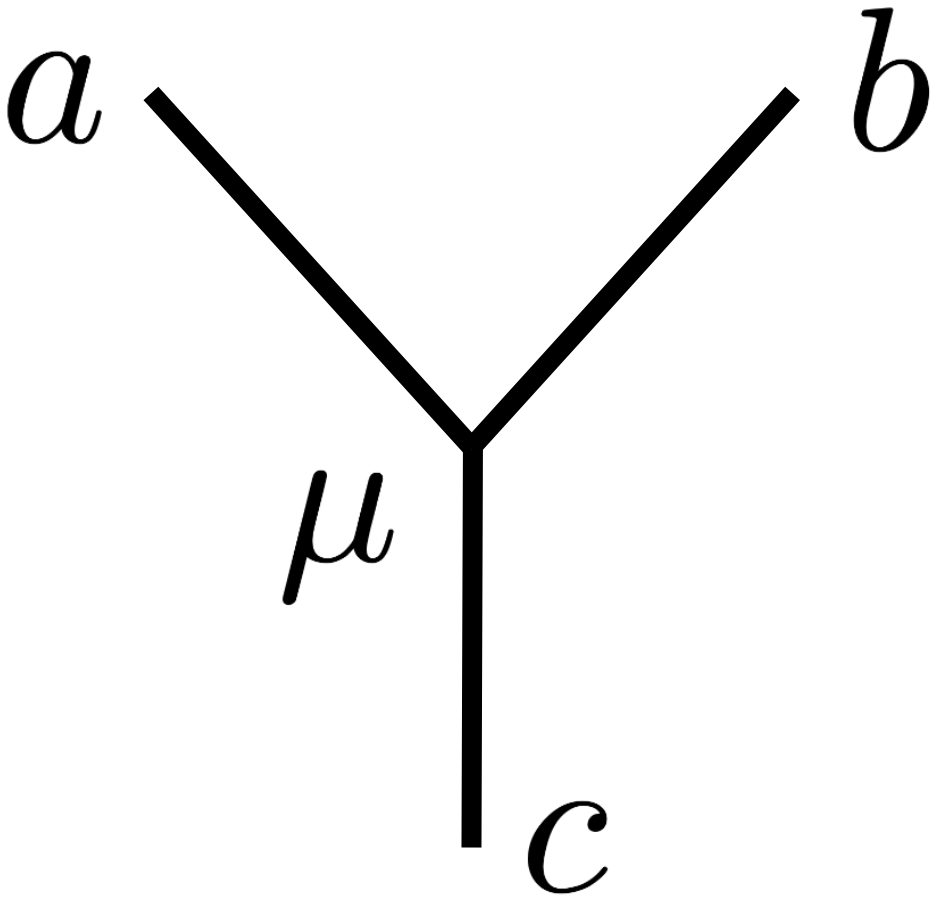}}\quad\sim \quad\raisebox{-20pt}{\includegraphics[scale=0.102]{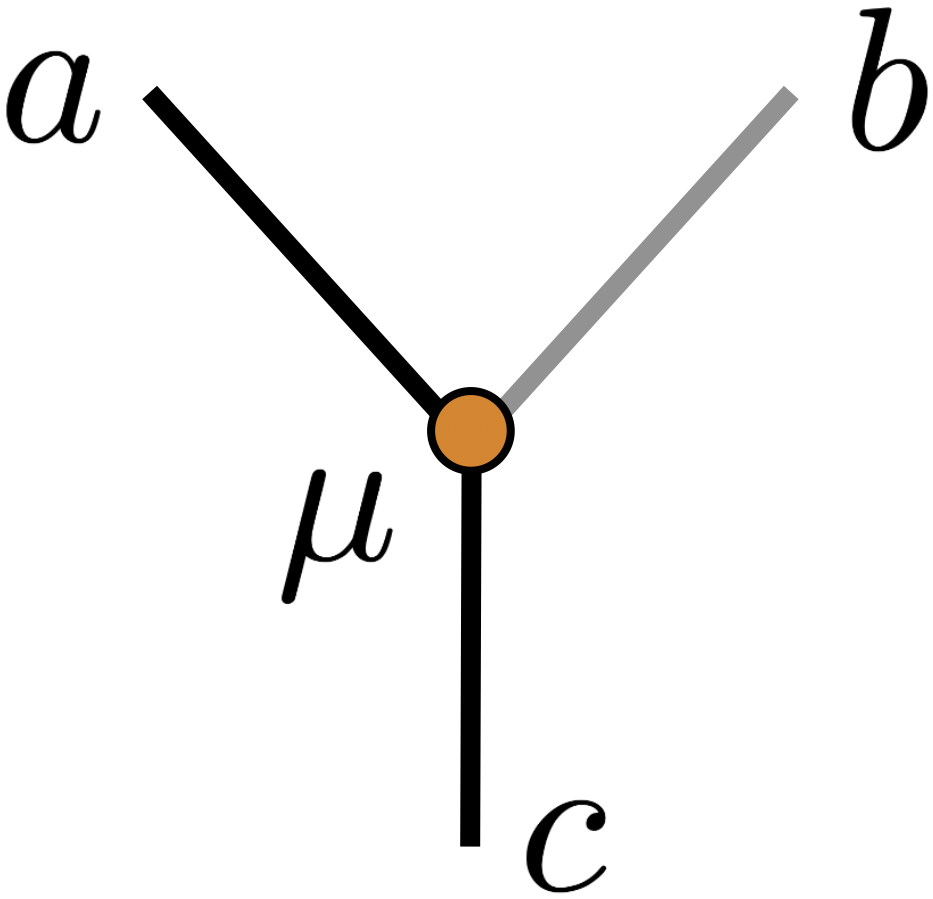}}
\end{align}
with the label $\mu$ accounting for possible degeneracies of the junction. Anticipating the circuit interpretation developed below, we will later regard one incoming leg as the gate label and the other as the input state, so that the junction acts as a local operation mapping an initial sector to a new one, as suggested by the right-hand diagram.

To realise fusion as a well-defined non-invertible circuit element, we will implement it via a Stinespring dilation, i.e. as an isometric map into an enlarged Hilbert space. The key structural input from the unitary modular tensor category is therefore a canonical Hermitian structure on each fusion space \eqref{eq:fusionspace}. For a fixed orthonormal choice of junction basis $\{t_{ab\to c}^{(\mu)}\}$, the pivotal structure\footnote{See appendix \ref{app:UMTC} for details on the relevant structure and properties of UMTCs.} implies the completeness relation
\begin{equation}\label{eq:junction_completeness_clean}
\sum_{c,\mu}\frac{d_c}{d_a d_b}\, \big(t_{ab\to c}^{(\mu)}\big)^{\dagger}\, t_{ab\to c}^{(\mu)} = \mathbf 1_{a\otimes b}\,,
\end{equation}
where $d_a$ denotes the quantum dimension of $X_a$, see eq. \eqref{eq:quantumdim}, and the adjoint is taken with respect to the pivotal structure. This identity is the categorical input for the Stinespring construction of the fusion gate in section \ref{sec:gateisometry}. Details of the trace conventions are summarised in appendix \ref{app:UMTC}.

The purpose of introducing these junction operators is to build non-invertible gates that map between different conformal families. Concretely, for fixed input sector $a$ and gate label $b$ we will construct in section \ref{sec:gateisometry} an isometry of the schematic form
\begin{align}
\mathcal H_a \xrightarrow{\;W_{ab}\;} \bigoplus_{c:\,N_{ab}^c\neq0}\,\mathcal H_c \otimes V_{ab}^{c}\,,
\end{align}
which retains all admissible fusion channels simultaneously. The precise composition law for successive fusion gates, and its relation to the associativity encoded in the $F$-symbols of the category, will be discussed in section \ref{sec:assoc}.

Finally, fusion gates do not act transitively on the set of sectors: from a given conformal family $a$, only those families $c$ with $N_{ab}^{\,c}\neq 0$ for some allowed gate label $b$ are reachable. We will return to the resulting graph-like optimisation problem, and to its role in defining total circuit cost, once the gate and selection costs have been introduced in section \ref{sec:extendedcosts}.

To make these constructions fully explicit, we now illustrate them in the Ising category, which will serve as a recurrent and basic example throughout the remainder of this work.

\paragraph{Example: Ising fusion.}
For concreteness, consider the Ising CFT. The corresponding unitary modular tensor category has three simple objects $\mathbf 1$, $\sigma$ and $\psi$, with fusion rule
\begin{align}
\sigma\otimes\sigma = \mathbf 1 \oplus \psi\,.
\end{align}
Both fusion spaces $V_{\sigma\sigma}^{\,\mathbf 1}$ and $V_{\sigma\sigma}^{\,\psi}$ are one-dimensional. Choosing orthonormal junction operators
\begin{align}
t_{\sigma\sigma\to \mathbf 1}\in \mathrm{Hom}(\sigma\otimes\sigma,\mathbf 1)\,,\qquad
t_{\sigma\sigma\to \psi}\in \mathrm{Hom}(\sigma\otimes\sigma,\psi)\,,
\end{align}
the completeness relation \eqref{eq:junction_completeness_clean} reduces to
\begin{align}
\frac{1}{2}\, t_{\sigma\sigma\to \mathbf 1}^{\dagger} t_{\sigma\sigma\to \mathbf 1} +\frac{1}{2}\, t_{\sigma\sigma\to \psi}^{\dagger} t_{\sigma\sigma\to \psi} = \mathbf 1_{\sigma\otimes\sigma}\,,
\end{align}
using $d_{\mathbf 1}=d_{\psi}=1$ and $d_\sigma=\sqrt2$.

\subsection{Fusion channels as quantum operations}\label{sec:gateisometry}

We now make the fusion structure described above precise by embedding it into the framework of quantum operations. In this language, a non-invertible process is realised as an isometric embedding into a larger Hilbert space, where auxiliary degrees of freedom keep track of information that is not retained in the final state. This provides a natural circuit-theoretic description of irreversible operations. In the present setting, these auxiliary degrees of freedom are identified with the fusion multiplicity spaces $V_{ab}^{\,c}$. Retaining them yields an isometric representation of fusion that keeps track of all admissible channels simultaneously, while discarding or projecting onto them produces completely positive, trace-preserving maps acting between the physical sector Hilbert spaces. We now construct this embedding explicitly and identify the resulting fusion gate.  For completeness, a brief review of the quantum-circuit and quantum-channel concepts used in this construction (such as Stinespring dilations, Kraus representations, and completely positive trace-preserving maps) is provided in appendix \ref{app:qc_basics}.

To represent fusion by a fixed defect label $b$ as a gate acting on $\mathcal H_a$, we introduce an ancilla space that records the fusion channel,
\begin{align}\label{eq:ancilla_space}
\mathcal A_{ab}=\bigoplus_{c} V_{ab}^{\,c}\,,
\end{align}
with orthonormal basis vectors $|\mu;c\rangle$. Let $t_{ab\to c}^{(\mu)} \in \mathrm{Hom}(a\otimes b,c)$, $\mu=1,\dots,N_{ab}^{\,c}$, denote an orthonormal basis of fusion intertwiners with respect to the pivotal inner product. We will repeatedly use the completeness relation \eqref{eq:junction_completeness_clean}, which is the categorical input ensuring that the Stinespring map defined below is an isometry.

\begin{figure}[t]
    \centering
    \includegraphics[width=0.42\linewidth]{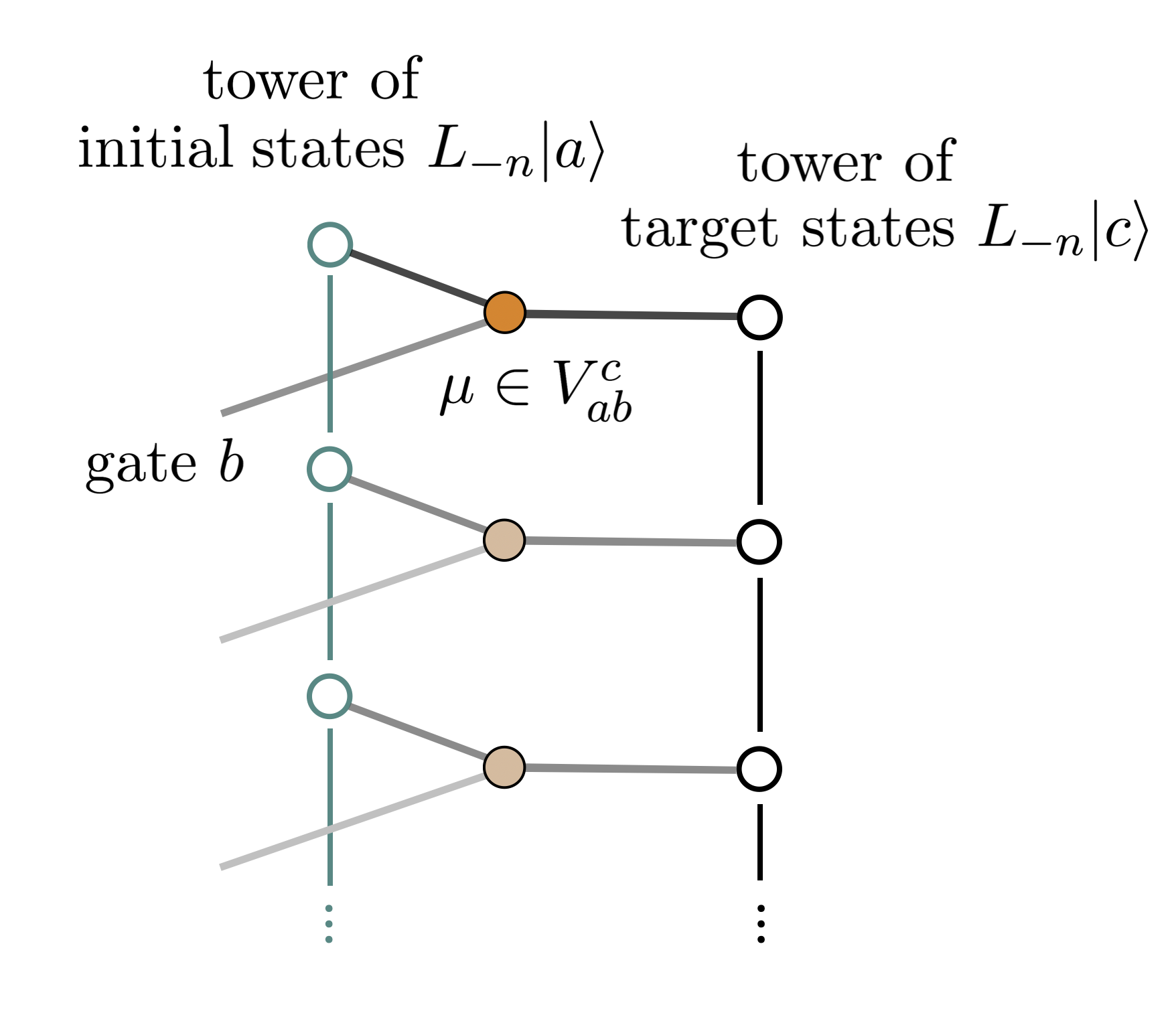}
    \caption{Schematic action of a fusion gate on Virasoro representations. A gate labelled by $b$ acts on an initial conformal family $a$, mapping it to admissible output families $c$ through fusion channels labelled by $\mu\in V^c_{ab}$. Vertical columns represent full Virasoro modules (entire towers of descendants), not a level-by-level map. The channel label $\mu$ encodes fusion data and is retained as an auxiliary degree of freedom in the Stinespring construction.}
    \label{fig:cartoon_fusion_process_small}
\end{figure}


We define the fusion gate as the Stinespring isometry
\begin{gather}
\begin{aligned}\label{eq:isometry}
&\qquad W_{ab}:\ \mathcal H_a \longrightarrow \bigoplus_c \mathcal H_c \otimes V_{ab}^{\,c}\,, \\ 
&W_{ab}|\psi\rangle_a = \sum_{c,\mu}\sqrt{\frac{d_c}{d_a d_b}}\, \big(t_{ab\to c}^{(\mu)}|\psi\rangle_a\big)\otimes|\mu;c\rangle\,.
\end{aligned}
\end{gather}
The choice of $|\phi_b\rangle$ is a fixed convention (analogous to fixing an ancilla initial state in Stinespring dilation) and will not affect any basis-independent quantities constructed later such as the channel weights \eqref{eq:expr_weigts}.

\begin{figure}[t]
    \centering
    \includegraphics[width=0.6\linewidth]{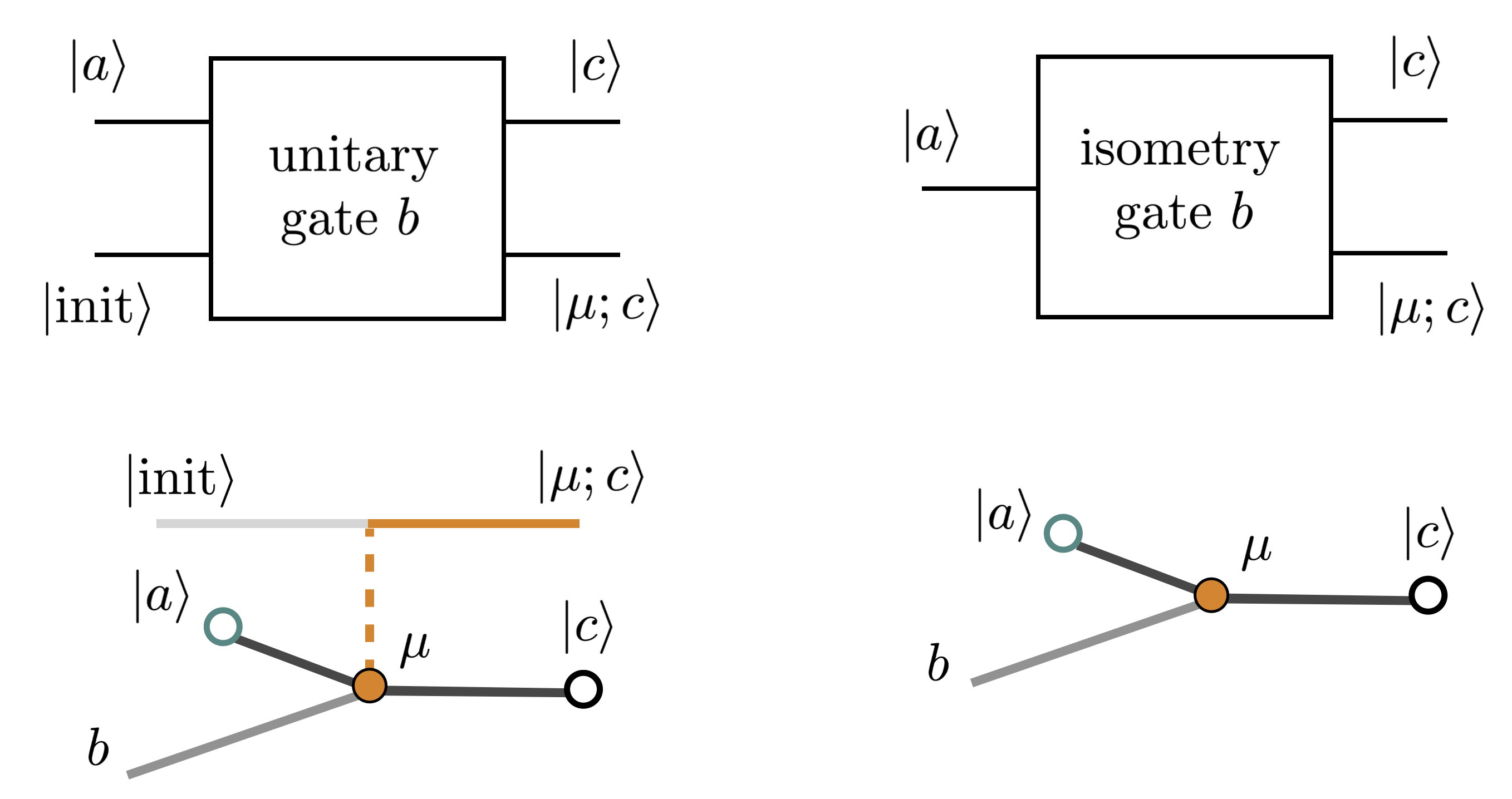}
    \caption{Schematic comparison between a unitary gate and a fusion-induced isometric gate. Fusion enlarges the Hilbert space by junction degrees of freedom, which become an ancilla in the Stinespring representation.}
    \label{fig:fig_unitary_isometry}
\end{figure}

At the circuit level, fusion with a simple object $b$ therefore defines a non-invertible gate acting on $\mathcal H_a$. The action of a fusion gate on entire Virasoro representations is illustrated schematically in figure \ref{fig:cartoon_fusion_process_small}. We denote this gate by $D_b$ and identify
\begin{align}
D_b\big|_{\mathcal H_a} \equiv W_{ab}\,.
\end{align}
The label $b$ specifies the gate type, while the index $a$ keeps track of the sector on which the gate acts. Using the orthonormality and completeness of the fusion intertwiners, \eqref{eq:junction_completeness_clean}, one finds
\begin{align}\label{eq:isometry_rel_W}
W_{ab}^\dagger W_{ab} = \mathbf 1_{\mathcal H_a}\,,
\end{align}
so $W_{ab}$ is an isometry. In general $W_{ab}W_{ab}^\dagger\neq\mathbf 1$ on the output space: fusion enlarges the Hilbert space by junction degrees of freedom and is not reversible. This intrinsic irreversibility is the circuit-theoretic reflection of the categorical fact that while $a\otimes b$ decomposes into simple objects, no single simple object canonically reconstructs $a\otimes b$. The contrast between an invertible unitary gate and a fusion-induced isometric gate with auxiliary junction degrees of freedom is illustrated schematically in figure \ref{fig:fig_unitary_isometry}.

Choosing the basis $\{|\mu;c\rangle\}$ of the ancilla space, the associated Kraus operators are obtained by projecting onto definite fusion channels,
\begin{align}\label{eq:Kraus}
K_{c,\mu}^{(a,b)} = (\mathbf 1\otimes\langle \mu;c|)\,W_{ab} = \sqrt{\frac{d_c}{d_a d_b}}\; t_{ab\to c}^{(\mu)} :\ \mathcal H_a\to\mathcal H_c\,.
\end{align}
They satisfy
\begin{align}\label{eq:krausid}
\sum_{c,\mu} (K_{c,\mu}^{(a,b)})^\dagger K_{c,\mu}^{(a,b)} =\mathbf 1_{\mathcal H_a}\,,
\end{align}
and therefore define a canonical completely positive, trace-preserving map associated with fusion by $b$ upon discarding the ancilla. Different orthonormal choices of the intertwiners $t_{ab\to c}^{(\mu)}$ correspond to unitarily equivalent Kraus representations of the same quantum channel. In particular, changes of basis in the fusion spaces $V_{ab}^{\,c}$ act by unitary
transformations on the Kraus index space and leave the underlying completely positive, trace-preserving map invariant. In the next subsection we use this structure to define channel selection and post-selection, which will play a central role in the definition of fusion costs.

\paragraph{Example: Ising category.}
For the Ising category, taking $a=b=\sigma$ gives $\sigma\otimes\sigma=\mathbf 1\oplus\varepsilon$, with $N_{\sigma\sigma}^{\mathbf 1}=N_{\sigma\sigma}^{\varepsilon}=1$. The corresponding ancilla space \eqref{eq:ancilla_space} is
\begin{align}
\mathcal A_{\sigma\sigma}\cong \mathbb C|\mathbf 1\rangle \oplus \mathbb C|\varepsilon\rangle\,.
\end{align}
The Stinespring isometry \eqref{eq:isometry} then acts as
\begin{align}
W_{\sigma\sigma}|\psi\rangle_\sigma = \frac{1}{\sqrt 2}\, t_{\sigma\sigma\to\mathbf 1}|\psi\rangle_\sigma \otimes|\mathbf 1\rangle + \frac{1}{\sqrt 2}\, t_{\sigma\sigma\to\varepsilon}|\psi\rangle_\sigma \otimes|\varepsilon\rangle\,.
\end{align}
The associated Kraus operators, see \eqref{eq:Kraus}, are
\begin{align}
K_{\mathbf 1} = \frac{1}{\sqrt 2}\,t_{\sigma\sigma\to\mathbf 1}\,,
\qquad K_{\varepsilon} = \frac{1}{\sqrt 2}\,t_{\sigma\sigma\to\varepsilon},
\end{align}
and the completeness relation \eqref{eq:krausid} reads
\begin{align}
K_{\mathbf 1}^\dagger K_{\mathbf 1}+K_{\varepsilon}^\dagger K_{\varepsilon}=\mathbf 1_{\mathcal H_\sigma}\,,
\end{align}
realising fusion by $\sigma$ as a canonical non-invertible quantum operation.

\subsection{Quantum channel interpretation and fusion outcome selection}
\label{sec:quantumchannels_sel}

Fusion in a unitary modular tensor category determines, for each pair of simple objects $(a,b)$, a finite set of output sectors $c$ with multiplicities $N_{ab}^{\,c}$. Because of this, at first sight as a circuit element, fusion is not unitary. Instead, implementing it
requires a quantum-channel realisation and, separately, a prescription for how one
conditions on a definite fusion outcome. Both structures are naturally provided by the Stinespring realisation of fusion gates constructed in section \ref{sec:gateisometry}.

Concretely, recall that fusion by $b$ is realised by the isometry $W_{ab}:\mathcal H_a\to\bigoplus_c \mathcal H_c\otimes V_{ab}^{\,c}$ of section \ref{sec:gateisometry}. If the auxiliary degrees of freedom are kept, the fusion operation naturally produces a superposition over all admissible fusion channels. By contrast, projecting onto a fixed auxiliary subspace isolates a definite fusion outcome $c$. The goal of this subsection is to make these two operations precise and extracts the canonical channel weights that will enter the definition of selection costs.

For two simple objects $a,b\in\mathcal I$, interpreted respectively as input sector and gate label, the auxiliary space $\mathcal A_{ab}$ defined in eq. \eqref{eq:ancilla_space} decomposes by fusion channel,
$\mathcal A_{ab}=\bigoplus_c V_{ab}^{\,c}$. The corresponding orthogonal projectors are
\begin{align}
\Pi_c=\sum_{\mu}|\mu;c\rangle\langle\mu;c|\,,
\end{align}
which resolve the identity on $\mathcal A_{ab}$, $\sum_c\Pi_c=\mathbf 1_{\mathcal A_{ab}}$. Composing these projectors with the Stinespring isometry defines a family of positive operators acting on the input sector,
\begin{align}\label{eq:pos_ops}
E_{ab}^{\,c}
= W_{ab}^\dagger(\mathbf 1\otimes\Pi_c)W_{ab}
= \frac{d_c}{d_a d_b}\sum_{\mu}
\big(t_{ab\to c}^{(\mu)}\big)^\dagger t_{ab\to c}^{(\mu)}\,.
\end{align}
By construction,
\begin{align}
E_{ab}^{\,c}\ge0\,,\qquad \sum_c E_{ab}^{\,c}=\mathbf 1_{\mathcal H_a}\,,
\end{align}
so the set $\{E_{ab}^{\,c}\}$ defines a positive operator-valued measure on $\mathcal H_a$ associated with fusion of $a$ and $b$.

The operators $E_{ab}^{\,c}$ encode the contribution of each fusion channel $c$ to the effective action of the fusion gate on $\mathcal H_a$, after the auxiliary junction degrees of freedom have been traced out. Taking the normalised categorical (quantum) trace on $\mathrm{End}(a)$, defined by\footnote{Here and in the following, $\mathrm{tr}$ denotes the categorical (quantum) trace on endomorphisms of the object a, normalised such that $\mathrm{tr}(1_a)=d_a$. This trace is intrinsic to the fusion category and should not be confused with the operator trace on the infinite-dimensional CFT Hilbert space. See appendix \ref{app:UMTC}.} $\mathrm{tr}(1_a)=d_a$, yields
\begin{align}\label{eq:expr_weigts}
p(c\,|\,a,b) = \frac{\mathrm{tr}(E_{ab}^{,c})}{\mathrm{tr}(\mathbf 1_{\mathcal H_a})} = \frac{N_{ab}^{\,c}\,d_c}{d_a d_b}\,,
\end{align}
which depends only on fusion multiplicities and quantum dimensions. Whenever a fusion process $(a,b)$ is fixed, we will write $p_c$ as shorthand for these canonical channel weights. We emphasise that $p(c\,|\,a,b)$ is a canonical weight determined by categorical data (multiplicities and quantum dimensions); it is not taken to be a probability.

It is important to distinguish basis-dependent from basis-independent structures. While the individual Kraus operators $K_{c,\mu}^{(a,b)}$ defined in eq. \eqref{eq:Kraus} depend on a choice of orthonormal basis in each fusion space $V_{ab}^{\,c}$, the operators $E_{ab}^{\,c}$ obtained by summing over $\mu$ are canonical. In particular, the weights $p(c\,|\,a,b)$ are invariant under changes of basis in the fusion spaces.

At this point, two conceptually distinct operations should be kept separate. One may either ignore the junction data entirely, leading to a fusion channel in which all admissible outcomes are retained and act collectively on $\mathcal H_a$, or one may explicitly select a definite fusion outcome $c$ by projecting onto the corresponding auxiliary subspace. The latter operation is intrinsically irreversible and will be assigned an additional cost in defining the complexity in section \ref{sec:costs}.

Finally, because the construction is expressed entirely in terms of categorical fusion data, it is compatible with the associativity structure of the unitary modular tensor category. In particular, different ways of composing successive fusion gates lead to equivalent quantum channels, with their equivalence implemented by the $F$-symbols. In the next subsection we make this statement explicit and show how associativity is realised at the level of non-invertible circuit elements.

\subsection{Associativity and ordering of non-invertible fusion gates}
\label{sec:assoc_order}

In this section, we consider two central properties of extended Nielsen circuits that include non-invertible gates, and their consistent embedding in the quantum circuit realised built in the previous sections. 

First, successive fusion operations must give rise to a well-defined quantum channel that is independent of how intermediate fusion steps are grouped. Although associativity of fusion is guaranteed at the level of categorical data, it is nontrivial that different fusion trees lead to the same completely positive trace-preserving map on the Hilbert space. This consistency must be verified explicitly. Second, fusion gates are isometric but not unitary, and therefore cannot be freely reordered with invertible circuit segments.

\begin{figure}[t]
    \centering
    \includegraphics[width=0.78\linewidth]{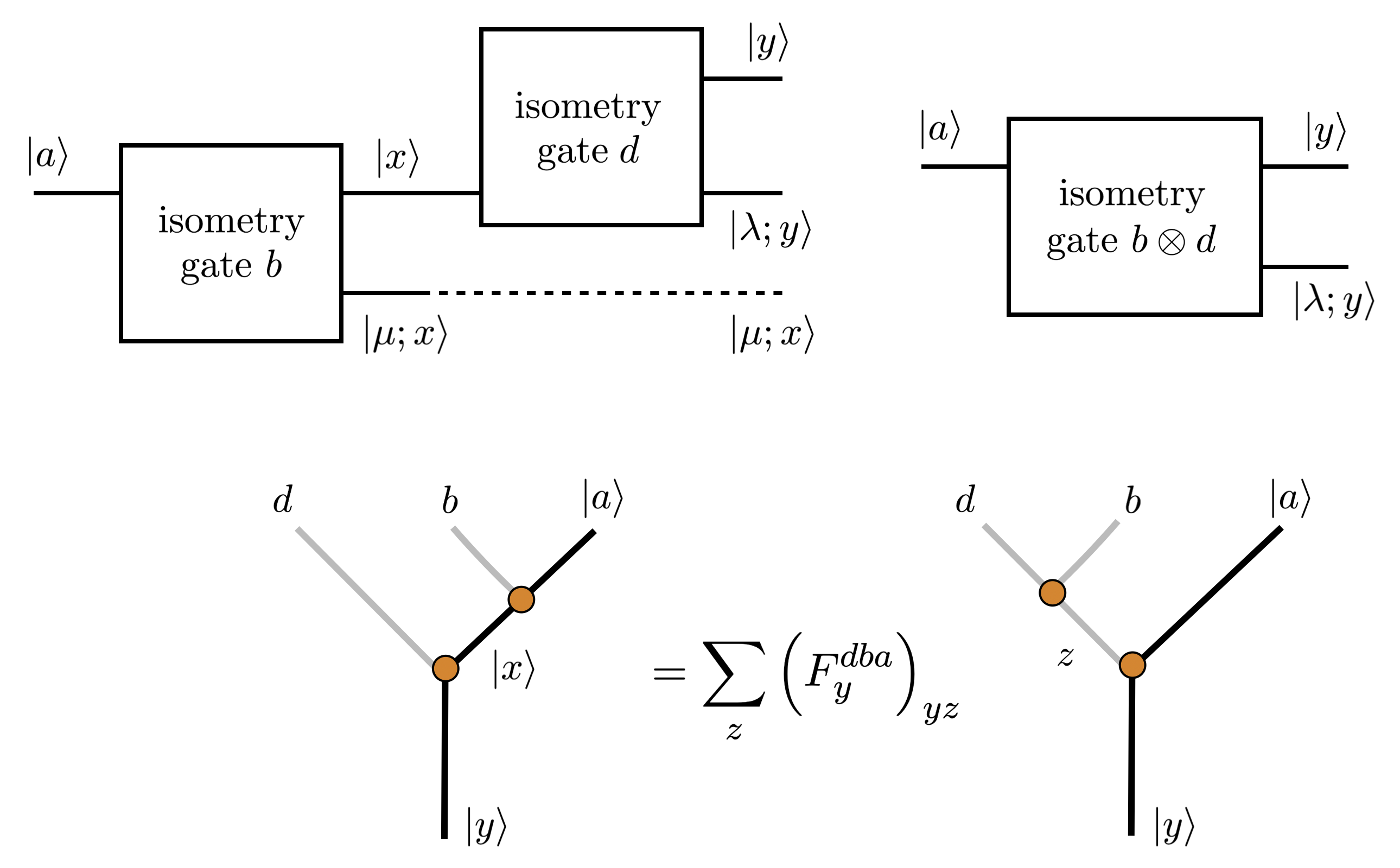}
    \caption{The dark black line is the ``evolution'' of the state, while the gray-scale lines denote gates. Every non-invertible action of a gate $b$ on a state $|a\rangle$ determines generically multiple outcomes $c$ such that $N_{ab}^c\neq 0$. Through the associators, one can construct two path from an initial state $|a\rangle$ to a target state $|y\rangle$. Note that on the RHS of the bottom figure, one first fuses two gates before applying one branch to the state $|a\rangle$. Above, the mirror circuit process.} 
    \label{fig:associator_circuits}
\end{figure}

\subsubsection{Associativity of non-invertible channels}\label{sec:assoc}

Successive fusion gates can be composed in different parenthesisations, and hence admit different fusion trees, see figure \ref{fig:associator_circuits}. For a circuit interpretation, the relevant consistency requirement is not merely associativity of fusion multiplicities, but the stronger statement that all parenthesisations define the same completely positive, trace-preserving (CPTP) map on the input sector Hilbert space. Each fusion tree determines a Stinespring isometry and therefore a specific Kraus representation of the induced quantum channel. Distinct parenthesisations correspond to different choices of intermediate channel labels carried by the ancilla. The nontrivial statement is that these different Kraus sets are related by a unitary change of basis on the Kraus index space, so that the resulting CPTP map is independent of how intermediate fusion steps are bracketed.

Concretely, consider two sequential fusion gates labelled by $b$ and $c$ acting on an input sector $a$, with total output $y$. The parenthesisation $((a\otimes b)\otimes c)\to y$ yields Kraus operators naturally indexed by a basis of $\bigoplus_x V_{ab}^{x}\otimes V_{x c}^{y}$, while the alternative $(a\otimes(b\otimes c))\to y$ yields Kraus operators indexed by a basis of $\bigoplus_z V_{bc}^{z}\otimes V_{a z}^{y}$. As shown in appendix \ref{app:assoc}, these two index spaces are related by the associator,
\begin{align}
F^{abc}_y:\ \bigoplus_x V_{ab}^{x}\otimes V_{x c}^{y}\;\xrightarrow{\ =\ }\;
\bigoplus_z V_{bc}^{z}\otimes V_{a z}^{y}\,,
\end{align}
which is unitary with respect to the pivotal inner product. As a result, the two Kraus representations are related by
\begin{align}\label{eq:Kraus_F_move}
M_{\beta} \;=\; \sum_{\alpha} F^{abc}_{y}[\beta,\alpha]\; L_{\alpha}\,,
\end{align}
where $\alpha$ and $\beta$ denote composite Kraus labels in the two parenthesisations (spelling these indices out explicitly is given in appendix \ref{app:assoc}). Since a unitary change of Kraus basis leaves the underlying CPTP map invariant, both fusion trees define the same quantum channel. Thus, successive non-invertible fusion gates compose associatively at the level of quantum channels, independently of how intermediate fusion steps are bracketed.

We emphasise that this statement concerns the fusion channel prior to any restriction to a specific outcome. The role of outcome selection and its associated cost will be discussed in section \ref{sec:costs}.

\subsubsection{Order dependence with respect to invertible evolution}
This subsection addresses a second feature of non-invertible fusion gates: their ordering relative to invertible circuit segments matters. While associativity ensures that successive fusion operations define a well-defined quantum channel, no such simplification occurs when fusion gates are interspersed with invertible evolutions.

Let $U(\tau)=\exp(-i Q\,\tau)$ denote an invertible circuit segment generated by some Hermitian generator $Q$, and let $D_a$ be a fusion gate realised as the isometric map defined in section \ref{sec:noninv_gates}. As an immediate consequence of eq. \eqref{eq:isometry_rel_W}, one has that 
\begin{align}
D_a^\dagger D_a=\mathbf 1\,,\qquad D_a D_a^\dagger\neq\mathbf 1\,.
\end{align}
In general, invertible and non-invertible operations do not commute:
\begin{align}
U(\tau)\,D_a \neq D_a\,U(\tau)\,.
\end{align}
The two circuits $U(\tau)D_a|\psi\rangle$ and $D_aU(\tau)|\psi\rangle$ therefore represent distinct boundary operations, and (when a bulk interpretation is available) induce inequivalent bulk data. Fusion gates cannot be absorbed into an invertible circuit segment, and their placement within a circuit is physically meaningful (see fig. \ref{fig:associator_circuits}). As a result, circuit equivalence under reordering breaks down once fusion gates are present, and circuit cost becomes sensitive to their precise placement along the invertible evolution.

Taken together with the associativity result of the previous subsection, this shows that fusion gates define operations that are well defined under composition but not group-like. Their intrinsic order dependence motivates a cost functional that distinguishes invertible circuit segments from non-invertible ones, and justifies treating fusion gates as genuinely new circuit elements rather than as special unitary deformations.

\section{Circuit geometry and cost for non-invertible gates}\label{sec:costs}

As reviewed in section \ref{sec:reviewNielsen}, in the standard Nielsen formulation applied to Virasoro circuits, the notion of cost admits a direct geometric interpretation: the circuit explores a continuous Virasoro coadjoint orbit, and complexity is measured by the length of a path with respect to a right-invariant metric. The extension to non-invertible gates constructed in the previous section necessarily departs from this picture. Fusion does not generate a continuous flow on a single orbit, but instead induces jumps between distinct coadjoint orbits, corresponding to changes of conformal family. A single gate application generically produces a direct-sum decomposition over several such transitions, rather than a single channel outcome. 

As a result, there is no underlying smooth geometry analogous to the Virasoro orbit geometry governing the full dynamics. Nevertheless, the categorical structure induces a different geometric structure. The action of a fusion gate is characterised by a finite set of fusion channels $c$, each associated with a multiplicity space and a canonical weight determined by quantum dimensions. Taken together, they define a vector in a projective space, on which the Fubini-Study metric provides a natural, basis-independent notion of distance between fusion outcomes.

Two additional sources of cost must then be distinguished. First, since Nielsen complexity is defined relative to fixed initial and final states, isolating a definite fusion outcome requires a post-selection operation, whose cost must be accounted for separately. Second, anticipating on the application in section \ref{sec:shocklike_def}, where we interpret fusion-induced jumps as shock-like defects in three-dimensional gravity, the cost must be refined to reflect the associated change in energy or conformal dimension. The following will introduce these three contributions in turn and clarify their respective roles.

Throughout this work we distinguish between single-channel fusion, where the fusion rules admit a unique outcome, and multi-channel fusion, where several outcomes are allowed. In the latter case, reaching a definite target sector requires an explicit channel-selection operation.

\subsection{Geometrical costs}\label{sec:geomcostfunction}\label{sec:proposed_cost}
We now define the intrinsic cost associated with a single non-invertible fusion gate. Since a fusion gate produces a finite set of channels rather than a unique outcome, its action is characterised by a vector of channel weights determined by categorical data. This induces a natural projective geometry on the space of fusion outcomes, allowing for an simple definition of gate cost.

Indeed, given a fusion process \eqref{eq:fusiondef} with channels labelled by $c$, one can associated to it a vector
\begin{align}\label{eq:ray}
    \varphi = (\sqrt{p_1},\sqrt{p_2},\dots,\sqrt{p_N}) \in \mathbb C^N\,,
\end{align}
where the weights $p_c$ are determined by the fusion data in eq. \eqref{eq:expr_weigts} and satisfy $\sum_c p_c = 1$. As stressed earlier, these weights should be understood as representation-theoretic measures associated with fusion channels, rather than as fundamental probabilities. By construction, $\|\varphi\|^2 = \sum_c (\sqrt{p_c})^2 = 1$, so $\varphi$ lies on the unit sphere in $\mathbb C^N$, restricted to the positive orthant. Since the overall phase of $\varphi$ is irrelevant, fusion outcomes are naturally identified with rays $[\varphi]$, and hence with points in complex projective space $ \mathbb C\mathbb P^{N-1} \cong U(N)\big/\left(U(1)\times U(N{-}1)\right).$. This projective space carries a canonical metric, the Fubini-Study (FS) metric, which provides a basis-independent notion of distance between rays. For a normalised vector $\varphi$, the FS line element reads
\begin{align}
\label{eq:FS}
    \mathrm ds_{\rm FS}^2 = \langle \mathrm d\varphi, \mathrm d\varphi \rangle  - |\langle \varphi, \mathrm d\varphi \rangle|^2\,.
\end{align}
Restricted to the positive real locus $\varphi_c=\sqrt{p_c}\in\mathbb R_{\ge0}$ with $\sum_c p_c=1$,  the Fubini-Study metric reduces to
\begin{align}\label{eq:metric_geom}
    \mathrm d\varphi_c = \frac{1}{2}\frac{\mathrm dp_c}{\sqrt{p_c}}\,,\qquad \mathrm ds_{\rm FS}^2 = \frac14 \sum_c \frac{(\mathrm dp_c)^2}{p_c}\,.
\end{align}

\begin{figure}[t]
    \centering
    \includegraphics[width=0.5\linewidth]{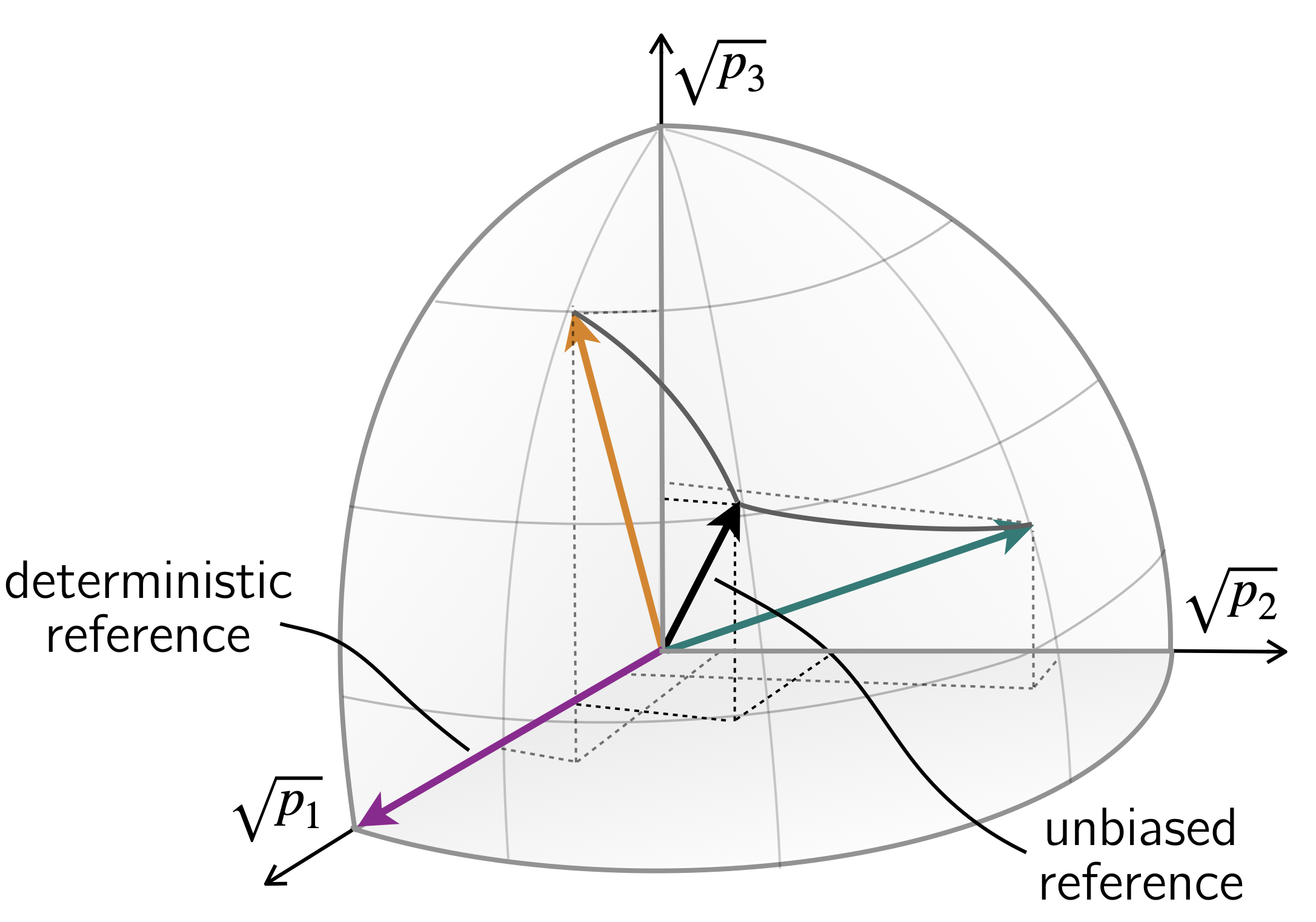}
    \caption{Depiction of the intrinsic gate cost for a fusion process with three channels. Vectors on the unit sphere represent normalised weight vectors $(\sqrt{p_1},\sqrt{p_2},\sqrt{p_3})$. The black unit vector denotes the unbiased reference ray corresponding to the uniform distribution over channels, while the purple vector indicates a deterministic reference in which a single channel is selected. The green and orange unit vectors illustrate two example fusion outcomes with non-uniform channel weights. The intrinsic gate cost is indicated by the solid arcs.}
    \label{fig:octant}
\end{figure}

More generally, the FS geometry allows one to define a finite distance between two fusion processes characterised by weight vectors $\{p_c\}$ and $\{q_c\}$. The Fubini-Study distance between two rays is defined as the length of the unique geodesic of this metric connecting them. On $\mathbb C\mathbb P^{N-1}$ this geodesic distance admits a closed-form expression in terms of representatives of the rays, given by the angle determined by their Hilbert-space overlap. The distance between two fusion processes characterised by weights $\{p_c\}$ and $\{q_c\}$ is
\begin{align}
    \mathrm{dist}_{\rm FS}([\varphi],[\phi]) =\arccos\left|\langle \varphi,\phi\rangle\right|\,.
\end{align}

To assign a gate cost to a single fusion process, one must specify a reference ray
with respect to which distances are measured.  For a non-selective fusion gate,
the natural reference configuration is the maximally unbiased distribution on the
allowed fusion channels.  Concretely, if fusion of $a\otimes b$ admits $N$ channels,
we take the reference weights to be $q_{\mathrm{ref},c} = \frac{1}{N}$, with $c=1,\dots,N$, corresponding to the ray
\begin{align}
    \phi_{\rm ref} = \left(\tfrac{1}{\sqrt N},\dots,\tfrac{1}{\sqrt N}\right)\,.
\end{align}
This reference ray represents a fusion gate that branches symmetrically among all
allowed channels and therefore carries no intrinsic geometric bias.  The geometric
gate cost is then defined as the Fubini-Study distance between the actual fusion
ray $[\varphi]$ and this unbiased reference ray.
We are thus led to the gate cost, in terms of fusion weights, given by 
\begin{align}
\label{eq:geomcost}
    \mathcal C_{\rm gate}^\mathrm{geom}(a,b) = \arccos\left(\sum_c \sqrt{p(c|a,b)\,q_\mathrm{ref,c}}\right)\,,
\end{align}
where the $c$'s label to fusion channel outcomes for $a\otimes b$, see fig. \eqref{fig:octant}. We take this distance as the geometric cost associated with jumping between two fusion-induced configurations. This choice of geometric gate cost quantifies the deviation of a fusion operation from uniform branching over its allowed channels, vanishing for unbiased fusion and increasing monotonically as the fusion weights become more asymmetric.

Several basic properties follow immediately. The cost is invariant under relabellings of fusion channels and under $F$-moves, as it depends only on the canonical fusion-channel weight distribution and not on the choice of fusion tree. It vanishes for deterministic fusion outcomes, where a single channel has unit weight, and it satisfies the triangle inequality by virtue of being a geodesic distance in projective space.

\paragraph{Example: Ising category.}
Consider the Ising UMTC with simple objects $\mathcal I=\{\mathbf 1,\sigma,\psi\}$, quantum dimensions $d_\sigma=\sqrt2$, $d_{\mathbf 1}=d_\psi=1$, and fusion rule $\sigma\otimes\sigma=\mathbf 1\oplus\psi$. For $a=b=\sigma$, the two fusion channels carry equal canonical weights
\begin{align}
p(\mathbf 1\,|\,\sigma,\sigma)=p(\psi\,|\,\sigma,\sigma)=\tfrac12\,.
\end{align}
We take as reference ray the uniform weighting $q_{\mathrm{ref}} (\tfrac12,\tfrac12)$, so
\begin{align}
\mathcal C^{\rm geom}_{\rm gate}(\sigma,\sigma) =\arccos \left(\sum_{c=\mathbf 1,\psi}\sqrt{p(c\,|\,\sigma,\sigma)\,q_{\mathrm{ref},c}}\right) =\arccos(1)=0\,.
\end{align}
Thus, in this convention, unbiased two-channel fusion has zero geometric gate cost: an energy-bias introduced in the next section will distinguish the channels.

\paragraph{Example: $\widehat{su(2)}_k$ WZW category.}
Another illustrative example is provided by the category associated with WZW models with chiral algebra $\widehat{su(2)}_k$. For such $\widehat{su(2)}_k$ WZW model, simple objects are labelled by $j \in \{0,\tfrac12,1,\dots,\tfrac{k}{2}\}$ with
\begin{align}\label{eq:labels_SU2k}
d_j=\frac{\sin((2j+1)\theta)}{\sin\theta}\,,\quad \theta=\frac{\pi}{k+2}\,,\quad h_j=\frac{j(j+1)}{k+2}\,.
\end{align}
Taking the gate $b=\tfrac12$ acting on sector $a=j$, the fusion is two-channel for $0<j<\tfrac{k}{2}$, with outputs $c=j\pm\tfrac12$ and canonical weights
\begin{align}
p_\pm=p \left(j\pm\tfrac12\,\Big|\,j,\tfrac12\right) =\frac{d_{j\pm1/2}}{d_{1/2}d_j} =\frac{\sin((2j{+}1\pm1)\theta)}{2\cos\theta\,\sin((2j{+}1)\theta)}\,, \quad p_++p_-=1\,.
\end{align}
With the uniform reference distribution $q_\mathrm{ref}=\tfrac12$, the geometric gate cost is
\begin{align}
\mathcal C^{\rm geom}_{\rm gate} \left(j,\tfrac12\right) =\arccos \left(\frac{1}{\sqrt2}\left(\sqrt{p_+}+\sqrt{p_-}\right)\right).
\end{align}
This vanishes when $p_+=p_-=\tfrac12$ and increases as fusion becomes more asymmetric, approaching $\arccos(1/\sqrt2)$ as $j$ approaches the boundary of the integrable alcove from the interior. We will return to this property in the explicit path optimisation examples of section \ref{sec:extendedcosts}.

\subsection{Energy-weighted costs}
\label{sec:energy_cost}

The geometric cost functional introduced in section \ref{sec:geomcostfunction} captures the kinematics of fusion-induced branching but is insensitive to this energetic information. In particular, it assigns the same cost to fusion channels that differ substantially in their effect on the stress tensor.  Indeed, in a two-dimensional conformal field theory however, the labels of irreducible representations carry direct energetic meaning. On the cylinder, the Hamiltonian is $H = L_0 + \bar L_0 - \frac{c}{12}$. For a conformal family labelled by $(h,\bar h)$, the minimal energy in that sector is $E_{\min} = h + \bar h - \frac{c}{12}$, 
independently of descendant excitations, which do not change the sector label. In the Virasoro-Nielsen framework reviewed earlier, each conformal family corresponds to a coadjoint orbit, and invertible circuits generate continuous trajectories within a fixed orbit. By contrast, non-invertible fusion gates induce discrete transitions between different orbits and therefore generically change the energy carried by the state.  For this reason, and motivated in particular by the gravitational interpretation of fusion-induced jumps as shock-like defects discussed in section \ref{sec:shocklike_def}, we introduce an energy-sensitive refinement of the geometric cost.

Let $E_c=\Delta h_c+\Delta\bar h_c$ denote the energy shift associated with a fusion channel $c$. Starting from the categorical channel weights $p_c$, we introduce energy-weighted\footnote{Note that the term ``energy-weighted'' refers to the direction in channel space selected by conformal weights, not to sensitivity to absolute energy differences.} channel weights
\begin{align}\label{eq:mod_probs}
p_c^{(\beta)}=\frac{p_c\, e^{-\beta E_c}}{\sum_d p_d\, e^{-\beta E_d}}\,,\qquad E_c=\Delta h_c+\Delta\bar h_c\,,
\end{align}
where $\beta\in[0,\infty)$ is an auxiliary parameter. No thermodynamic interpretation is assumed: $\beta$ is not a physical temperature and does not characterise the state produced by the fusion process. Rather, it parametrises a one-parameter family of deformations of the categorical weights used to induce a metric structure on the space of fusion channels.  We associate to the one-parameter family $p^{(\beta)}$ with the same metric as in \eqref{eq:metric_geom}. Writing
\begin{align}
p_c^{(\beta)}=\frac{p_c e^{-\beta E_c}}{Z(\beta)}\,,\quad Z(\beta)=\sum_d p_d e^{-\beta E_d}\,,
\end{align}
one finds
\begin{align}\label{eq:variance}
\mathrm ds^2 = \frac14\,\mathrm{Var}_\beta(E)\,\mathrm d\beta^2\,,\quad  \mathrm{Var}_\beta(E) = \sum_c p_c^{(\beta)}(E_c-\langle E\rangle_\beta)^2\,.
\end{align}
Note that although this expression is written in terms of the auxiliary parameter $\beta$, it may equivalently be rewritten in terms of the weighted channel energy $\mathcal E\equiv\langle E\rangle_\beta$ using
\begin{align}
\frac{d}{d\beta}\langle E\rangle_\beta=-\mathrm{Var}_\beta(E)\,.
\end{align}
This yields the equivalent form
\begin{align}\label{eq:metric_cost_energy}
\mathrm ds^2=\frac{1}{4\,\mathrm{Var}_\beta(E)}\,\mathrm d\mathcal E^2\,.
\end{align}

The intrinsic cost of a non-invertible fusion gate is then defined as the length of the curve $\beta\mapsto p_c^{(\beta)}$ in the space of channel weights, measured with the FS metric \eqref{eq:metric_geom} using the energy-weighted channel weights \eqref{eq:metric_geom}. Explicitly, we define the gate cost for fusing a gate $b$ with a state corresponding to a simple object $a$ as
\begin{align}\label{eq:gate_cost_def}
\mathcal C_{\mathrm{gate}}^\mathrm{energy}(a,b) =\frac12\int_{0}^{\infty}\mathrm d\beta\; \sqrt{\mathrm{Var}_\beta(E)}\,.
\end{align}
The lower limit $\beta=0$ corresponds to the geometrical regime of the previous section in which all fusion channels are weighted only by their quantum dimensions, see eq. \eqref{eq:expr_weigts}, while the limit $\beta\to\infty$ suppresses all but the energetically minimal channel, yielding a single or deterministic outcome. Note that one could alternatively, introducing a tunable penalisation strength, be invoking instead a truncated integration upper bound $\beta_\star$ in \eqref{eq:gate_cost_def} instead of integrating to infinity. Finally, for any fusion process with finitely many channels and strictly positive categorical weights, the integral in \eqref{eq:gate_cost_def} is finite. A proof is given in appendix \ref{app:finiteness_energy_cost}.

Before proceeding to  introducing the selection cost and the explicit path optimisation examples, it is useful to derive a closed-form expression for the energy-weighted gate cost in the simplest nontrivial situation, namely when a fusion step admits exactly two output channels. 

\paragraph{Two-channel case.}
Consider a fusion step $(a,b)$ with exactly two allowed output channels, labelled by $c_{\rm low}$ and $c_{\rm high}$, and assume their energies satisfy
\begin{align}
E_{\rm low} < E_{\rm high}, \qquad \Delta = E_{\rm high}-E_{\rm low}>0\,,
\end{align}
with canonical weights $p_{\rm low}=p(c_{\rm low}\,|\,a,b)$ and $p_{\rm high}=p(c_{\rm high}\,|\,a,b)$, where $p_{\rm low}+p_{\rm high}=1$. The $\beta$-deformed channel weights \eqref{eq:mod_probs} take the form
\begin{align}
p^{(\beta)}_{\rm high} =\frac{p_{\rm high}e^{-\beta \Delta}}{p_{\rm low}+p_{\rm high}e^{-\beta \Delta}}\,,
\qquad p^{(\beta)}_{\rm low}=1-p^{(\beta)}_{\rm high}\,.
\end{align}
Since the energy takes only the two values $E_{\rm low}$ and $E_{\rm high}$, the variance appearing in \eqref{eq:gate_cost_def} reduces to
\begin{align}
\mathrm{Var}_\beta(E) =p^{(\beta)}_{\rm low}\,p^{(\beta)}_{\rm high}\,(E_{\rm high}-E_{\rm low})^2 =\Delta^2\,p^{(\beta)}_{\rm low}\,p^{(\beta)}_{\rm high}\,.
\end{align}
The energy-weighted gate cost therefore admits the closed form
\begin{align}\label{eq:energy_cost_2ch}
\mathcal C_{\rm gate}^{\rm energy}(a,b) =\frac{\Delta}{2}\int_0^\infty \mathrm d\beta\; \sqrt{p^{(\beta)}_{\rm low}\,p^{(\beta)}_{\rm high}} =\arctan \sqrt{\frac{p_{\rm high}}{p_{\rm low}}}\,.
\end{align}
This last equation is independent of the magnitude of the energy gap $\Delta$ (as long as $\Delta\neq0$) and depends only on the ratio of categorical weights. For a deterministic fusion step ($p_{\rm high}=0$), the cost vanishes identically.  This cost characterises the difficulty of biasing a fusion operation toward specific outcomes and should not be interpreted as measuring the magnitude of any physical observable associated with those outcomes.

\subsection{Selection costs}\label{sec:select_cost} 

In Nielsen's geometric approach to circuit complexity, one defines the minimal cost of transforming a fixed reference state into a fixed target state for an allowed set of gates. Extending this framework to non-invertible gates raises an immediate issue: a single gate application generically produces multiple outcomes, as dictated by the fusion relation \eqref{eq:fusion}, and the notion of a definite target state must therefore be refined. In the present context where we have multiple outcomes as discussed in the context of the quantum circuit realisation in section \ref{sec:quantumchannels_sel}, this leads to a distinction between pre-selection and  post-selection. Pre-selection refers to the choice of an initial superselection sector or conformal family, i.e. fixing a simple object $a$ and working within the corresponding Hilbert space $\mathcal H_a$. As in conventional Nielsen complexity, this choice is part of the problem definition and carries no intrinsic cost.

By contrast, non-invertible fusion gates generically produce outputs supported on multiple superselection sectors, corresponding to the different fusion channels allowed by the categorical data. A single gate application therefore does not map a given initial state to a unique target Hilbert space, but rather to a multi-channel output distributed over several $\mathcal H_c$.

Post-selection corresponds to restricting this multi-channel output to a specific fusion outcome $c_\ast$, thereby selecting a definite target Hilbert space $\mathcal H_{c_\ast}$. Unlike pre-selection, this operation is not part of the intrinsic action of the fusion gate itself and must be implemented as an additional, irreversible step. It is therefore natural to associate a separate cost to post-selection, reflecting the extra resources required to enforce a definite fusion outcome. Physically, selection costs quantify the irreversibility associated with discarding the other fusion channels. Geometrically, we will measure this cost by how broadly the fusion output is distributed over multiple channels, relative to the extremal situation in which a single channel is isolated.

To fix this cost, we compare the canonical fusion output associated with a fixed process $(a,b)$ to the extremal configuration corresponding to a definite outcome $c_\ast$. Concretely, recall that fusion defines a vector \eqref{eq:ray} built from the canonical channel weights $p_c=p(c\,|\,a,b)$ in \eqref{eq:expr_weigts}, with $\sum_c p_c=1$. A selection onto $c_\ast$ corresponds to the ray represented by $\phi_c=\delta_{c,c_\ast}$, see fig. \ref{fig:octant}. Treating both $\varphi$ and $\phi$ as unit vectors in the positive orthant, a natural, basis-independent measure of distinguishability is provided by the Fubini-Study distance.

We therefore choose the post-selection cost as
\begin{align}\label{eq:selection_cost}
\mathcal C_{\mathrm{sel}}(c_\ast\,|\,a,b) = \arccos \big(\sqrt{p(c_\ast\,|\,a,b)}\big)\,.
\end{align}
This cost vanishes precisely when the fusion outcome is deterministic, $p(c_\ast\,|\,a,b)=1$, and increases as the selected outcome has smaller weight in the full fusion output. By construction, it depends only on the categorical fusion data, is invariant under relabellings and associator moves, and remains finite whenever the selected channel is categorically allowed.

\subsection{Extended gate sets and total circuit cost}\label{sec:extendedcosts}

The total cost of the action of a shock-like defect is thus decomposed into an intrinsic gate cost, capturing the effect of the non-invertible gate itself, and a selection cost, accounting for the additional resource required to isolate a definite target sector.  We emphasise again that  we  restrict attention to a unitary modular tensor subcategory of the full (infinite) Virasoro category, capturing the rational subsector that arises naturally via the Bershadsky-Ooguri reduction \cite{Bershadsky:1989mf}.

Before moving to the most general circuits including both invertible and non-invertible gates, let us first apply the constructed cost for a sequence of non-invertible gates. That is, fix an initial conformal family, i.e. a simple object of the pertinent UMTC, $a$ and a target family $c$. A sequence of non-invertible fusion gates $b_1,\dots,b_m$ induces a discrete fusion path
\begin{align}\label{eq:fusion_path_def}
a=a_0 \xrightarrow{\,b_1\,} a_1 \xrightarrow{\,b_2\,} a_2 \xrightarrow{\,\cdots\,} a_m=c\,,
\qquad\text{with}\quad N_{b_j\,a_{j-1}}^{\,a_j}\neq 0\ \ \forall j\,,
\end{align}
where at each step, if there are multiple fusion channel outcomes, we select for $a_i$, throwing away the other outcomes. The condition $N_{b_j a_{j-1}}^{\,a_j}\neq 0$ enforces that each step is an allowed sector transition. In particular, reaching $c$ from $a$ requires selecting, after each fusion gate, a specific channel outcome $a_j$ compatible with \eqref{eq:fusion_path_def}. This is the non-invertible analogue of steering a circuit toward a prescribed target state.

For a fixed path $\Gamma=(a_0\to a_1\to\cdots\to a_m)$ and chosen gate labels $b_1,\dots,b_m$, we define its total non-invertible cost additively by
\begin{align}\label{eq:path_cost_def}
\mathcal C_{\rm noninv}[\Gamma;\{b_j\}] =\sum_{j=1}^{m}\Big(\mathcal C_{\rm gate}(a_{j-1},b_j) +\mathcal C_{\rm sel}(a_j\,|\,a_{j-1},b_j)\Big)\,,
\end{align}
where $\mathcal C_{\rm gate}(a,b)$ is the gate cost in eq. \eqref{eq:geomcost} or eq. \eqref{eq:gate_cost_def} associated with fusion by $b$ acting on sector $a$, with or without energy bias depending on case under consideration, and $\mathcal C_{\rm sel}(c\,|\,a,b)$ is the additional cost \eqref{eq:selection_cost} of selecting the fusion output onto the specific channel $c$. If the fusion outcome is deterministic (a unique $c$ with $N_{ba}^{\,c}\neq0$), then $\mathcal C_{\rm sel}=0$ by construction. The complexity of preparing the target sector $c$ from $a$ using fusion gates is the minimal path cost
\begin{align}\label{eq:min_cost_discrete}
\mathcal C_{\rm noninv}(a\to c) =\inf_{\Gamma}
\ \mathcal C_{\rm noninv}[\Gamma;\{b_j\}]\,,
\end{align}
with minimisation over all paths $\Gamma$ as in eq. \eqref{eq:fusion_path_def}. Equivalently, \eqref{eq:min_cost_discrete} defines a shortest-path problem on the directed graph whose vertices are simple objects and whose directed weighted edges $a\to c$ exist whenever $N_{ba}^{\,c}\neq0$ for some allowed gate label $b$, with edge weight $\mathcal C_{\rm gate}(a,b)+\mathcal C_{\rm sel}(c\,|\,a,b)$. We will return to this concretely in the examples provided at the end of this section.

Consider now a circuit where the the set of allowed gates is enlarged beyond purely invertible operations to circuits of the form given in eq. \eqref{eq:new_circuit}. That is an extended circuit generated by two classes of gates: first, the usual invertible gates corresponding to unitary conformal transformations, whose cost is measured by the standard Nielsen geometry on the Virasoro coadjoint orbits; second, a finite set of non-invertible gates associated with fusion with simple objects of a unitary modular tensor category describing a (rational subsector) of non-invertible symmetries of the theory. A general circuit may then be written as an alternating composition of invertible and non-invertible gates,
\begin{align}
\mathcal U= U(\tau, \tau_m) D_{b_m}D_{b_{m-1}}\cdots U(\tau_m, \tau_{m-1})\cdots D_{b_3}D_{b_2}D_{b_1} U(\tau_1,\tau_0)\,,
\end{align}
where $\tau$ is the circuit paramater and each $D_{b_j}$ is the quantum channel associated with fusion by a simple object $b_j$, and each $U(\tau_{i},\tau_{i-1})$ is a unitary gate acting before, respectively after, the insertion of the non-invertible gate $D_{b_i}$, respectively $D_{b_{i+1}}$. The total cost of such a circuit is then defined additively as 
\begin{align}
\mathcal C[\mathcal U]=\sum_{i=0}^{m}\mathcal C_{\mathrm{inv}}[U_i]+\sum_{j=1}^{m}\Big(\mathcal C_{\mathrm{gate}}[D_{b_j}]+\chi_j\,\mathcal C_{\mathrm{sel}}(c^\star_{j})\Big)\,,
\end{align}
where we have simplified the arguments of the different costs to declutter the expresion. In this expression $\mathcal C_\mathrm{inv}\equiv S_\mathrm{SA}$ denotes the standard Nielsen cost on the unitary subgroup corresponding to the Alekseev-Shatashvili action\footnote{Note that for simplicity we restrict to the $F_1$ cost functional for the invertible part of the Nielsen circuit of the 2d CFT; since for $F_2$ the identification with the AS-action holds only approximately at large central charge \cite{Caputa:2018kdj}.} on the co-adjoint orbit determined by $b_i$ \cite{Caputa:2018kdj} and $\mathcal C_\mathrm{ens}$ is the energy-weighted geometric cost associated with the fusion channel data. Here, since this choice must be specified in addition to the circuit itself, the post-selection cost $\mathcal C_\mathrm{sel}$ is included only when the fusion gate is explicitly projected onto a definite outcome, with $\chi_j=1$ in that case and $\chi_j=0$ otherwise. Finally, as mentioned earlier, pre-selection of the reference sector is treated as part of the circuit specification and carries no intrinsic cost. This master formula therefore defines a cost functional for circuits combining continuous conformal transformations with discrete, non-invertible operations arising from rational fusion data. The associated circuit complexity is obtained, as usual in the Nielsen framework, by minimising $\mathcal C[\mathcal U]$ over all admissible circuits implementing the same overall transformation.

In the purely invertible Nielsen framework, circuit complexity is formulated as a geodesic optimisation problem on a smooth configuration space, typically a group manifold or coadjoint orbit generated by the allowed symmetry transformations. In particular, any two states within a fixed conformal family are connected by continuous paths, and complexity measures the minimal cost among them. Once non-invertible fusion gates are included, this picture changes
qualitatively. The space of accessible configurations becomes discrete and directed, with vertices given by superselection sectors and directed edges determined by admissible fusion transitions. Circuit optimisation is therefore no longer a geodesic problem in a differential-geometric sense, but a shortest-path problem on a directed graph with edge weights given by gate cost $\mathcal C_{\rm gate}$ and selection cost $\mathcal C_{\rm sel}$. Crucially, reachability itself now depends on the chosen gate set and fusion rules: only sectors connected by admissible fusion paths are accessible from a given initial sector\footnote{
For example, for $\widehat{su(2)}_k$ WZW models, simple objects are labelled by spins $j\in\{0,\tfrac12,1,\dots,\tfrac{k}{2}\}$, and fusion rules impose strict selection constraints on which spins can appear in tensor products. If the allowed non-invertible gate set is restricted to integer-spin representations (for example $b=1$), then repeated fusion starting from an integer-spin sector can only produce integer-spin sectors. Half-integer sectors never appear at any intermediate step and are therefore inaccessible from integer ones using such a gate set. The directed fusion graph thus decomposes into disconnected components, and reachability of target sectors depends explicitly on the choice of allowed fusion gates.
}. In this sense, the extended configuration space is not connected in general, and the existence of a path to a prescribed target sector becomes a nontrivial constraint rather than a given.

Turning to concrete examples of non-invertible circuits and their associated complexity, the Ising category is too small to exhibit genuinely distinct multi-step routes between fixed simple objects. A minimal example with nontrivial path dependence is provided by the $\widehat{su(2)}_k$ WZW model when $k\ge3$.\\

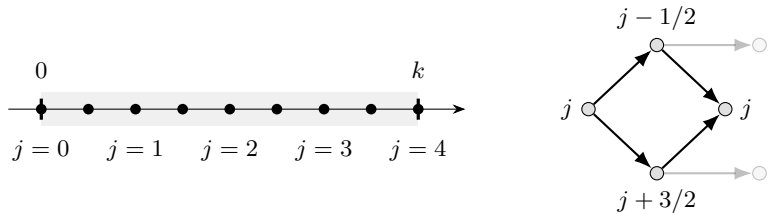
\begin{figure}[t]
\centering
\begin{tikzpicture}[x=1cm,y=1cm]

\begin{scope}[x=0.62cm,y=1cm]
  \def\k{8}
  \fill[gray!12] (0,-0.22) rectangle (\k,0.22);
  \draw[->] (-0.7,0) -- (\k+1.0,0) node[right] { };

  \draw[very thick] (0,0.14) -- (0,-0.14);
  \draw[very thick] (\k,0.14) -- (\k,-0.14);
  \node[above] at (0,0.28) {\footnotesize$0$};
  \node[above] at (\k,0.28) {\footnotesize$k$};

  \foreach \lam in {0,...,\k} { \fill (\lam,0) circle (2pt); }
  \foreach \lam/\lab in {0/0,2/1,4/2,6/3,8/4} {
    \node[below] at (\lam,-.28) {\footnotesize $j=\lab$};
  }

\end{scope}

\definecolor{nodefillfaint1}{gray}{0.92}
\definecolor{nodefill1}{gray}{0.88}
\begin{scope}[shift={(7.2,0)},x=0.9cm,y=0.85cm,>=Latex] 
  \node[circle,draw,minimum size=5pt,inner sep=0pt,fill=nodefill1] (L) at (0,0) {};
  \node[circle,draw,minimum size=5pt,inner sep=0pt,fill=nodefill1] (T) at (1,1) {};
  \node[circle,draw,minimum size=5pt,inner sep=0pt,fill=nodefill1] (B) at (1,-1) {};
  \node[circle,draw,minimum size=5pt,inner sep=0pt,fill=nodefill1] (R) at (2,0) {};

  \node[left=2pt]  at (L) {\footnotesize $j$};
  \node[above=2pt] at (T) {\footnotesize $j-1/2$};
  \node[below=2pt] at (B) {\footnotesize $j+3/2$};
  \node[right=2pt] at (R) {\footnotesize $j$};
  \node[right=2pt] at (2.175,1.45) {\footnotesize };
  \node[right=2pt] at (2.175,-1.405) {\footnotesize };
  
  \draw[->,line width=0.8pt] (L) -- (T);
  \draw[->,line width=0.8pt] (L) -- (B);
  \draw[->,line width=0.8pt] (T) -- (R);
  \draw[->,line width=0.8pt] (B) -- (R);
  
      \node[circle,draw,inner sep=1.8pt,opacity=0.25,fill=nodefill1] (zero) at (2.5,1) {};
      \node[circle,draw,inner sep=1.8pt,opacity=0.25,fill=nodefill1] (two)  at (2.5,-1) {};
      \draw[->,line width=0.8pt,opacity=0.25] (T) -- node[above,opacity=0.35] { } (zero);
      \draw[->,line width=0.8pt,opacity=0.25] (B) -- node[below,opacity=0.35] { } (two);

\end{scope}

\end{tikzpicture}
\caption{\textit{Left}: The set of integrable $\widehat{su(2)}_k$ representations, labelled by spin $j = 0, \tfrac12, \dots, \tfrac{k}{2}$. The finite interval reflects the integrability constraint; fusion gates act by inducing transitions between these labels, as illustrated on the right. \textit{Right}: Two distinct competing length-two fusion paths from $j=1$, which sets at the boundary of the integrable alcove, back to $j=1$ under repeated fusion with $b=\tfrac12$: the intermediate channel can be $j_1=\tfrac12$ or $j_1=\tfrac32$. Faint paths and dots indicate alternative fusion outcomes. }
\label{fig:su2k_side_by_side}
\end{figure}

\paragraph{Example: $\widehat{su(2)}_3$ return paths $1\to1$.}
We illustrate the path optimisation problem in the simplest nontrivial case, $\widehat{su(2)}_3$, where, using \eqref{eq:labels_SU2k}, $\theta=\pi/5$ and the simple objects are $j\in\{0,\tfrac12,1,\tfrac32\}$. The corresponding quantum dimensions and conformal weights are $d_0=1$, $d_{1/2}=d_1=\varphi$, $d_{3/2}=1$ with $\varphi=\tfrac{1+\sqrt5}{2}$ the golden ratio and conformal weights $h_j=\frac{j(j+1)}{5}$. We fix the non-invertible gate label $b=\tfrac12$ and initial and final sectors $a=c=1$. Since $1\otimes\tfrac12=\tfrac12\oplus\tfrac32$, two distinct two-step fusion paths exist,
\begin{align}
\Gamma_-:\quad 1 \xrightarrow{\,1/2\,} \tfrac12 \xrightarrow{\,1/2\,} 1\,, \qquad
\Gamma_+:\quad 1 \xrightarrow{\,1/2\,} \tfrac32 \xrightarrow{\,1/2\,} 1\,,
\end{align}
as shown in figure \ref{fig:su2k_side_by_side}. The associated weights, see \eqref{eq:expr_weigts}, (and here all multiplicities are $1$) are
\begin{align}
p\left(\tfrac12\,\middle|\,1,\tfrac12\right)&=\frac{d_{1/2}}{d_1 d_{1/2}}=\frac{1}{\varphi}\,, \qquad p\left(\tfrac32\,\middle|\,1,\tfrac12\right)=\frac{d_{3/2}}{d_1 d_{1/2}}=\frac{1}{\varphi^2}\,,
\label{eq:su2_3_step1_weights}\\
p\left(1\,\middle|\,\tfrac12,\tfrac12\right)&=\frac{d_1}{d_{1/2}^2}=\frac{1}{\varphi}\,, \qquad\;\;\;\, p\left(0\,\middle|\,\tfrac12,\tfrac12\right)=\frac{d_0}{d_{1/2}^2}=\frac{1}{\varphi^2}\,, \label{eq:su2_3_step2_weights}
\end{align}
while $\tfrac32\otimes\tfrac12=1$ is deterministic, so $p(1\,|\,\tfrac32,\tfrac12)=1$. 

\vspace{8pt}

\noindent
\textit{Geometric cost.}
In the purely geometric regime, the total cost receives contributions both from the intrinsic geometric gate cost associated with multi-channel fusion and from selection whenever a definite fusion outcome is specified. For a fusion step $(a,b)$ with more than one allowed channel, the geometric gate cost is given by \eqref{eq:geomcost}, while deterministic steps carry zero gate cost.

For the first step and second step, $(a,b)=(1,\tfrac12)$ and $(\tfrac12,\tfrac12)$ are two-channel fusion, given the  weights in \eqref{eq:su2_3_step1_weights}, the corresponding geometric gate cost coincide
\begin{align}
\mathcal C_{\rm gate}^{\rm geom}(1,\tfrac12)=\mathcal C_{\rm gate}^{\rm geom}(\tfrac12,\tfrac12)
=\arccos \left(\frac{1}{\sqrt2}\big(\varphi^{-1/2}+\varphi^{-1}\big)\right)\,.
\end{align}
By contrast, the fusion $(\tfrac32,\tfrac12)\to 1$ is deterministic and therefore has $\mathcal C_{\rm gate}^{\rm geom}(\tfrac32,\tfrac12)=0$. The total geometric path cost, including both gate and selection contributions, with the latter given by \eqref{eq:selection_cost}, is then
\begin{align}
\mathcal C_{\rm geom}(\Gamma_-) &=\mathcal C_{\rm gate}^{\rm geom}(1,\tfrac12)
+\mathcal C_{\rm sel} \left(\tfrac12\,\middle|\,1,\tfrac12\right) +\mathcal C_{\rm gate}^{\rm geom}(\tfrac12,\tfrac12) +\mathcal C_{\rm sel} \left(1\,\middle|\,\tfrac12,\tfrac12\right) \nonumber\\
&=2\,\arccos \left(\frac{1}{\sqrt2}\big(\varphi^{-1/2}+\varphi^{-1}\big)\right) +2\,\arccos \left(\varphi^{-1/2}\right), \\ 
\mathcal C_{\rm geom}(\Gamma_+)
&=\mathcal C_{\rm gate}^{\rm geom}(1,\tfrac12) +\mathcal C_{\rm sel} \left(\tfrac32\,\middle|\,1,\tfrac12\right) \nonumber\\
&=\arccos \left(\frac{1}{\sqrt2}\big(\varphi^{-1/2}+\varphi^{-1}\big)\right) +\arccos \left(\varphi^{-1}\right).
\label{eq:su2_3_geom_costs}
\end{align}
The presence of a deterministic second step favours the path $\Gamma_+$ in the purely geometric regime.

\vspace{8pt}

\noindent
\textit{Energy-weighted cost.}
We now turn on an energy bias for the gate cost by assigning to each channel $c$ the conformal-weight shift $E_c=\Delta h_c=h_c-h_a$,  and using the $\beta$-dependent weights $p_c^{(\beta)}\propto p(c\,|\,a,b)\,e^{-\beta E_c}$ inside \eqref{eq:gate_cost_def}. For the first step $(a,b)=(1,\tfrac12)$ one has $E_{1/2}=h_{1/2}-h_1<E_{3/2}=h_{3/2}-h_1$ so the lower-energy channel is $c_{\rm low}=\tfrac12$ and the higher-energy channel is $c_{\rm high}=\tfrac32$. The corresponding canonical weights are those in \eqref{eq:su2_3_step1_weights},
\begin{align}
p_{\rm low}=p \left(\tfrac12\,\middle|\,1,\tfrac12\right)=\varphi^{-1}, \quad p_{\rm high}=p \left(\tfrac32\,\middle|\,1,\tfrac12\right)=\varphi^{-2}\,.
\end{align}
Since this is a two-channel fusion, the energy-weighted gate cost reduces to \eqref{eq:energy_cost_2ch},
\begin{align}
\mathcal C_{\rm gate}^{\rm energy}(1,\tfrac12) =\arctan\sqrt{\frac{p_{\rm high}}{p_{\rm low}}} =\arctan \left(\varphi^{-1/2}\right)\,.
\end{align}
Similarly, for the second step $(a,b)=(\tfrac12,\tfrac12)$ the two allowed channels are $c=1,0$ with weights given in \eqref{eq:su2_3_step2_weights}. In this case $E_0=h_0-h_{1/2}<E_1=h_1-h_{1/2}$ so $c_{\rm low}=0$ and $c_{\rm high}=1$, and \eqref{eq:energy_cost_2ch} gives
\begin{align}\label{eq:gatecostE_return}
\mathcal C_{\rm gate}^{\rm energy}(\tfrac12,\tfrac12)=\arctan\sqrt{\frac{p(1\,|\,\tfrac12,\tfrac12)}{p(0\,|\,\tfrac12,\tfrac12)}}=\arctan \left(\varphi^{1/2}\right)\,.
\end{align}
Finally, $(\tfrac32,\tfrac12)$ is deterministic and hence $\mathcal C_{\rm gate}^{\rm energy}(\tfrac32,\tfrac12)=0$.

The total energy-weighted path cost in this regime, namely the sum of energy-weighted gate costs and energy-less selection costs \eqref{eq:selection_cost} are
\begin{align}
\mathcal C_{\rm energy}(\Gamma_-)&=\mathcal C_{\rm gate}^{\rm energy}(1,\tfrac12)+\mathcal C_{\rm sel} \left(\tfrac12\,\middle|\,1,\tfrac12\right)+\mathcal C_{\rm gate}^{\rm energy}(\tfrac12,\tfrac12)+\mathcal C_{\rm sel} \left(1\,\middle|\,\tfrac12,\tfrac12\right) \nonumber\\
&=\frac{\pi}{2}+2\,\arccos \left(\varphi^{-1/2}\right)\,, \label{eq:su2_3_energy_costsmin}\\
\mathcal C_{\rm energy}(\Gamma_+)
&=\mathcal C_{\rm gate}^{\rm energy}(1,\tfrac12)+\mathcal C_{\rm sel} \left(\tfrac32\,\middle|\,1,\tfrac12\right)
=\arctan \left(\varphi^{-1/2}\right)+\arccos \left(\varphi^{-1}\right)\,.
\label{eq:su2_3_energy_costsplus}
\end{align}
In particular, the deterministic second step again favours $\Gamma_+$.

Two features are worth noting. First, deterministic fusion steps are favoured, since they contribute neither an intrinsic gate cost nor a selection cost. Second, in this example the preferred route is the same in the purely geometric and energy-weighted regimes: the $\beta$-deformation changes the intrinsic gate contributions but does not alter the advantage gained by funnelling into a deterministic return.

\begin{figure}[t]
\centering

\begin{minipage}[t]{0.49\textwidth}
\centering

\definecolor{softgray}{gray}{0.65}
\definecolor{midgray}{gray}{0.55}
\definecolor{lightgray}{gray}{0.75}
\definecolor{nodefill}{gray}{0.88}
\definecolor{nodefillfaint}{gray}{0.92}
\begin{tikzpicture}[x=1.25cm,y=0.95cm,>=Latex]

 \tikzset{
  vtx/.style={
    circle,
    draw=black,
    fill=nodefill,
    minimum size=6pt,
    inner sep=0pt
  },
  vtxfaint/.style={
    circle,
    draw=softgray,
    fill=nodefillfaint,
    minimum size=6pt,
    inner sep=0pt
  },
    half/.style={
      ->,
      line width=0.8pt,
      draw=black
    },
    one/.style={
      ->,
      line width=0.8pt,
      draw=black
    },
    halffaint/.style={
      half,
      draw=softgray,
      opacity=0.65
    },
    onefaint/.style={
      one,
      dashed,
      draw=softgray,
      opacity=0.85
    },
    guide/.style={
      line width=0.6pt,
      draw=lightgray
    }
  }

  \def\yzero{0}
  \def\yhalf{-1}
  \def\yone{-2}
  \def\ythreehalf{-3}
  \def\ytwo{-4}

  \draw[guide] (-0.1,\yzero) -- (4.3,\yzero);
  \draw[guide] (-0.1,\yhalf) -- (4.3,\yhalf);
  \draw[guide] (-0.1,\yone) -- (4.3,\yone);
  \draw[guide] (-0.1,\ythreehalf) -- (4.3,\ythreehalf);
  \draw[guide] (-0.1,\ytwo) -- (4.3,\ytwo);

  \node[left] at (-0.15,\yzero) {\footnotesize $0$};
  \node[left] at (-0.15,\yhalf) {\footnotesize $1/2$};
  \node[left] at (-0.15,\yone) {\footnotesize $1$};
  \node[left] at (-0.15,\ythreehalf) {\footnotesize $3/2$};
  \node[left] at (-0.15,\ytwo) {\footnotesize $2$};

  \node[vtx]      (j0L)  at (0,\yzero) {};
  \node[vtxfaint] (j0M)  at (2,\yzero) {};
  \node[vtxfaint] (j0R)  at (4,\yzero) {};

  \node[vtx]      (j12a) at (1,\yhalf) {};
  \node[vtxfaint] (j12b) at (3,\yhalf) {};

  \node[vtxfaint] (j1L)  at (0,\yone) {};
  \node[vtx]      (j1M)  at (2,\yone) {};
  \node[vtxfaint] (j1R)  at (4,\yone) {};

  \node[vtxfaint] (j32a) at (1,\ythreehalf) {};
  \node[vtx]      (j32b) at (3,\ythreehalf) {};

  \node[vtxfaint] (j2L)  at (0,\ytwo) {};
  \node[vtxfaint] (j2M)  at (2,\ytwo) {};
  \node[vtx]      (j2R)  at (4,\ytwo) {};

  \draw[halffaint] (j1M) -- (j12b);
  \draw[half]      (j0L) -- (j12a);
  \draw[halffaint] (j12a) -- (j0M);

  \draw[half]      (j12a) -- (j1M);
  \draw[half]      (j1M)  -- (j32b);
  \draw[half]      (j32b) -- (j2R);
  \draw[halffaint] (j32b) -- (j1R);

  \draw[one,dashed]
    (j0L) to[out=5,in=105,looseness=0.70] (j1M);

  \draw[one,dashed]
    (j1M) to[out=5,in=105,looseness=0.70] (j2R);

  \draw[onefaint]
    (j1M) to[out=75,in=180,looseness=0.70] (j0R);
    
  \draw[onefaint]
 	(j1M) to[out=-80,in=210,looseness=20] (j1M);
\end{tikzpicture}
\end{minipage}
\hspace{10pt}
\begin{minipage}[b]{0.41\textwidth}
\centering
\includegraphics[width=\linewidth]{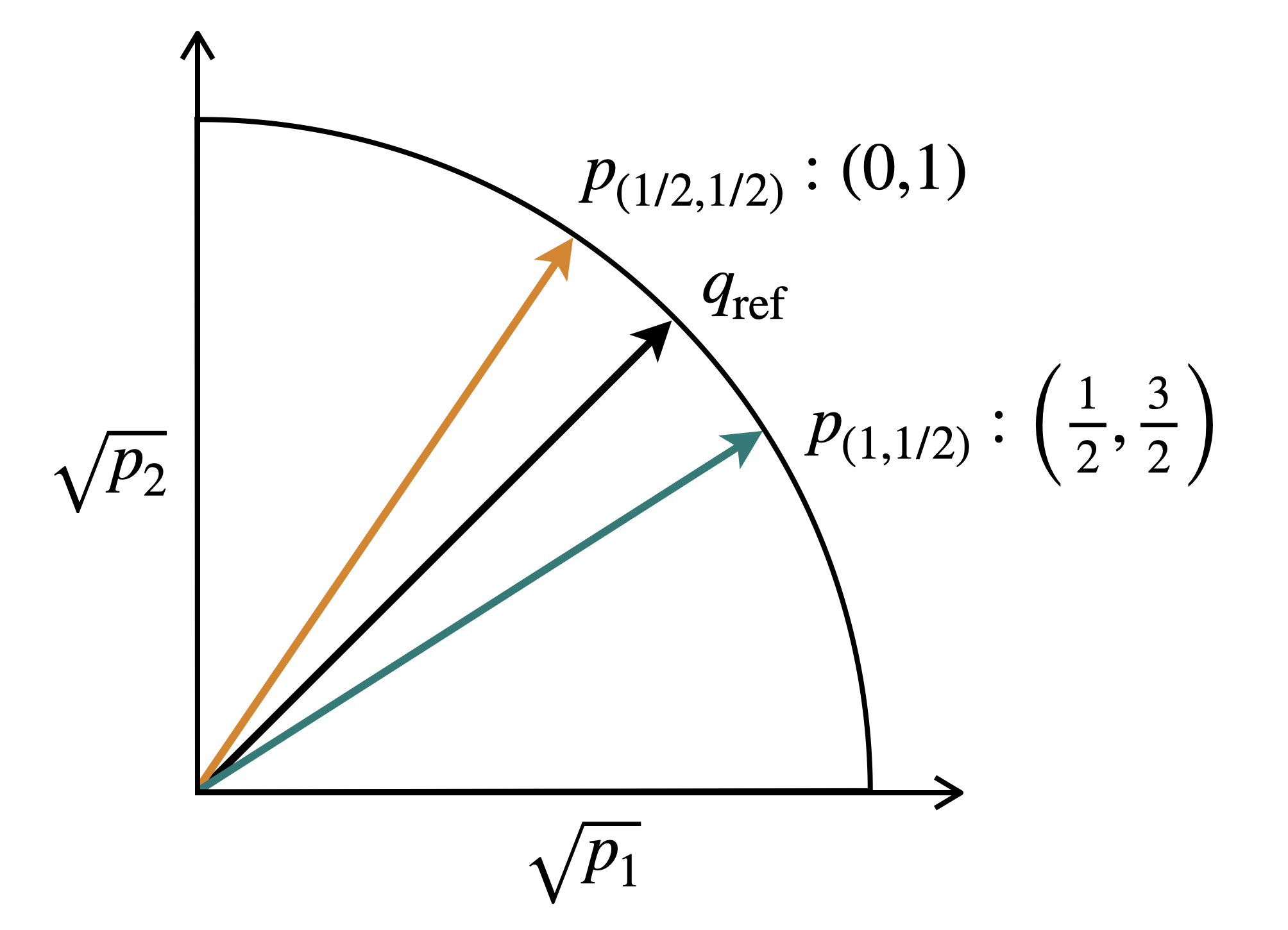}
\end{minipage}
\caption{\textit{Left:} Fusion graph for the $\widehat{su(2)}_4$ example illustrating two admissible paths from sector $j=0$ to $j=2$.
Vertices are arranged by spin label $j=0,\tfrac12,1,\tfrac32,2$. Solid arrows denote fusion steps with gate label $b=\tfrac12$ ($j\to j\pm\tfrac12$), while dashed arrows denote unselected steps. The solid path represents a longer sequence of $\tfrac12$-steps, whereas the dashed path illustrates a shorter route involving a $b=1$ fusion. \textit{Right}: Weight vectors for two-channel fusion steps lie on the unit quarter-circle in $(\sqrt{p_1},\sqrt{p_2})$ coordinates. The gate cost is the Fubini-Study angle to a fixed reference vector $q_{\rm ref}$. The labels $p_(a,b):(p_1,p_2)$ indicate the ordered fusion weights for the two outcomes of  $a\otimes b$. In a slight abuse of geometry, since  the outcome labels differ between fusion steps, we depict them in the same unit quarter-circle. }\label{fig:su24_jumps}
\end{figure}
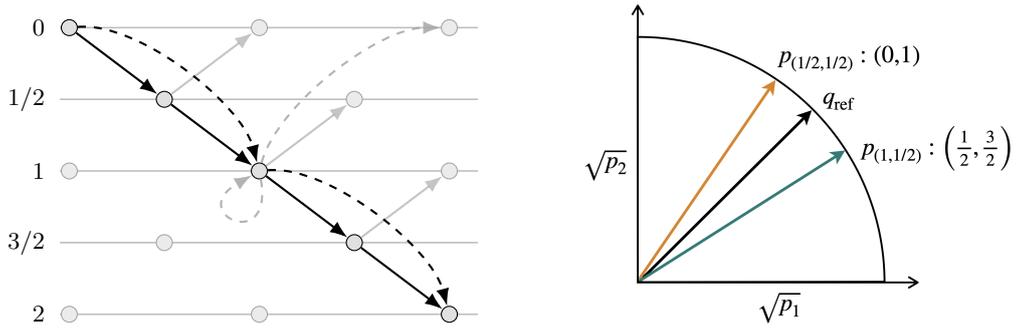

\paragraph{Example: $\widehat{su(2)}_4$ unequal length paths $0\to2$.}
Now we turn to $\widehat{su(2)}_k$ with $k=4$, hence using \eqref{eq:labels_SU2k}, we have that $\theta=\pi/6$ and the quantum dimensions are $d_0=1$, $d_{1/2}=\sqrt3$, $d_1=2$, $d_{3/2}=\sqrt3$, $d_2=1$ while conformal weights are $h_j=\frac{j(j+1)}{k+2}=\frac{j(j+1)}{6}$. This case allows for two admissible routes from $0$ to $2$ of unequal length (see fig. \ref{fig:su24_jumps}).

\vspace{8pt}
\noindent
\textit{Geometric cost.}
In the geometric regime, the total cost of a path receives contributions from (i) the intrinsic geometric gate cost associated with multi-channel fusion and (ii) the selection cost whenever a definite fusion outcome is specified. Deterministic fusion steps carry neither contribution.

\begin{itemize}
\item[(\textit{i})] The short path is:
\begin{align}
\Gamma_{\rm short}:\quad 0\xrightarrow{\,1\,}1\xrightarrow{\,1\,}2\,.
\end{align}
The first step is deterministic ($0\otimes1=1$) and contributes no cost. The second step is the three-channel fusion $1\otimes1=0\oplus1\oplus2$ with canonical weights
\begin{align}
p_0=\tfrac14\,,\quad p_1=\tfrac12\,,\quad p_2=\tfrac14\,.
\end{align}
Then the total geometric cost of the short path including the seleecting for the $c=2$-outcome, is
\begin{align}
\mathcal C_{\rm geom}(\Gamma_{\rm short})
=\mathcal C^{\rm geom}_{\rm gate}(1,1)+\mathcal C^{\rm geom}_{\rm sel}(2\,|\,1,1)
=\arccos \left(\frac{2+\sqrt2}{2\sqrt3}\right)+\frac{\pi}{3}\,.
\end{align}

\item[(\textit{ii})] The long path is:
\begin{align}
\Gamma_{\rm long}:\quad
0\xrightarrow{\,1/2\,}\tfrac12\xrightarrow{\,1/2\,}1\xrightarrow{\,1/2\,}\tfrac32\xrightarrow{\,1/2\,}2\,.
\end{align}
The first step is deterministic and contributes no cost. The remaining three steps are all two-channel fusions and therefore each contribute both a geometric gate cost and a selection cost. Using the geometric gate cost expression \eqref{eq:energy_cost_2ch} for  two-channel fusion  and adding the geometric selection costs computed in the previous example, the total geometric cost of the long path is
\begin{align}
\mathcal C_{\rm geom}(\Gamma_{\rm long})
=\sum \mathcal C^{\rm geom}_{\rm gate}+\mathcal C^{\rm geom}_{\rm sel}(\Gamma_{\rm long})
=2\,\arccos \left(\frac{\sqrt{1/3}+\sqrt{2/3}}{\sqrt2}\right)+\frac{3\pi}{4}\,.
\end{align}
\end{itemize}
In the purely geometric regime the shorter route is therefore still favoured, $\mathcal C_{\rm geom}(\Gamma_{\rm short})<\mathcal C_{\rm geom}(\Gamma_{\rm long})$, but now for a different reason: the long path accumulates additional geometric gate cost from repeated multi-channel fusion.

\vspace{8pt}

\noindent
\textit{Energy-weighted cost.}
We now consider the energy-weighted gate contribution $\mathcal C_{\rm gate}(a,b)$ defined in eq. \eqref{eq:gate_cost_def}.
Throughout this item we keep the post-selection cost geometric, i.e. we continue to use $\mathcal C^{\rm geom}_{\rm sel}$ from \eqref{eq:selection_cost} without any $\beta$-deformation, and only the gate contribution is energy-weighted.

\begin{itemize}
    \item[(\textit{i})] Short path.
The first step $0\otimes 1= 1$ is deterministic, hence $\mathcal C_{\rm gate}(0,1)=0$. For the second step $(a,b)=(1,1)$ with channels $c\in\{0,1,2\}$ and weights $p_0=\tfrac14$, $p_1=\tfrac12$, $p_2=\tfrac14$, the conformal-weight shifts are
\begin{align}
\Delta h_0=h_0-h_1=-\frac13\,,\quad \Delta h_1=h_1-h_1=0\,,\quad \Delta h_2=h_2-h_1=\frac23\,,
\end{align}
so $\mathrm{Var}_\beta(E)$ is nonzero and gives a strictly positive gate cost. Evaluating \eqref{eq:gate_cost_def} yields
\begin{align}
\mathcal C_{\rm gate}(1,1) =\frac12\int_0^\infty \mathrm d\beta\;\sqrt{\mathrm{Var}_\beta(E)} \approx1.1740\,.
\end{align}
The total cost of the short path is therefore
\begin{align}
\mathcal C(\Gamma_{\rm short}) =\mathcal C_{\rm gate}(1,1)+\mathcal C^{\rm geom}_{\rm sel}(2|1,1) =\mathcal C_{\rm gate}(1,1)+\frac{\pi}{3} \approx 2.2212\,.
\end{align}

    \item[(\textit{ii})] Long path.
The first step $0\otimes \tfrac12= \tfrac12$ is deterministic, hence $\mathcal C_{\rm gate}(0,\tfrac12)=0$. The remaining three steps are all two-channel fusions and we can use \eqref{eq:energy_cost_2ch} again. The different steps give
\begin{align}
\tfrac12\otimes\tfrac12= 0\oplus 1:&\quad  E_0=h_0-h_{1/2}<E_1=h_1-h_{1/2}\,,\nonumber\\
&\quad p_{\rm low}=p(0|\tfrac12,\tfrac12)=\tfrac13\,, 
p_{\rm high}=p(1|\tfrac12,\tfrac12)=\tfrac23\,,\\[4pt]
1\otimes\tfrac12= \tfrac12\oplus\tfrac32:&\quad
E_{1/2}=h_{1/2}-h_{1}<E_{3/2}=h_{3/2}-h_{1}\,,\nonumber\\
&\quad p_{\rm low}=p(\tfrac12|1,\tfrac12)=\tfrac12\,,\quad
p_{\rm high}=p(\tfrac32|1,\tfrac12)=\tfrac12\,,\\[4pt]
\tfrac32\otimes\tfrac12= 1\oplus 2\,:&\quad
E_{1}=h_{1}-h_{3/2}<E_{2}=h_{2}-h_{3/2}\,,\nonumber\\
&\quad p_{\rm low}=p(1|\tfrac32,\tfrac12)=\tfrac23\,,\quad
p_{\rm high}=p(2|\tfrac32,\tfrac12)=\tfrac13\,,
\end{align}
leading, respectively, to $\mathcal C_{\rm gate}(\tfrac12,\tfrac12)=\arctan \sqrt2$,
$\mathcal C_{\rm gate}(1,\tfrac12)=\arctan(1)=\frac{\pi}{4}$ and
$\mathcal C_{\rm gate}(\tfrac32,\tfrac12)=\arctan \frac{1}{\sqrt2}$.
Summing the different contributions together with the geometric selection cost computed in earlier, the full long-path cost is
\begin{align}
\mathcal C(\Gamma_{\rm long})=\mathcal C_{\rm gate}(\Gamma_{\rm long})+\mathcal C^{\rm geom}_{\rm sel}(\Gamma_{\rm long})=\frac{3\pi}{2}\approx 4.7124\,.
\end{align}
\end{itemize}
In particular, even after including the energy-weighted gate cost, the unequal-length comparison still favours the short route: $\mathcal C(\Gamma_{\rm short})<\mathcal C(\Gamma_{\rm long})$.
This example thus highlights that cost optimisation is not controlled simply by path length or by the number of fusion channels at a given step. A single multi-channel fusion followed by one selection can be cheaper than a longer route built from repeated two-channel branchings, because the latter accumulates selection overhead at each enforced intermediate choice. The ordering persists after energy-weighting.

\section{Shock-like defects in Ising gravity}\label{sec:shocklike_def}

In this section we embed the non-invertible gate set constructed in section \ref{sec:noninv_gates} into a piecewise Virasoro circuit, interpret the resulting discrete changes of conformal family in the $SL(2,\mathbb R)\times SL(2,\mathbb R)$ Chern-Simons description of AdS$_3$ gravity, and evaluate the corresponding cost functionals.

Following \cite{Erdmenger:2021wzc}, we identify the generator of the invertible part of the Nielsen circuit in \eqref{eq:inv_Nielsen_gate} with the CFT Hamiltonian,
\begin{align}\label{eq:nielsenHamevol}
Q(\tau)=H\,,\qquad U(\tau)=\mathrm e^{-i\int_0^\tau H\,\mathrm d\tau'} =\mathrm e^{-iH\tau}\,.
\end{align}
With this identification, the circuit parameter $\tau$ coincides with physical boundary time. The invertible segments of the circuit therefore describe time-sliced evolution within a fixed Virasoro coadjoint orbit, or equivalently within a fixed conjugacy class of $SL(2,\mathbb R)\times SL(2,\mathbb R)$ holonomies. In the bulk description, this corresponds to the evolution of a single Ba\~nados configuration.

Non-invertible operations are then inserted at isolated values of $\tau$. An elementary circuit segment thus takes the form
\begin{align}\label{eq:atomic_path}
U(\tau')\,D_a\,U(\tau)\,|\psi_0\rangle\,,\qquad \tau'>\tau_0>\tau\,,
\end{align}
where $\tau_0$ denotes the insertion time and $D_a$ is a non-invertible fusion gate.

Throughout this section we restrict attention to a rational subsector of the boundary theory whose topological defect operators close under fusion and are labelled by simple objects of a unitary modular tensor category. This provides a natural setting for the non-invertible fusion gates constructed in section \ref{sec:noninv_gates}, and allows for a controlled comparison between categorical data and bulk Chern--Simons variables. In particular, the Bershadsky-Ooguri construction relates special choices of $SL(2,\mathbb R)\times SL(2,\mathbb R)$ Chern-Simons boundary conditions to Virasoro minimal-model representation data at specific values of the couplings \cite{Bershadsky:1989mf}. Moreover, \cite{Castro:2011zq} showed that at a special, strongly coupled point, the three-dimensional gravity path integral reproduces the Ising modular invariant on the torus, pointing to a distinguished rational sector rather than a complete gravitational dual. More recently, dynamical studies of the Ising CFT have further highlighted the usefulness of formulations based on stress-tensor data and holonomies, rather than semiclassical bulk metrics \cite{Janik:2025zji}. Adopting this perspective, a non-invertible fusion gate induces a discrete change of Virasoro data at a definite circuit time, and therefore a discrete transition between admissible holonomy sectors of the Chern-Simons description.

\subsection{From Virasoro data to Ba\~nados geometries}\label{sec:CS_Banados_hol}

Before analysing the effect of non-invertible gate insertions on bulk holonomy data, we briefly review how Virasoro coadjoint-orbit data are encoded in classical solutions of three-dimensional AdS gravity. In particular, we recall how boundary stress-tensor data specify Ba\~nados geometries in the Chern-Simons formulation. This will allow us to translate the discrete changes in conformal data induced by fusion gates into precise statements about the corresponding bulk holonomy classes. For pedagogical reviews and additional details see e.g. \cite{Castro:2016tlm,Campoleoni:2024ced,Banados:1998gg}.

Three-dimensional gravity with negative cosmological constant admits a formulation as an $SL(2,\mathbb R)\times SL(2,\mathbb R)$ Chern-Simons theory at level $k=\frac{\ell}{4G}=\frac{c}{6}$ \cite{Achucarro:1986uwr,Witten:1987ty}, where the Brown-Henneaux relation identifies the central charge as $c=\frac{3\ell}{2G}$ \cite{Brown:1986nw}. The Chern-Simons connections are written as
\begin{align}
A=\omega+\frac{1}{\ell}e\,,\qquad \bar A=\omega-\frac{1}{\ell}e\,,
\end{align}
where $e=e^a L_a$ is the dreibein,, $\omega=\omega^a L_a$ the spin connection and $L_a$ are the usual $sl(2, \mathbb R)$-generators. Flatness of $A$ and $\bar A$ is equivalent to torsionless constant-curvature gravity, and the spacetime metric is reconstructed as $g_{\mu\nu}=\frac12\langle e_\mu e_\nu\rangle$.

Since three-dimensional AdS gravity has no local propagating degrees of freedom, all solutions obeying Brown-Henneaux boundary conditions are locally AdS$_3$, and the physical information is encoded in global data (boundary charges or holonomies). In Fefferman-Graham gauge the bulk metric is reconstructed from the boundary metric together with $\langle T_{++}\rangle$ and $\langle T_{--}\rangle$. The most general such solutions can be written in terms of two functions $\mathcal L(x^+)$ and $\bar{\mathcal L}(x^-)$, which are identified with the chiral components of the boundary stress tensor \cite{Banados:1998gg,Banados:1998sm}. In this sense, Ba\~nados geometries parametrise the space of classical solutions with these boundary conditions: specifying $\mathcal L(x^+)$ and $\bar{\mathcal L}(x^-)$ (up to the residual large diffeomorphisms) determines the corresponding classical AdS$_3$ metric, and conversely the metric determines the associated Virasoro data.

We specialise to the Ba\~nados family of solutions, written in light-cone coordinates $(\rho,x^+,x^-)$, with $x^\pm=t\pm x$ and $\rho$ the radial coordinate, as 
\begin{gather}
\begin{aligned}\label{eq:radial_gaugeA}
A &= b^{-1}\left(L_1-\frac{2\pi}{k}\,\mathcal L(x^+)\,L_{-1}\right)b\,\mathrm dx^+ + L_0\,\mathrm d\rho\,,\\
\bar A &= b\left(L_{-1}-\frac{2\pi}{k}\,\bar{\mathcal L}(x^-)\,L_{1}\right)b^{-1}\mathrm dx^- + L_0\,\mathrm d\rho\,,
\end{aligned}    
\end{gather}
where $L_0,L_{\pm1}$ generate $sl(2,\mathbb R)$ and satisfy $[L_0,L_{\pm1}]=\mp L_{\pm1}$, $[L_1,L_{-1}]=2L_0$ and  with $b(\rho)=e^{\rho L_0}$. The functions $\mathcal L(x^+)$ and $\bar{\mathcal L}(x^-)$ encode the expectation values of the boundary stress tensor via \cite{Banados:1998gg,Coussaert:1995zp}
\begin{gather}
\begin{aligned}\label{eq:EMT_to_CS_calL}
T_{++}=\frac{c}{12\pi}\,\mathcal L(x^+)\,,\qquad T_{--}=\frac{c}{12\pi}\,\bar{\mathcal L}(x^-)\,.
\end{aligned}    
\end{gather}
For completeness, we recall that these Chern-Simons connections correspond to the Ba\~nados metrics, which in Fefferman-Graham coordinates take the universal form
\begin{align}\label{eq:metric_EMT_to_CS_calL}
\mathrm ds^2=\frac{\mathrm dr^2}{r^2} +r^2\,\mathrm dx^+\mathrm dx^- +\mathcal L(x^+)\,\mathrm dx^{+2} +\bar{\mathcal L}(x^-)\,\mathrm dx^{-2} +\frac{1}{r^2}\mathcal L(x^+)\bar{\mathcal L}(x^-)\,\mathrm dx^+\mathrm dx^-\,.
\end{align}
Thus specifying the functions $\mathcal L(x^+)$ and $\bar{\mathcal L}(x^-)$ is
equivalent to specifying the full classical bulk geometry.
Of particular interest for our purposes are constant coadjoint-orbit representatives,
\begin{align}\label{eq_h_to_mL}
\mathcal L = h-\frac{c}{24}\,,\qquad \bar{\mathcal L}=\bar h-\frac{c}{24}\,,
\end{align}
which correspond to highest-weight Virasoro representations with conformal weights $(h,\bar h)$.  Different values of $(h,\bar h)$ therefore label distinct Ba\~nados geometries, ranging from global AdS$_3$ and conical defects to BTZ black holes.

Because three-dimensional AdS gravity is topological and admits no local propagating bulk degrees of freedom, classical solutions are characterised by flat Chern-Simons connections up to gauge, and in particular by their global holonomies together with boundary data fixed by the asymptotic conditions \cite{Banados:1992wn,Carlip:1995qv}. For any closed contour $\gamma$ in the bulk (or at the boundary at fixed $\rho$), the holonomy of the flat connection $A$ is the path-ordered exponential
\begin{align}\label{eq:holonomy_def}
\mathrm{Hol}_\gamma[A]= \mathcal P\exp \left(\oint_\gamma A\right)\ \in SL(2,\mathbb R)\,,
\end{align}
and similarly $\mathrm{Hol}_\gamma[\bar A]\in SL(2,\mathbb R)$ for $\bar A$. Under a gauge transformation $A\mapsto g^{-1}Ag+g^{-1}\mathrm dg$, one has $\mathrm{Hol}_\gamma[A]\mapsto g^{-1}(x_0)\mathrm{Hol}_\gamma[A]\,g(x_0)$, so the gauge-invariant information is the conjugacy class of $\mathrm{Hol}_\gamma[A]$ (equivalently its trace in the fundamental representation). For constant representatives $\mathcal L,\bar{\mathcal L}$, the Ba\~nados geometry is therefore characterised by the conjugacy classes of the spatial holonomies around the boundary circle $\gamma_\phi$.

In particular, for constant representatives $\mathcal L,\bar{\mathcal L}$, the associated Ba\~nados geometry is completely characterised by the conjugacy class of the $SL(2,\mathbb R)\times SL(2,\mathbb R)$ holonomies around the spatial circle. Three  distinct cases occur:
\begin{itemize}
    \item If $\mathcal L,\bar{\mathcal L} > 0$ (equivalently $h,\bar h > \tfrac{c}{24}$), the holonomies are hyperbolic and the geometry corresponds to a BTZ black hole. 
    \item If $\mathcal L=\bar{\mathcal L}=0$, the holonomies are parabolic, corresponding to the massless BTZ geometry. 
    \item If $-\,\tfrac{c}{24} \leq \mathcal L,\bar{\mathcal L} < 0$, the holonomies are elliptic and the geometry is global $\mathrm{AdS}_3$ or a conical defect thereof. The vacuum $\mathcal L=\bar{\mathcal L}=-\,\tfrac{c}{24}$ gives global $\mathrm{AdS}_3$. 
\end{itemize}

At this stage, the distinction between invertible and non-invertible circuit elements admits a simple bulk interpretation. Invertible Virasoro circuits act within a fixed coadjoint orbit and therefore correspond to continuous deformations of a single Ba\~nados geometry, leaving the associated holonomy conjugacy class invariant. By contrast, a non-invertible fusion gate changes the Virasoro highest-weight data $(h,\bar h)$ and hence moves the state between distinct coadjoint orbits. In the Chern-Simons formulation of AdS$_3$ gravity, this implies that fusion gates necessarily induce discrete changes in the allowed holonomy data. We will analyse the precise realisation of these jumps, and their interpretation as shock-like defects, in section \ref{sec:defect_to_shocks}.

\paragraph{Ba\~nados data for Ising primaries.}
The Ising model has three Virasoro primaries $\mathbf 1$, $\sigma$, and $\varepsilon$ with conformal weights $h_{\mathbf 1}=0$, $h_\sigma=\tfrac{1}{16}$, and $h_\varepsilon=\tfrac{1}{2}$. Using the elements reviewed in section \ref{sec:CS_Banados_hol} and in particular eq. \eqref{eq_h_to_mL}, the corresponding constant coadjoint-orbit representatives are
\begin{align}
\mathcal L_0(\mathbf 1)=-\frac{c}{24}=-\frac{1}{48}\,,\quad \mathcal L_0(\sigma)=\frac{1}{24}\,,\quad \mathcal L_0(\varepsilon)=\frac{23}{48}\,.
\end{align}
Since the theory is diagonal, the same values hold for the anti-holomorphic sector.

These values determine the holonomy classes of the associated Ba\~nados configurations:
\begin{align}
\mathbf 1 &: \text{elliptic $\times$ elliptic (global AdS holonomy class)}\,,\label{eq:Ising_glAdS}\\
\sigma &: \text{hyperbolic $\times$ hyperbolic}\,,\label{eq:IsingBTZ}\\
\varepsilon &: \text{hyperbolic $\times$ hyperbolic (larger mass parameter)}\,.\label{eq:IsingHeavyBTZ}
\end{align}
The same classification applies to all descendants within each conformal family. A parabolic holonomy would require $h=\tfrac{c}{24}=\tfrac{1}{48}$, which is not realised by any Ising primary.

Transitions due to shock-like defects between these sectors depend on the specific gate and input sector. For example, acting with a $\sigma$-gate on a state in the $\sigma$ sector admits precisely two output channels, $\sigma \times \sigma = \mathbf 1 \oplus \varepsilon$, corresponding to discrete jumps between the holonomy classes listed above. More generally, admissible jumps are fully determined by the fusion rules of the Ising category and do not allow arbitrary transitions between sectors.

\subsection{Fusion-induced energy jumps as shock-like defects}\label{sec:defect_to_shocks}

We now apply the Chern-Simons to Ba\~nados dictionary, eqs. \eqref{eq:EMT_to_CS_calL} and \eqref{eq:metric_EMT_to_CS_calL}, to non-invertible fusion gates. We first derive the discrete jumps in conformal weight and energy induced by fusion at the level of Virasoro representations, and then analyse the conditions under which these jumps admit a consistent bulk interpretation. This provides an interpretation of fusion-induced transitions between conformal families on the boundary as shock-like insertions in three-dimensional AdS gravity.

\subsubsection{Energy jumps from fusion intertwiners}
\label{subsec:fusion_energy_jump}

We now make precise how fusion gates induce discrete energy shifts at the level of Virasoro representations, and associated Ba\~nados geometries. The key input is that fusion intertwiners act nontrivially on highest-weight states, mapping primaries between different conformal families. This allows one to compute explicitly the change in Virasoro zero modes associated with each fusion channel, and hence the corresponding jump in boundary energy. These representation-theoretic energy jumps will then be translated into discontinuous Ba\~nados data, providing the basis for their interpretation as shock-like defects in three-dimensional AdS gravity.

For one chiral half of the theory, the Hamiltonian on the cylinder of circumference $2\pi R$ is
\begin{align}
H_L=\frac{2\pi}{R}\left(L_0-\frac{c}{24}\right)\,.
\end{align}
Let $\ket{h_a}$ denote the highest-weight vector of conformal weight $h_a$ in $\mathcal H_a$. Fusion with $b$ is implemented at the circuit level by the Stinespring isometry $W^{(a)}_b$, whose channel-resolved form involves intertwiners $t^{(\mu)}_{ba\to c}\in\mathrm{Hom}(a\otimes b,c)$.
On highest-weight states, these intertwiners map primaries to primaries, $t^{(\mu)}_{ba\to c}\ket{h_a}\propto\ket{h_c}$ and satisfy
\begin{align}
L_0\, t^{(\mu)}_{ba\to c}\ket{h_a} &= h_c\, t^{(\mu)}_{ba\to c}\ket{h_a}\,,\qquad t^{(\mu)}_{ba\to c}\,L_0\ket{h_a} = h_a\, t^{(\mu)}_{ba\to c}\ket{h_a}\,,
\end{align}
so that, on such states,
\begin{align}
[L_0,\,t^{(\mu)}_{ba\to c}]\,\ket{h_a} = (h_c-h_a)\,t^{(\mu)}_{ba\to c}\ket{h_a}\,.
\end{align}
Accordingly, the left-moving energy jump associated with a fusion outcome $c$ is
\begin{align}
\Delta H_L^{(c)}=\frac{2\pi}{R}\,(h_c-h_a)\equiv\frac{2\pi}{R}\,\Delta h_c\,.
\end{align}
To translate this into bulk language, we work directly with the Ba\~nados data. The holomorphic boundary stress tensor is related to the Chern-Simons connection by \eqref{eq:EMT_to_CS_calL}.
Identifying the circuit parameter $\tau$ with physical boundary time, a fusion gate inserted at $\tau=\tau_0$ corresponds to a localised boundary insertion at $t=\tau_0$. Projected onto a single chiral sector, this induces a localised profile in $x^+$ at fixed $x^-$. At the level of Ba\~nados data, the effect of a fusion outcome $c$ may therefore be modelled by a distributional jump
\begin{align}\label{eq:jumpL}
\Delta\mathcal L(x^+) =\frac{\Delta h_c}{R}\,\delta(x^+-u_0)\,,
\end{align}
corresponding to a localised modification of the Chern-Simons connection. The factor of $R^{-1}$ follows from the cylinder normalisation and ensures that $\mathcal L$ remains dimensionless. This representation should be viewed as a convenient parametrisation of a sudden change in conformal data, rather than as a statement about local bulk sources.

\begin{figure}[t]
    \centering
    \includegraphics[width=0.92\linewidth]{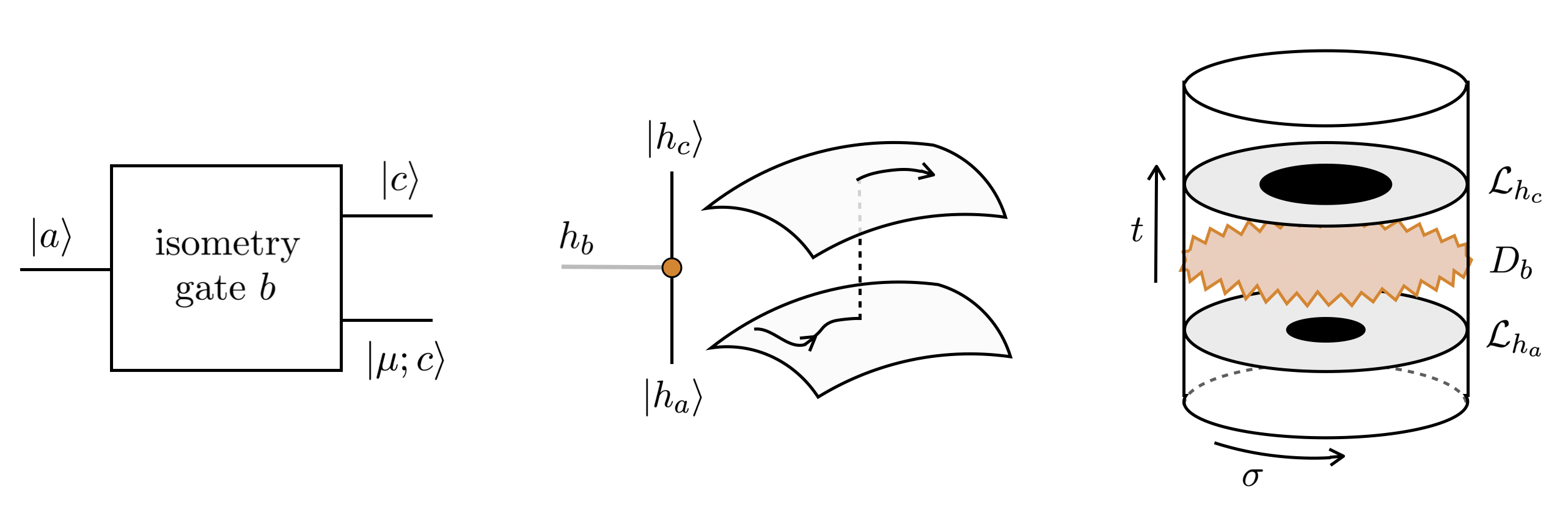}
    \caption{Illustration of a non-invertible fusion gate at three complementary levels. \textit{Left:} circuit-level view, in which fusion with a simple object $b$ acts as an isometric gate mapping an input sector $|a\rangle$ to a superposition of output sectors $|\mu;c\rangle$, with $\mu$ labelling junction degeneracies. \textit{Middle:} representation-theoretic picture, where fusion induces discrete jumps between conformal families labelled by their conformal weights $h_a,h_b,h_c$, rather than continuous motion along a single Virasoro coadjoint orbit. \textit{Right:} bulk cartoon in AdS$_3$, in which the fusion gate is interpreted as an instantaneous insertion of a defect $D_b$ at fixed boundary time, producing a discontinuous change in the boundary stress-tensor data and separating regions associated with different highest-weight representations. }
    \label{fig:cartoon_defectshock_AdS3}
\end{figure}

From the boundary perspective, a fusion gate thus implements a non-invertible operation that induces a discrete jump in Virasoro data, as illustrated in figure \ref{fig:cartoon_defectshock_AdS3}. In the Chern-Simons formulation, this corresponds to a transition between distinct flat connections characterised by different holonomy conjugacy classes. After the insertion, the circuit continues as in \eqref{eq:atomic_path}, with invertible Nielsen evolution proceeding on the new coadjoint orbit determined by the updated conformal weight.

Importantly, individual fusion channels may carry either positive or negative energy shifts \eqref{eq:jumpL}. The question of physical consistency concerns only the effective state obtained after summing over the fusion outcomes; restricting to a specific fusion channel represents an additional, resource-dependent operation.

Finally, we note that fusion-induced family changes share a feature with standard  ``shock'' constructions in AdS$_3$ such as Vaidya-type shells \cite{vaidya1999external} (see also e.g. \cite{Chapman:2018dem,Chapman:2018lsv,Jiang:2018tlu,Balasubramanian:2019stt}) or Shenker-Stanford shocks \cite{dray1985gravitational,Shenker:2013pqa,Shenker:2013yza,Roberts:2014ifa}) in that they implement an effectively instantaneous change in boundary stress-tensor data. In  the present setting, however, the defect is not a propagating gravitational disturbance but  a localised modification of Chern-Simons holonomy data in a topological theory. For this  reason, we will refer to these insertions as shock-like defects.

\paragraph{Example: Ising fusion channel.}
As a concrete illustration, consider the Ising fusion rule $\sigma\times\sigma=\mathbf 1\oplus\varepsilon$ with conformal weights $h_\sigma=\tfrac{1}{16}$, $h_{\mathbf 1}=0$ and $h_\varepsilon=\tfrac12$. Acting with a $\sigma$-gate on a highest-weight state in the $\sigma$ sector admits two possible fusion outcomes. Selecting the channel $\sigma\to\varepsilon$ or $\sigma\to\mathbf 1$ produces the conformal-weight shifts
\begin{align}\label{eq:jumps_Ising}
\Delta h_\varepsilon=\tfrac{7}{16}\,,\qquad \Delta h_{\mathbf 1}=-\tfrac{1}{16}\,,
\end{align}
corresponding respectively to positive- and negative-energy jumps.

If no specific fusion outcome is selected, one instead sums over both admissible channels with their canonical categorical weights. The resulting expectation value of the stress tensor is then an average over the two contributions. In this unselected case, the averaged stress tensor satisfies the averaged null energy condition, as discussed in section \ref{subsec:ANEC}. By contrast, negative-energy contributions arise only when a definite fusion channel is explicitly selected. This example thus makes clear that energy-violating effects are tied to the selection of a particular fusion outcome, rather than to the fusion operation itself.

At this point it is important to distinguish between two conceptually different quantities that appear in the discussion. The fusion-induced jump in the Ba\~nados function, and hence the strength of the corresponding shock-like defect, is controlled by the channel-weighted energy shift $\sum_c p_c\,\Delta h_c$. This is the quantity that will be constrained below by the averaged null energy condition. By contrast, the energy-weighted gate cost introduced in section \ref{sec:costs} does not quantify the magnitude of the energy injected into the bulk. Rather, as highlighted by eq. \eqref{eq:energy_cost_2ch}, it measures the circuit complexity required to bias a fusion gate toward particular energy-carrying channels relative to the canonical fusion weights. In this sense, the cost characterises how difficult it is to favour certain shock profiles, not how energetic those profiles are.

\subsubsection{Energy conditions in non-invertible circuits} \label{subsec:ANEC}

A basic consistency check for interpreting fusion-induced discontinuities in boundary stress-tensor data as admissible shock-like defects in AdS$_3$ gravity is compatibility with boundary energy conditions. In three dimensions, where gravity is topological and has no local propagating
degrees of freedom, the relevant constraint is the averaged null energy condition (ANEC), imposed on stress-tensor expectation values integrated along complete null generators \cite{Hofman:2008ar,Hartman:2016lgu}; see also \cite{Kundu:2021nwp}.

In the present setting, a fusion gate does not produce a single outcome, but rather a finite set of allowed fusion channels with fixed categorical weights. In the circuit description of section \ref{sec:gateisometry}, an unselected fusion gate is described by a quantum channel obtained by tracing over the channel register $\mathcal A_{ab}$. Expectation values of CFT observables, such as the stress tensor, are therefore computed in this reduced state and do not depend on which fusion channel is realised. In what follows we analyse the fusion operation at this level, where all allowed channels contribute with their canonical weights $p_c$ defined in \eqref{eq:expr_weigts}. Selecting a definite fusion outcome corresponds to an additional operation, discussed already in section \ref{sec:select_cost}. When assessing energy conditions, we therefore test the stress-tensor profile associated with the fusion gate itself, obtained by summing over all admissible fusion channels. The role of the averaged null energy condition is then to constrain whether this fusion-induced stress-tensor profile admits a consistent causal interpretation in the bulk.

From section \ref{sec:shocklike_def}, a fusion gate inserted at fixed boundary time induces a localised jump in the Ba\~nados function $\mathcal L(x^+)$ of the form\footnote{We work on the cylinder with angular coordinate $\sigma\sim\sigma+2\pi$ and
dimensionless time $\tau=t/R$, so that the lightcone coordinates $x^\pm=\tau\pm\sigma$ are dimensionless. 
}
\begin{align}\label{eq:shockL}
\Delta\mathcal L(x^+) = \frac{1}{R}\sum_c p_c\,\Delta h_c\;\delta(x^+-u_0)\,,
\end{align}
where $p_c$ are the categorical channel weights and $\Delta h_c=h_c-h_a$ are the conformal-weight shifts associated with the fusion outcomes. Using the standard relation between $\mathcal L$ and the boundary stress tensor, eq.\eqref{eq:EMT_to_CS_calL}, and choosing the future-directed null generator $\gamma_+: x^-=\mathrm{const.}$, the ANEC integral reduces in the shock limit to
\begin{align}
\int_{\gamma_+}\mathrm dx^+\, T_{++}(x^+) \propto \sum_c p_c\,\Delta h_c\,.
\end{align}
ANEC therefore imposes the simple constraint
\begin{align}\label{eq:ANEC_constraint}
\sum_c p_c\,\Delta h_c \ge 0\,.
\end{align}
In this sense, compatibility with ANEC constrains the fusion operation itself: the channel-weighted sum $\sum_c p_c\,\Delta h_c$ appearing in eq. \eqref{eq:ANEC_constraint} must be non-negative. Possible negative-energy contributions arise only when a specific fusion channel is explicitly selected, which constitutes an additional operation beyond the fusion gate.

\paragraph{Example: Ising fusion.}
For the Ising fusion rule $\sigma\times\sigma=\mathbf 1\oplus\varepsilon$ with $p_{\mathbf 1}=p_\varepsilon=\tfrac12$ and $\Delta h_{\mathbf 1}=-\tfrac{1}{16}$, $\Delta h_\varepsilon=\tfrac{7}{16}$, one finds
\begin{align}
\sum_c p_c\,\Delta h_c=\tfrac{3}{16}>0\,,
\end{align}
so the fusion gate, when all allowed channels are taken into account, obeys the ANEC constraint. By contrast, post-selecting the vacuum channel alone yields a negative-energy excitation and violates \eqref{eq:ANEC_constraint}.

\subsection{Costs for shock-like defects in rational sectors}
We now interpret fusion-induced circuit costs in gravitational terms, focusing on rational subsectors where discrete changes of Virasoro data admit a controlled bulk description in terms of Ba\~nados geometries. In this regime, fusion gates do not generate continuous deformations of a fixed solution but instead induce discrete transitions between distinct holonomy classes, which we interpret as shock-like defects in three-dimensional AdS gravity.

A purely geometric gate cost, sensitive only to the branching structure of fusion, is blind to the physical distinction between fusion outcomes corresponding to light and heavy bulk geometries. In particular, it assigns comparable cost to transitions that differ substantially in their conformal weights. The energy-weighted gate cost introduced in section \ref{sec:energy_cost} refines this by incorporating the conformal-weight shifts carried by the fusion channels, thereby penalising jumps according to the associated change in Virasoro zero modes and bulk holonomy data. Selecting a definite fusion outcome (thus enforcing a specific target geometry) is treated as a separate operation with its own cost, reflecting the additional control required to isolate a particular branch among the allowed shock-like transitions.

The examples below illustrate three complementary aspects of the gravitational interpretation of fusion costs: a fully explicit Ising consistency check, a single fusion gate branching across holonomy classes, and a multi-step circuit in which different intermediate geometries lead to different total costs despite identical endpoints.

\paragraph{Minimal example: the Ising $\sigma$-gate.}

As a simplest consistency check, consider the Ising CFT with primaries $\mathbf 1$, $\sigma$, $\varepsilon$. Acting with the $\sigma$ fusion gate on the $\sigma$ sector gives $\sigma\otimes\sigma=\mathbf 1\oplus\varepsilon$ corresponding to a branch between an elliptic (conical) and a heavier hyperbolic (BTZ) sector. The channel weights are equal, and the energy-weighted gate cost is $\pi/4$, while the geometric selection cost is the same for both outcomes.

This example illustrates that selection costs are taken to be energy-insensitive by design: they quantify the control required to isolate a specific fusion channel, while the energetic asymmetry between outcomes is captured entirely by the intrinsic, energy-weighted gate cost. This separation is a modelling choice, but it is internally consistent and sufficient for grading fusion-induced shock-like transitions in the bulk.

\paragraph{Fusion gate branching branching across holonomy classes.}

A simple instance where a single gate branches between different holonomy classes is provided by the $\widehat{su(2)}_3$ WZW category. 
We take the fusion gate label $b=\tfrac12$ acting on the input sector $a=\tfrac12$, for which
\begin{align}
\tfrac12\otimes\tfrac12 = 0\oplus 1\,.
\end{align}
Using \eqref{eq_h_to_mL}, one obtains
\begin{align}
\mathcal L_0(0)= -\frac{3}{40}<0\quad (\text{elliptic})\,,\quad
\mathcal L_0(1)=\frac{13}{40}>0\quad (\text{hyperbolic})\,,
\end{align}
while the input sector is already hyperbolic $\mathcal L_0(\tfrac12)=\frac{3}{40}>0$. Thus the same fusion gate allows either a conical AdS-branch ($c=0$) or a heavier BTZ branch ($c=1$).

The two fusion outcomes correspond to opposite shifts of the Virasoro zero mode: the $c=0$ channel lowers $\mathcal L_0$, while $c=1$ raises it. In this sense, the gate branches between lighter and heavier bulk geometries, a distinction that is invisible to purely geometric costs but directly enters the energy-weighted gate contribution.

At the level of the fusion gate itself, the channel-averaged conformal-weight shift is positive, so the induced stress tensor satisfies the averaged null energy condition. As discussed in section \ref{subsec:ANEC}, violations arise only upon post-selection onto a specific channel, which constitutes an additional operation beyond the gate.

Using the details of the same path considered in section \eqref{sec:extendedcosts}, see \eqref{eq:gatecostE_return} for the energy-wieghted gate cost. Forcing a definite outcome incurs the geometric selection costs
\begin{align}
\mathcal C_{\rm sel} \left(0\,\middle|\,\tfrac12,\tfrac12\right) =\arccos \left(\varphi^{-1}\right)\,,\qquad
\mathcal C_{\rm sel} \left(1\,\middle|\,\tfrac12,\tfrac12\right) =\arccos \left(\varphi^{-1/2}\right)\,.
\end{align}

\paragraph{The cost of an elliptic detour.}

In the $\widehat{su(2)}_3$ category, consider the two-step return $\tfrac12 \xrightarrow{1/2} x \xrightarrow{1/2} \tfrac12$, already considered in section \ref{sec:extendedcosts}, which admits two routes,
\begin{align}
\Gamma_{\rm ell}:\ \tfrac12\to 0\to \tfrac12\,,\qquad
\Gamma_{\rm hyp}:\ \tfrac12\to 1\to \tfrac12\,.
\end{align}
The intermediate sector $x=0$ is elliptic, while $x=1$ is hyperbolic. Using the
costs computed in \eqref{eq:su2_3_energy_costsmin} and \eqref{eq:su2_3_energy_costsplus}, identifying $\Gamma_-=\Gamma_+$ and $\Gamma_-=\Gamma_+$, one finds the cost of the selected paths
\begin{align}
\mathcal C_{\rm energy}(\Gamma_{\rm hyp})
&=\frac{\pi}{2}+2\,\arccos \left(\varphi^{-1/2}\right)\,,\\
\mathcal C_{\rm energy}(\Gamma_{\rm ell})
&=\arctan \big(\sqrt{\varphi}\big)+\arccos \left(\varphi^{-1}\right)\,.
\end{align}
Although both routes begin and end in the same sector, they carry different total costs. In bulk terms, forcing a brief excursion through an elliptic (conical AdS) geometry and returning deterministically is cheaper than remaining entirely within hyperbolic (BTZ-like) geometries, which requires additional branching and selection steps.

\section{Conclusions and future directions}\label{sec:conclusions}

In this work we extended Nielsen's geometric formulation of quantum circuit complexity to include intrinsically non-invertible gate operations. Enlarging the gate set to include categorical fusion operations removes a basic limitation of symmetry-based circuits, namely their inability to change conformal families or superselection sectors. We realised fusion operations as completely positive, trace-preserving quantum channels acting between sectors, with consistency and associativity ensured by the fusion rules and associators of an underlying unitary modular tensor category. This led to a well-defined notion of circuit composition that is associative but non-invertible, and whose action depends intrinsically on its ordering relative to invertible symmetry evolution.

A central departure from the standard Nielsen framework is that fusion gates do not act transitively on the space of conformal families. Only sectors connected by nonzero fusion coefficients are reachable, and the resulting circuit configuration space is no longer a smooth manifold but a discrete structure determined by the fusion graph of simple objects. Circuit optimisation therefore ceases to be a geodesic problem on a continuous orbit space, and instead becomes a shortest-path problem on a weighted, directed graph. Reachability of target states and associated cost depend explicitly on the available gate set and on the branching structure determined by the admissible fusion paths.

The cost assignments chosen track two distinct tasks: applying a fusion gate that opens several allowed channels, and perform the optional extra step that selects a single target channel.  This naturally leads to a decomposition of fusion costs into an intrinsic gate cost, associated with opening the set of allowed channels, and a separate selection cost incurred when isolating a definite target sector. For physical applications in which fusion changes conformal weights, we further refined the intrinsic gate cost by incorporating energy data through a controlled energy-weighting. Together with the standard Nielsen cost for invertible symmetry gates, these ingredients combine into a single additive expression assigning a well-defined complexity to circuits built from both invertible and non-invertible operations. We illustrated this structure in rational conformal field theories, focusing on the categories associated to the Ising model and WZW models with chiral algebra $\widehat{su(2)}_k$.  

Finally, we interpreted fusion-induced transitions in the context of three-dimensional AdS gravity formulated as $SL(2,\mathbb R)\times SL(2,\mathbb R)$ Chern-Simons theory. In rational sectors, a fusion insertion can be modelled as a shock-like defect on the boundary: a localised, discrete jump in Virasoro data and hence transitions between admissible Ba\~nados geometries, characterised by changes in holonomy data rather than by continuous deformations of a fixed solution. From this perspective, the intrinsic gate cost quantifies the energetic imbalance between allowed transitions, while the selection cost measures the additional resources required to enforce a specific geometric outcome.  


\vspace{15pt}

\noindent
Finally, let us mention a number of interesting future directions: 

\vspace{-5pt}

\paragraph{Large fusion graphs and discrete optimisation.}
Non-invertible fusion gates replace Nielsen's geodesic optimisation problem by a shortest-path problem on a directed, weighted graph: vertices correspond to superselection sectors, while edges encode admissible fusion transitions, weighted by the gate costs and, when imposed, selection costs. For small unitary modular tensor categories such as the Ising model this structure is rather simple, but for categories with many simple objects and nontrivial fusion multiplicities the space of admissible paths grows rapidly and exhibits genuinely combinatorial complexity. In those more extended cases, circuit optimisation becomes a  genuine problem of navigating a discrete, resulting notion of complexity. Since for invertible Nielsen complexity the appearance of cut-locus or conjugate-point how been shown to lead to obstructions to global optimality, see e.g. \cite{Balasubramanian:2019wgd,Balasubramanian:2021mxo,Auzzi:2020idm},  it would be interesting how global features of such fusion graph rather than on local metric data alone, would influence the complexity growth.

\paragraph{Which Nielsen-circuit phenomena survive non-invertibility?}
In the invertible Nielsen framework, circuit complexity is formulated as a geodesic or optimal-control problem whose qualitative behaviour depends on penalty factors and cost assignments \cite{Nielsen:2005mkt,Jefferson:2017sdb,Chapman:2021jbh}. Well-known features include the switchback effect in chaotic evolutions \cite{Stanford:2014jda,Roberts:2014isa,Brown:2015bva,Brown:2015lvg}.
Allowing intrinsically non-invertible fusion gates necessarily modifies or removes these mechanisms. For example, exact forward-backward cancellations rely on invertibility and are generically obstructed once a fusion step is inserted, so that circuit optimisation becomes a hybrid problem combining continuous Nielsen evolution with discrete shortest-path optimisation on the fusion graph. A natural direction is to revisit which qualitative features of Nielsen complexity survive in this setting, and which are intrinsically tied to invertibility.

\paragraph{Beyond UMTCs.}
Our construction is deliberately restricted to rational subsectors, where fusion data are finite and organised by a unitary modular tensor category. A natural direction for future work is to ask whether an analogous channel-based circuit framework can be developed beyond this setting. The relevant representation data is then continuous and fusion is described by integral kernels rather than finite sums. Such structures are known to arise in non-rational two-dimensional conformal field theories, most notably in Liouville CFT, where associativity is governed by Virasoro fusion kernels instead of finite $F$-symbols \cite{Ponsot:1999uf}. Continuous fusion data also appear naturally in the Chern-Simons formulation of AdS$_3$ gravity and in related settings involving non-compact quantum groups, where the discrete anyon types of rational theories are replaced by continuous families of representations \cite{McGough:2013gka}, see also \cite{Mertens:2022ujr,Wong:2022eiu}. It would be interesting to exploring whether such continuous fusion structures admit a similar circuit interpretation as the one constructed here.

\paragraph{SymTFT reformulation.}
A natural question raised by our construction is whether fusion circuits admit a reformulation directly within the framework of symmetry topological field theories \cite{muger2003subfactors,Freed:2022qnc,Freed:2009qp,Kaidi:2021gbs,Bhardwaj:2023idu}. Since fusion gates are implemented by defect junctions, it is tempting to ask whether the circuit-level distinction between ``opening'' all fusion channels and ``selecting'' a definite outcome has an intrinsic SymTFT avatar. One possibility is that selection amounts to choosing a simple summand by inserting an idempotent (e.g. a minimal projector in the tube algebra), which canonically organises superselection sectors and defect composition. Making this correspondence precise would require specifying how such topological projections implement the operational notion of projecting onto a definite fusion outcome used in the circuit description. If so, this would give a purely topological characterisation of our non-invertible circuit operations, independent of auxiliary Hilbert-space constructions.

\paragraph{Higher-dimensional generalisations.}
It is natural to ask whether aspects of the present framework admit extensions beyond two-dimensional rational conformal field theory. In higher dimensions, non-invertible symmetries and topological defects are no longer organised by unitary modular tensor categories, but by more general higher fusion or braided tensor-categorical structures, often lacking finiteness, rigidity, or modularity. While some elements of the circuit construction, such as the interpretation of the non-invertible gates constructed here as completely positive maps and the resulting shortest-path optimisation on a directed graph of superselection sectors, may persist. Moreover, any holographic or gravitational interpretation is inherently tied to AdS$_3$ and does not extend directly to higher-dimensional gravity, where local degrees of freedom and different energy conditions play a central role. It would be interesting to clarifying which aspects of non-invertible circuit complexity are genuinely two-dimensional and which reflect more general principles.

\subsection*{Acknowledgments}
I would like to thank Stefano Baiguera, Shira Chapman, Rathindra Nath Das and  Arnab Kundu for interesting discussions and comments. I would also like to thank  DESY and Universit\"at Hamburg, where part of this work was carried out, for the kind hospitality.

\newpage
\appendix

\section{Basics of unitary modular tensor categories}\label{app:UMTC}

In a rational CFT (e.g. WZW models or minimal models), the chiral data (representations of the chiral algebra, fusion rules, braiding, etc.) are naturally encoded by a modular tensor category $\mathscr C$.

A modular tensor category is, in particular, a ribbon fusion category with a finite set of simple objects, associative fusion, braiding, a twist, and a non-degenerate modular $S$-matrix. A  unitary MTC has, in addition, a $\star$-structure (an involution on morphisms) compatible with the monoidal, braiding, and ribbon structures, such that all morphism spaces are finite-dimensional Hilbert spaces with a positive-definite inner product compatible with composition.

We summarise the main ingredients we will use. See e.g. the book \cite{bakalov2001lectures}, appendix A of \cite{Kaidi:2021gbs}, section~5 of \cite{Benini:2018reh}, or appendix~E of \cite{Kitaev:2005hzj}. For a Chern--Simons theory, one associates a category $\mathscr C$ whose objects may be thought of as Wilson lines (equivalently, integrable representations of the chiral algebra), and whose morphisms are maps between Wilson lines (intertwiners between representations). We will assume throughout that the category is semisimple: simple objects have only scalar endomorphisms, and every object decomposes as a finite direct sum of simples. There is a distinguished unit object $\mathbf 1$, corresponding physically to the vacuum or trivial representation.

\paragraph{Simple objects and fusion rules.}
The category consists of objects or labels $a,b,\dots$ obeying fusion rules
\begin{align}\label{eq:fusion_anyons}
a\otimes b=\sum_c N^c_{ab}\,c\,,\qquad N^c_{ab}\in\mathbb Z_{\ge 0}\,.
\end{align}
The non-negative integers $N^c_{ab}$ are the fusion multiplicities. An anyon or label $a$ is called Abelian if $a\otimes b$ contains a single simple object for every $b$, i.e. $\sum_c N^c_{ab}=1$ for all $b$. If there exists at least one $b$ for which the fusion contains more than one simple summand, then $a$ is  non-Abelian.

Physically, the labels are charges of anyons or quasiparticles in the topological phase described by the Chern--Simons theory (equivalently, highest weights of irreducible representations of the gauge group $G$ in the CS description). One may view the coefficients $N^c_{ab}$ as the analogue of Clebsch--Gordan multiplicities in the decomposition of tensor products. The space of morphisms from $a\otimes b$ to $c$ is
\begin{align}
V^c_{ab}=\mathrm{Hom}(a\otimes b,c)\,,
\end{align}
and its dimension is precisely the fusion coefficient: $N^c_{ab}=\dim V^c_{ab}$ (symmetric in $a,b$ in the unitary MTC examples we consider). In a unitary MTC each $V^c_{ab}$ carries a natural positive-definite inner product, so one can choose orthonormal bases of junction states in these fusion spaces.

\paragraph{Associator and braiding data: $F$ and $R$.}
The recoupling of fusion channels and the braiding of (simple) anyons are captured by the $F$-symbols and $R$-symbols, respectively: they give linear relations between  isomorphic spaces of morphisms corresponding to different fusion trees and different orderings. These data satisfy consistency conditions (pentagon ir hexagon identities). When these relations are consistently defined, the category is a braided fusion category.

Diagrammatically, self-braiding corresponds to exchanging two anyons twice, for example
\begin{align}\label{eq:pivotal}
a\;\,\includegraphics[valign=c,scale=0.16]{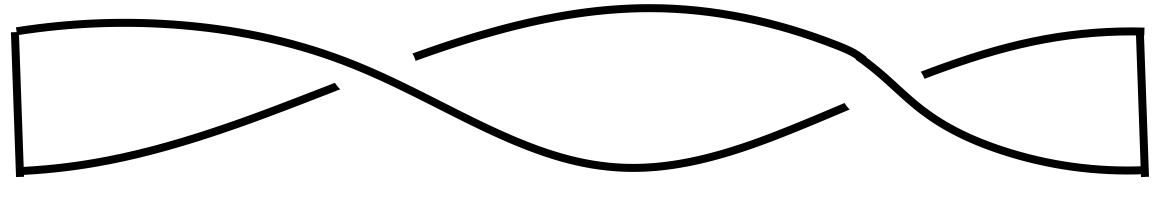}\;\,a\,,
\end{align}
where time runs horizontally. To depict a $2\pi$ rotation in a ribbon category one thickens worldlines into ribbons.

\paragraph{Duals and pivotal structure.}
We assume the category is pivotal: for every label $a$ there is a dual $\bar a$ such that
$a\otimes\bar a$ contains the vacuum,
\begin{align}
a\otimes\bar a=\mathbf 1+\cdots\,.
\end{align}
Equivalently, any anyon can be created from the vacuum together with its dual (anti-anyon). This structure is the starting point for defining evaluation and coevaluation maps and trace operations later in the appendix.

\paragraph{Ribbon structure, twist, and quantum dimensions.}
To obtain a ribbon braided fusion category, one introduces an isomorphism from $a$ to $\bar a$ (compatible with the braiding and pivotal structure). These isomorphisms can be encoded by phases $\epsilon_a$ for each label $a$. Physically, this is the topological spin $\theta(a)$: the phase picked up when the anyon is rotated by $2\pi$. This phase is constrained by relations involving the $F$-symbols. (In many unitary examples one may choose conventions in which $\epsilon_a = \pm 1$, we will not need this choice explicitly.)

The quantum dimension of an anyon is defined by a trace-like operation,
\begin{align}\label{eq:quantumdim}
d_a=\mathrm{tr}(\mathrm{id}_a)=\mathcircled{\phantom{\tfrac{\mathrm df}{\mathrm d x}}}\;a\,.
\end{align}
One has $d_a=d_{\bar a}$ and $d_{\mathbf 1}=1$. Abelian anyons have $d_a=1$. The total dimension of the category is
\begin{align}
D=\sqrt{\sum_a d_a^2}\,.
\end{align}
In a unitary theory the quantum dimensions are positive, and the modular $S$-matrix is unitary.

\paragraph{Modularity and the $S$-matrix.}
Finally, one does not want `transparent' anyons: every nontrivial anyon (i.e. not the vacuum $\mathbf 1$) should braid nontrivially with at least one other anyon. This is equivalent to non-degeneracy (invertibility) of the modular $S$-matrix. One convenient expression is
\begin{align}\label{eq:ModularS_from_Rmatrix}
S_{ab}=\frac{1}{D}\sum_c d_c\,\mathrm{tr} \left(R^c_{ab}R^c_{ba}\right) =\frac{1}{D}\,\tilde S_{ab}\,,
\end{align}
with a corresponding diagrammatic representation
\begin{align}
    \raisebox{-0.51cm}{\includegraphics[scale=0.22]{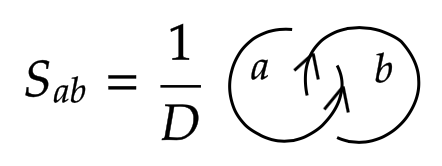}}
\end{align}
While the $F$- and $R$-symbols depend on basis choices in the fusion spaces, quantities such as $\theta_a$ and the modular $S$-matrix are basis-independent.

\paragraph{Modular data, Verlinde formula, and $c_-$.}
When $S_{ab}$ is invertible, one may define
\begin{align}
T_{aa}=d_a^{-1}\sum_c d_c\,R^c_{aa}=\theta_a\,,
\end{align}
where $\theta_a$ is the topological spin introduced above. Then $S$ and $T$ furnish a unitary projective representation of the modular group $SL(2,\mathbb Z)$. The fusion coefficients can be recovered from $S$ via the Verlinde formula
\begin{align}\label{eq:Verlinde_matrix}
N^c_{ab}=\sum_e\frac{S_{ae}S_{be}S_{ec}}{S_{0e}}\,.
\end{align}
Finally, the chiral central charge $c_-$ (defined modulo $8$) is determined by
\begin{align}
\frac{1}{D}\sum_a \theta_a d_a^2 =e^{2\pi i c_-/8}\,.
\end{align}

\paragraph{Encircling, monodromy scalar, and Abelian braiding.}
Before moving on, let us summarise a couple of useful properties encoded by the $S$-matrix. The modular $S$ controls the effect of encircling one anyon by another, or equivalently can remove an encircling anyon:
\begin{align}\label{eq:diag_action}
\raisebox{-1.2cm}{\includegraphics[scale=0.2]{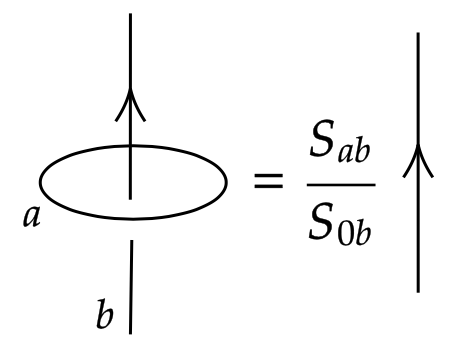}}
\end{align}
In particular, $S_{00}=1/D$ and $S_{0a}=d_a/D$. One can also define the ``monodromy scalar component''
\begin{align}
M_{ab}=\frac{S_{ab}^\ast S_{00}}{S_{0a}S_{0b}}\,.
\end{align}
When this quantity is a phase, $M_{ab}=e^{i\phi(a,b)}$, the braiding of $a$ with $b$ is Abelian:
\begin{align}
\raisebox{-1.1cm}{\includegraphics[scale=0.22]{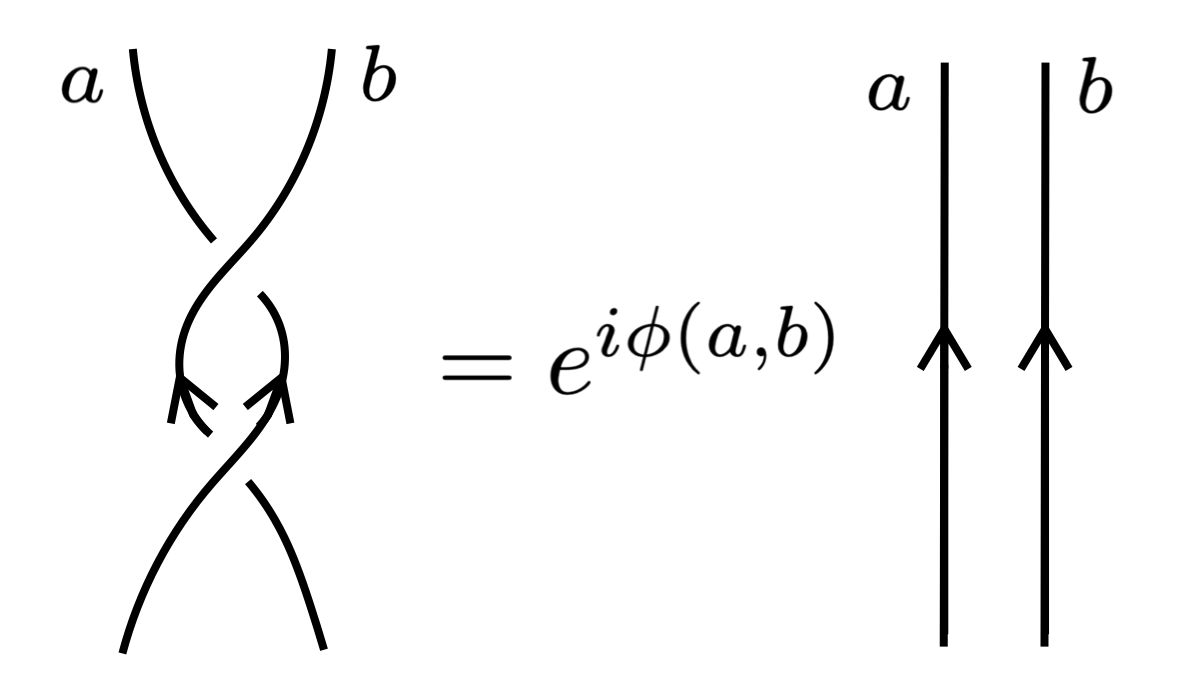}}
\end{align}
A useful consequence is that if an anyon $a$ has Abelian braiding with all other charges, meaning $M_{ab}=e^{i\phi(a,b)}$ for all $b$ in the category, then $a$ has quantum dimension $d_a=1$:
\begin{align}
d_a^2 &=\sum_b\left|\frac{d_a d_b}{D}e^{i\phi(a,b)}\right|^2 =\sum_b\left|\frac{S_{0a}S_{0b}}{S_{00}}M_{ab}\right|^2 =\sum_b |S_{ab}|^2 =1\,,
\end{align}
where the last step uses unitarity of the $S$-matrix. This implies that non-Abelian anyons have $d_a>1$, or equivalently must have non-Abelian braiding in a unitary modular tensor category.

\paragraph{Categorical trace and partial trace.}
Using the previous ingredients one can define a ``categorical trace'' by introducing evaluation and coevaluation maps. In a unitary (spherical) fusion category one has
\begin{align}\label{eq:def_coev_ev}
\mathrm{ev}_X:X\otimes X^*\to\mathbf 1\,,\qquad \mathrm{coev}_X:\mathbf 1\to X^*\otimes X\,,
\end{align}
and the categorical trace of $F\in\mathrm{End}(X)$ is
\begin{align}
\mathrm{tr}_X(F) =\mathrm{ev}_X\circ(\mathrm{id}_{X^*}\otimes F)\circ\mathrm{coev}_X \in \mathrm{End}(\mathbf 1)\cong\mathbb C\,.
\end{align}
One usually normalises so that $\mathrm{tr}_X(\mathrm{id}_X)=d_X$.

From this trace one defines a partial trace. For $F\in\mathrm{End}(B\otimes A)$ the partial trace over $A$ is
\begin{align}\label{eq:partial_cat_tr}
\mathrm{tr}_A(F) =(\mathrm{id}_B\otimes\mathrm{ev}_A)\circ(F\otimes\mathrm{id}_{A^*})\circ(\mathrm{id}_B\otimes\mathrm{coev}_A) \in \mathrm{End}(B)\,.
\end{align}
``Sphericality'' implies cyclicity:
\begin{align}\label{eq:spherical_impl}
\mathrm{tr}_B \left(\mathrm{tr}_A(F)\right)=\mathrm{tr}_{B\otimes A}(F)\,.
\end{align}
We emphasise that this categorical trace should not be confused with the usual Hilbert-space inner product (trace) on the vector space of intertwiners $\mathrm{Hom}(a\otimes b,c)$.

\section{Elements of quantum-circuit theory and quantum channel theory}\label{app:qc_basics}

This appendix fixes notation and recalls a minimal set of notions from quantum information theory that are used implicitly in the main text. The discussion is deliberately restricted to those ingredients required to describe non-invertible gate operations in a circuit language and is not intended as a general review. Standard introductions may be found in \cite{NielsenChuang,PreskillQI}.

\paragraph{Quantum circuits and CPTP maps.}
A quantum circuit will be understood as a composition of elementary operations acting on a Hilbert space $\mathcal H$. In the standard Nielsen formulation, these operations are unitary and therefore invertible. More generally, and in particular for the purposes of this work, it is necessary to allow operations that are not invertible. The appropriate framework is that of completely positive, trace-preserving (CPTP) maps, also known as quantum channels.

A quantum channel $\mathcal E$ acts on density operators $\rho \in \mathcal B(\mathcal H)$ and preserves both positivity and unit trace. Any CPTP map admits a Kraus representation
\begin{align}
\mathcal E(\rho) = \sum_\alpha K_\alpha \rho K_\alpha^\dagger\,,\qquad \sum_\alpha K_\alpha^\dagger K_\alpha = \mathbf 1_{\mathcal H}\,,
\end{align}
where the operators $K_\alpha : \mathcal H \to \mathcal H$ are called Kraus operators. Different sets of Kraus operators related by a unitary rotation of the index $\alpha$ describe the same physical channel. A unitary gate is recovered as the special case in which a single Kraus operator appears.

\paragraph{Stinespring dilation and ancilla degrees of freedom.}
A result the will underly the use of non-invertible gates is the Stinespring dilation theorem. It states that any CPTP map can be realised by embedding the system into a larger Hilbert space, applying an isometry, and discarding auxiliary degrees of freedom.

Concretely, there exists an auxiliary Hilbert space $\mathcal A$ and a linear map
\begin{align}
W : \mathcal H \longrightarrow \mathcal H \otimes \mathcal A\,, \quad W^\dagger W = \mathbf 1_{\mathcal H}\,,
\end{align}
such that, tracing over the ancilla space, one obtains the quantum channel
\begin{align}
\mathcal E(\rho) = \mathrm{tr}_{\mathcal A} \left[ W \rho W^\dagger \right]\,.
\end{align}
The map $W$ is called a Stinespring isometry. It preserves inner products on the input space but need not be surjective, so that $W W^\dagger \neq \mathbf 1$ in general. Irreversibility arises precisely because the auxiliary system is traced out.

Physically, this construction  is based on simple idea: any non-unitary operation can be realised as a unitary evolution on a larger system (system plus ancilla), followed by discarding the ancilla. Equivalently, one may use an isometry $W$ that is unitary on its image, and the irreversible character enters only upon tracing out the auxiliary record $\mathcal A$.

Given a Kraus decomposition of $\mathcal E$, a corresponding Stinespring isometry may be
chosen as
\begin{align}
W = \sum_\alpha K_\alpha \otimes |\alpha\rangle_{\mathcal A}\,,
\end{align}
where $\{|\alpha\rangle_{\mathcal A}\}$ is an orthonormal basis of the auxiliary space. Projecting $W$ onto a basis element of $\mathcal A$ recovers the corresponding Kraus operator. From this perspective, the isometry $W$ is the elementary circuit operation, while the CPTP map arises only after discarding the auxiliary system.

\paragraph{Cross-sector Kraus operators.}
Often, including those studied in the main text, the Hilbert space admits a decomposition into sectors (e.g. conformal families or topological charge sectors),
$\mathcal H=\bigoplus_a \mathcal H_a$.
\begin{align}
\mathcal H = \bigoplus_a \mathcal H_a\,.
\end{align}
It is then natural to allow Kraus operators that map between different sectors. We will refer to such operators as cross-sector Kraus operators and denote them by
\begin{align}
K_{c,\mu} : \mathcal H_a \longrightarrow \mathcal H_c\,,
\end{align}
where $\mu$ labels a finite degeneracy. The trace-preservation condition becomes
\begin{align}
\sum_{c,\mu} K_{c,\mu}^\dagger K_{c,\mu} = \mathbf 1_{\mathcal H_a}\,.
\end{align}
In the main text, these operators arise from categorical fusion data, but no categorical input is required for the abstract circuit discussion given here.

\paragraph{Positive operator-valued measures (POVMs).}
Closely related to Kraus representations is the notion of a positive operator-valued measure. Given a set of Kraus operators $\{K_\alpha\}$, the operators
\begin{align}
E_\alpha = K_\alpha^\dagger K_\alpha
\end{align}
are positive and sum to the identity. The collection $\{E_\alpha\}$ therefore defines a POVM on $\mathcal H$.

Operationally, POVMs describe generalised measurements or coarse-grainings of a quantum system. In the present work, they are used to encode how different outcomes are weighted when auxiliary degrees of freedom are not resolved, for example when fusion-channel data carried by junction degrees of freedom are not retained.

\section{Associativity of non-invertible quantum channels}\label{app:assoc}

In this appendix we verify that the non-invertible fusion channels defined in the main text are compatible with the associator of a unitary modular tensor category\footnote{Where we emphasise that  ``associativity'' here does not refer to the abstract composition of CPTP maps, which is always associative, but to the independence of the resulting quantum channel from the choice of fusion-tree parenthesisation.}.  Concretely, the composite channel obtained by applying two fusion gates admits Kraus decompositions associated with the two fusion trees $((a\otimes b)\otimes c)\to y$ and $(a\otimes(b\otimes c))\to y$. These decompositions are related by an isometry on the Kraus index space (in fact unitary, because the $F$-move is unitary in a unitary MTC), and therefore define the same completely positive trace-preserving map.

We begin by recalling the implementation of fusion with a simple object $b$ as a quantum channel acting on the sector $a$. Let $V_{ab}^{x}=\mathrm{Hom}(a\otimes b,x)$ denote the fusion multiplicity spaces. Fusion with $b$ is implemented by an isometry
\begin{align}
W_{ab}:\ \mathcal H_a \longrightarrow \bigoplus_x \mathcal H_x \otimes V_{ab}^{x}\,,
\end{align}
with Kraus operators
\begin{align}
K^{(a,b)}_{x,\mu}
=\sqrt{\frac{d_x}{d_a d_b}}\,t^{(\mu)}_{ab\to x}\,,\quad t^{(\mu)}_{ab\to x}\in V_{ab}^{x}\,,
\end{align}
where $\{t^{(\mu)}_{ab\to x}\}$ is an orthonormal basis of intertwiners.

We now consider the sequential application of two fusion gates $b$ and $c$ to an input sector $a$, and construct the associated quantum channel for the parenthesisation $((a\otimes b)\otimes c)\to y$. Applying first $b$ and then $c$ yields the composite isometry
\begin{align}
W^{((ab)c)} = \widetilde W_c \circ W_{ab}\,, \qquad \widetilde W_c = \bigoplus_x (W_{xc}\otimes \mathbf 1_{V_{ab}^{x}})\,.
\end{align}
The corresponding Kraus operators are labelled by $(x,\mu,\lambda)$ and read
\begin{align}
L^{(a;b,c)}_{y,(x,\mu,\lambda)}=K^{(x,c)}_{y,\lambda}\, K^{(a,b)}_{x,\mu}\,.
\end{align}
The resulting channel is
\begin{align}
\Phi^{((ab)c)}(\rho_a)=\sum_{y,x,\mu,\lambda}L^{(a;b,c)}_{y,(x,\mu,\lambda)}\,\rho_a\,\left(L^{(a;b,c)}_{y,(x,\mu,\lambda)}\right)^\dagger\,,
\end{align}
and is naturally indexed by a basis of $\bigoplus_x V_{ab}^{x}\otimes V_{x c}^{y}$.

Next, we describe an equivalent Kraus decomposition of the same composite channel obtained above, but written in the alternative fusion-tree basis. The combined ancilla space for two successive fusions admits two canonical decompositions,
\begin{align}
\bigoplus_x V_{ab}^{x}\otimes V_{x c}^{y} \qquad \text{and}\qquad \bigoplus_z V_{bc}^{z}\otimes V_{a z}^{y}\,,
\end{align}
corresponding to the two parenthesisations of the fusion tree with fixed external labels $(a,b,c;y)$. Choosing an orthonormal basis in the second decomposition defines an alternative set of Kraus operators, which we denote by $M^{(a;b,c)}_{y,(z,\nu,\sigma)}$ with indices $(z,\nu,\sigma)$ specifying an element in $\bigoplus_z V_{bc}^{z}\otimes V_{a z}^{y}$, and hence an alternative Kraus representation of the composite channel,
\begin{align}
\Phi^{(a(bc))}(\rho_a)=\sum_{y,z,\nu,\sigma}M^{(a;b,c)}_{y,(z,\nu,\sigma)}\,\rho_a\,
\left(M^{(a;b,c)}_{y,(z,\nu,\sigma)}\right)^\dagger\,.
\end{align}
The equivalence of these two channel constructions ultimately follows from the associator of the unitary modular tensor category, which acts at the level of fusion intertwiners. Concretely, the associator provides a canonical unitary isomorphism
\begin{align}
F^{abc}_y:\ \bigoplus_x V_{ab}^{x}\otimes V_{x c}^{y}\;\xrightarrow{\ =\ }\;
\bigoplus_z V_{bc}^{z}\otimes V_{a z}^{y}\,,
\end{align}
relating the two spaces of fusion trees corresponding to the parenthesisations $((a\otimes b)\otimes c)\to y$ and $(a\otimes(b\otimes c))\to y$. At the level of intertwiners, this means that the composed maps
\begin{align}
t^{(\lambda)}_{x c\to y}\, t^{(\mu)}_{ab\to x}\in \mathrm{Hom}((a\otimes b)\otimes c,\, y)
\end{align}
can be expanded in the alternative fusion basis as
\begin{align}
t^{(\lambda)}_{x c\to y}\, t^{(\mu)}_{ab\to x}=\sum_{z,\nu,\sigma}F^{abc}_y \left[(z,\nu,\sigma),(x,\mu,\lambda)\right]\,t^{(\sigma)}_{a z\to y}\, t^{(\nu)}_{b c\to z}\,,
\end{align}
where the coefficients are precisely the $F$-symbols of the category. Here we use the canonical Hilbert-space inner products on the fusion spaces $V_{ab}^{x}$ of a unitary MTC (see appendix \ref{app:UMTC}), and we choose the bases $\{t^{(\mu)}_{ab\to x}\}$ orthonormal with respect to these inner products. With this convention, the $F$-move acts unitarily on $\bigoplus_x V_{ab}^{x}\otimes V_{xc}^{y}$.

To now compare Kraus decompositions, note that the overall dimension prefactor is independent of the intermediate label. Indeed,
\begin{align}
\sqrt{\frac{d_y}{d_x d_c}}\,\sqrt{\frac{d_x}{d_a d_b}} =\sqrt{\frac{d_y}{d_a d_b d_c}} =\sqrt{\frac{d_y}{d_z d_a}}\,\sqrt{\frac{d_z}{d_b d_c}}\,.
\end{align}
Thus the $F$-move on intertwiners lifts directly to a unitary change of Kraus basis. To then pass to the quantum-channel description, recall that the Kraus operators are defined as dimension-weighted intertwiners,
\begin{align}
L^{(a;b,c)}_{y,(x,\mu,\lambda)} &=\sqrt{\frac{d_y}{d_x d_c}}\,\sqrt{\frac{d_x}{d_a d_b}}\,t^{(\lambda)}_{x c\to y}\, t^{(\mu)}_{ab\to x}\,, \\
M^{(a;b,c)}_{y,(z,\nu,\sigma)} &= \sqrt{\frac{d_y}{d_z d_a}}\,\sqrt{\frac{d_z}{d_b d_c}}\,t^{(\sigma)}_{a z\to y}\, t^{(\nu)}_{b c\to z}\,.
\end{align}
Using the fusion-dimension identity and the unitarity of the $F$-symbols, the previous intertwiner relation lifts directly to a unitary transformation between the two Kraus sets,
\begin{align}
M^{(a;b,c)}_{y,(z,\nu,\sigma)}=\sum_{x,\mu,\lambda} F^{abc}_y \left[(z,\nu,\sigma),(x,\mu,\lambda)\right]\, L^{(a;b,c)}_{y,(x,\mu,\lambda)}\,.
\end{align}
Since two Kraus representations related by a unitary transformation on the index space define the same completely positive trace-preserving map, the two parenthesisations yield identical quantum channels. This establishes that non-invertible channels are associative, with associativity controlled precisely by the categorical $F$-symbols.

\paragraph{Example: Ising category.}
\begin{figure}[t]
    \centering
    \includegraphics[width=0.71\linewidth]{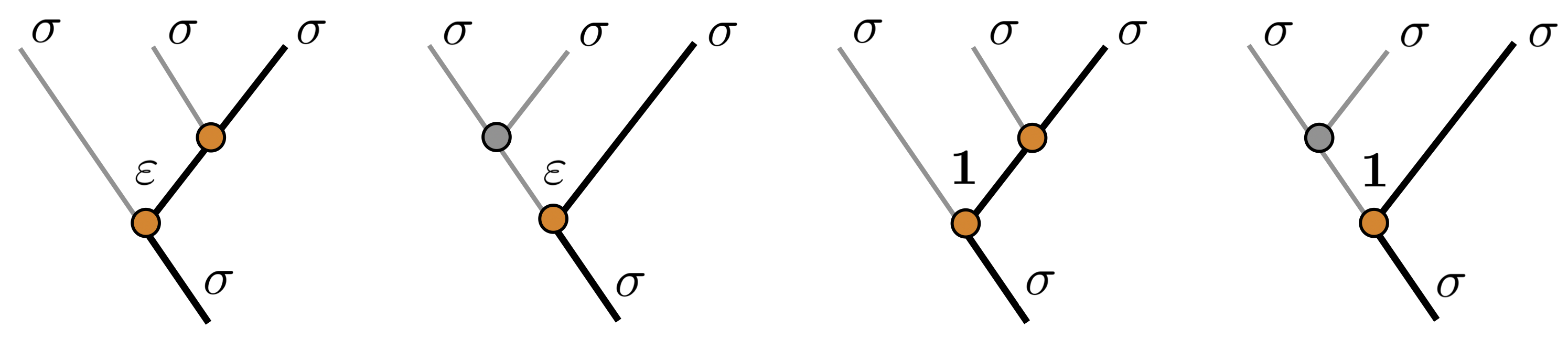}
    \caption{Two fusion trees for three $\sigma$ defects in the Ising category with total output $\sigma$. The two possible parenthesisations, $(\sigma\otimes\sigma)\otimes\sigma$ and $\sigma\otimes(\sigma\otimes\sigma)$, admit intermediate channels $1$ and $\varepsilon$ (shown at the trivalent vertices). These trees are related by the nontrivial $F$-move of the Ising UMTC.}
    \label{fig:example_Ising_assoc}
\end{figure}
In the Ising UMTC one has $\sigma\otimes\sigma=\mathbf 1\oplus\varepsilon$, with all fusion multiplicities equal to $0$ or $1$. Consider three non-invertible gates $\sigma,\sigma,\sigma$ with total output $y=\sigma$. Both fusion trees admit two intermediate channels,
\begin{align}
(\sigma\otimes\sigma)\otimes\sigma\to\sigma \quad\text{via }x\in\{\mathbf 1,\varepsilon\}\,,\qquad \sigma\otimes(\sigma\otimes\sigma)\to\sigma \quad\text{via }z\in\{\mathbf 1,\varepsilon\}\,,
\end{align}
as shown in fig. \ref{fig:example_Ising_assoc}. The nontrivial $F$-matrix is
\begin{align}
F^{\sigma\sigma\sigma}_{\sigma}= \frac{1}{\sqrt 2}
\begin{pmatrix}
1 & 1\\
1 & -1
\end{pmatrix}\,,
\end{align}
in the basis $(\mathbf 1,\varepsilon)$. The Kraus operators associated with the two parenthesisations may be written as
\begin{align}
L_{\sigma;x} &=\sqrt{\frac{d_\sigma}{d_\sigma^3}}\; t_{x\sigma\to\sigma}\,t_{\sigma\sigma\to x}\,,\quad
M_{\sigma;z} =\sqrt{\frac{d_\sigma}{d_\sigma^3}}\; t_{\sigma z\to\sigma}\,t_{\sigma\sigma\to z}\,,
\end{align}
and the $F$-move identity implies
\begin{align}
M_{\sigma;z} =\sum_{x\in\{\mathbf 1,\varepsilon\}} F^{\sigma\sigma\sigma}_{\sigma}[z,x]\; L_{\sigma;x}\,.
\end{align}
Since $F^{\sigma\sigma\sigma}_{\sigma}$ is unitary, both fusion trees define the same CPTP channel. 

\section{Finiteness of the energy-weighted gate cost} \label{app:finiteness_energy_cost}

In this appendix we explain why the energy-weighted gate cost defined in eq. \eqref{eq:gate_cost_def} is finite whenever a fusion process admits only finitely many channels with strictly positive categorical weights.

We consider a fusion process with a finite set of allowed channels labelled by $c$, each appearing with base weight $p_c>0$, $\sum_c p_c=1$, and associated energy shift $E_c$. Introducing the $\beta$-deformed weights
\begin{align}
p_c^{(\beta)}=\frac{p_c e^{-\beta E_c}}{Z(\beta)}\,,\quad Z(\beta)=\sum_c p_c e^{-\beta E_c}\,,
\end{align}
the variance entering the cost is the variance of $E$ with respect to the probability distribution $p_c^{(\beta)}$. To analyse the behaviour of the integrand at large $\beta$, it is convenient to separate the minimal-energy channels from the rest. Let
\begin{align}
E_{\min}=\min_c E_c\,, \quad
\Delta_c=E_c-E_{\min}\ge 0\,, \end{align}
and denote by $M=\{c:\Delta_c=0\}$ the set of channels saturating the minimal energy. Since all base weights are
strictly positive, the total weight carried by these channels,
\begin{align}
P_{\min}=\sum_{m\in M}p_m\,,
\end{align}
is non-zero. We also define the energy gap to the first excited channel, $\Delta_*=\min_{c\notin M}\Delta_c$, with the understanding that $\Delta_*=\infty$ if all channels are energetically degenerate.

\begin{itemize}
	\item Large-$\beta$ behaviour: As $\beta\to\infty$, the Gibbs reweighting suppresses all channels whose energy
exceeds $E_{\min}$. More precisely, for any $c\notin M$ one has
\begin{align}
p_c^{(\beta)} =\frac{p_c e^{-\beta\Delta_c}}{\sum_{m\in M}p_m+\sum_{d\notin M}p_d e^{-\beta\Delta_d}} \le \frac{p_c}{P_{\min}}\,e^{-\beta\Delta_c} \le \frac{p_c}{P_{\min}}\,e^{-\beta\Delta_*}\,.
\end{align}
Summing over all non-minimal channels shows that the total probability weight outside $M$ is exponentially suppressed,
\begin{align}
\sum_{c\notin M}p_c^{(\beta)} \le \frac{1-P_{\min}}{P_{\min}}\,e^{-\beta\Delta_*}\,.
\end{align}
 
Because the number of channels is finite, the maximal energy separation $\Delta_{\max}=\max_c \Delta_c$ is also finite. Using $E_c-E_{\min}=\Delta_c\in[0,\Delta_{\max}]$, the variance can be bounded as
\begin{align}
\mathrm{Var}_\beta(E) &=\sum_c p_c^{(\beta)}\bigl(E_c-\langle E\rangle_\beta\bigr)^2
\le \sum_c p_c^{(\beta)}(E_c-E_{\min})^2 \nonumber\\
&=\sum_c p_c^{(\beta)}\Delta_c^2 \le \Delta_{\max}^2\sum_{c\notin M}p_c^{(\beta)}\,.
\end{align}
Combining this with the exponential suppression above yields
\begin{align}
\sqrt{\mathrm{Var}_\beta(E)} \le \Delta_{\max}\sqrt{\frac{1-P_{\min}}{P_{\min}}}\; e^{-\beta\Delta_*/2}\,,
\end{align}
which is manifestly integrable at large $\beta$. In the special case where all channels are energetically degenerate ($\Delta_*=\infty$), the variance vanishes identically for all $\beta$, and the energy-weighted cost is zero.

	\item Behaviour near $\beta=0$: On any finite interval $\beta\in[0,B]$, the weights $p_c^{(\beta)}$ are smooth and uniformly bounded. Since the energies take values in the finite interval $[E_{\min},E_{\max}]$, with $E_{\max}=\max_c E_c$, the variance satisfies
\begin{align}
\mathrm{Var}_\beta(E)\le (E_{\max}-E_{\min})^2\,,
\end{align}
and therefore
\begin{align}
\int_0^\infty\mathrm d\beta\;\sqrt{\mathrm{Var}_\beta(E)}<\infty\,.
\end{align}
\end{itemize}
The integral defining the energy-weighted gate cost thus receives a finite contribution from any finite interval in $\beta$, while its large-$\beta$ tail is exponentially convergent whenever only finitely many fusion channels are present.

\bibliographystyle{JHEP}
\bibliography{biblio}   

@article{Kaidi:2021gbs,
    author = "Kaidi, Justin and Komargodski, Zohar and Ohmori, Kantaro and Seifnashri, Sahand and Shao, Shu-Heng",
    title = "{Higher central charges and topological boundaries in 2+1-dimensional TQFTs}",
    eprint = "2107.13091",
    archivePrefix = "arXiv",
    primaryClass = "hep-th",
    doi = "10.21468/SciPostPhys.13.3.067",
    journal = "SciPost Phys.",
    volume = "13",
    number = "3",
    pages = "067",
    year = "2022"
}

@article{Benini:2018reh,
    author = "Benini, Francesco and C{\'o}rdova, Clay and Hsin, Po-Shen",
    title = "{On 2-Group Global Symmetries and their Anomalies}",
    eprint = "1803.09336",
    archivePrefix = "arXiv",
    primaryClass = "hep-th",
    reportNumber = "SISSA 10/2018/FISI, SISSA-10-2018-FISI",
    doi = "10.1007/JHEP03(2019)118",
    journal = "JHEP",
    volume = "03",
    pages = "118",
    year = "2019"
}

@article{Kitaev:2005hzj,
    author = "Kitaev, Alexei",
    title = "{Anyons in an exactly solved model and beyond}",
    eprint = "cond-mat/0506438",
    archivePrefix = "arXiv",
    doi = "10.1016/j.aop.2005.10.005",
    journal = "Annals Phys.",
    volume = "321",
    number = "1",
    pages = "2--111",
    year = "2006"
}

@article{Caputa:2018kdj,
    author = "Caputa, Pawel and Magan, Javier M.",
    title = "{Quantum Computation as Gravity}",
    eprint = "1807.04422",
    archivePrefix = "arXiv",
    primaryClass = "hep-th",
    reportNumber = "YITP-18-75",
    doi = "10.1103/PhysRevLett.122.231302",
    journal = "Phys. Rev. Lett.",
    volume = "122",
    number = "23",
    pages = "231302",
    year = "2019"
}

@article{Witten:1987ty,
    author = "Witten, Edward",
    title = "{Coadjoint Orbits of the Virasoro Group}",
    reportNumber = "PUPT-1061",
    doi = "10.1007/BF01218287",
    journal = "Commun. Math. Phys.",
    volume = "114",
    pages = "1",
    year = "1988"
}

@article{Janik:2025zji,
    author = "Janik, Romuald A.",
    title = "{Ising model as a window on quantum gravity with matter}",
    eprint = "2502.19015",
    archivePrefix = "arXiv",
    primaryClass = "hep-th",
    doi = "10.1103/PhysRevD.111.106016",
    journal = "Phys. Rev. D",
    volume = "111",
    number = "10",
    pages = "106016",
    year = "2025"
}

@article{Bershadsky:1989mf,
    author = "Bershadsky, Michael and Ooguri, Hirosi",
    editor = "Bouwknegt, P. and Schoutens, K.",
    title = "{Hidden SL(n) Symmetry in Conformal Field Theories}",
    reportNumber = "IASSNS-HEP-89-09",
    doi = "10.1007/BF02124331",
    journal = "Commun. Math. Phys.",
    volume = "126",
    pages = "49",
    year = "1989"
}

@article{Erdmenger:2020sup,
    author = "Erdmenger, Johanna and Gerbershagen, Marius and Weigel, Anna-Lena",
    title = "{Complexity measures from geometric actions on Virasoro and Kac-Moody orbits}",
    eprint = "2004.03619",
    archivePrefix = "arXiv",
    primaryClass = "hep-th",
    doi = "10.1007/JHEP11(2020)003",
    journal = "JHEP",
    volume = "11",
    pages = "003",
    year = "2020"
}

@article{Erdmenger:2021wzc,
    author = "Erdmenger, Johanna and Flory, Mario and Gerbershagen, Marius and Heller, Michal P. and Weigel, Anna-Lena",
    title = "{Exact Gravity Duals for Simple Quantum Circuits}",
    eprint = "2112.12158",
    archivePrefix = "arXiv",
    primaryClass = "hep-th",
    doi = "10.21468/SciPostPhys.13.3.061",
    journal = "SciPost Phys.",
    volume = "13",
    number = "3",
    pages = "061",
    year = "2022"
}

@book{bakalov2001lectures,
  title={Lectures on tensor categories and modular functors},
  author={Bakalov, Bojko and Kirillov, Alexander A and others},
  volume={21},
  year={2001},
  publisher={American Mathematical Society Providence, RI}
}

@article{Campoleoni:2024ced,
    author = "Campoleoni, Andrea and Fredenhagen, Stefan",
    title = "{Higher-Spin Gauge Theories in Three Spacetime Dimensions}",
    eprint = "2403.16567",
    archivePrefix = "arXiv",
    primaryClass = "hep-th",
    doi = "10.1007/978-3-031-59656-8_2",
    journal = "Lect. Notes Phys.",
    volume = "1028",
    pages = "121--267",
    year = "2024"
}

@article{Balasubramanian:2019stt,
    author = "Balasubramanian, Vijay and Craps, Ben and De Clerck, Marine and Nguyen, K{\'e}vin",
    title = "{Superluminal chaos after a quantum quench}",
    eprint = "1908.08955",
    archivePrefix = "arXiv",
    primaryClass = "hep-th",
    doi = "10.1007/JHEP12(2019)132",
    journal = "JHEP",
    volume = "12",
    pages = "132",
    year = "2019"
}

@article{Fuchs:2002cm,
    author = "Fuchs, Jurgen and Runkel, Ingo and Schweigert, Christoph",
    title = "{TFT construction of RCFT correlators 1. Partition functions}",
    eprint = "hep-th/0204148",
    archivePrefix = "arXiv",
    primaryClass = "hep-th",
    doi = "10.1016/S0550-3213(02)00744-7",
    journal = "Nucl. Phys. B",
    volume = "646",
    pages = "353--497",
    year = "2002"
}

@article{Castro:2011zq,
    author = "Castro, Alejandra and Gaberdiel, Matthias R. and Hartman, Thomas and Maloney, Alexander and Volpato, Roberto",
    title = "{The Gravity Dual of the Ising Model}",
    eprint = "1111.1987",
    archivePrefix = "arXiv",
    primaryClass = "hep-th",
    reportNumber = "AEI-2011-079, NSF-KITP-11-221",
    doi = "10.1103/PhysRevD.85.024032",
    journal = "Phys. Rev. D",
    volume = "85",
    pages = "024032",
    year = "2012"
}

@article{Chagnet:2021uvi,
    author = "Chagnet, Nicolas and Chapman, Shira and de Boer, Jan and Zukowski, Claire",
    title = "{Complexity for Conformal Field Theories in General Dimensions}",
    eprint = "2103.06920",
    archivePrefix = "arXiv",
    primaryClass = "hep-th",
    doi = "10.1103/PhysRevLett.128.051601",
    journal = "Phys. Rev. Lett.",
    volume = "128",
    number = "5",
    pages = "051601",
    year = "2022"
}

@article{Wong:2022eiu,
    author = "Wong, Gabriel",
    title = "{A note on the bulk interpretation of the quantum extremal surface formula}",
    eprint = "2212.03193",
    archivePrefix = "arXiv",
    primaryClass = "hep-th",
    doi = "10.1007/JHEP04(2024)024",
    journal = "JHEP",
    volume = "04",
    pages = "024",
    year = "2024"
}

@article{Mertens:2022ujr,
    author = "Mertens, Thomas G. and Sim{\'o}n, Joan and Wong, Gabriel",
    title = "{A proposal for 3d quantum gravity and its bulk factorization}",
    eprint = "2210.14196",
    archivePrefix = "arXiv",
    primaryClass = "hep-th",
    doi = "10.1007/JHEP06(2023)134",
    journal = "JHEP",
    volume = "06",
    pages = "134",
    year = "2023"
}

@article{McGough:2013gka,
    author = "McGough, Lauren and Verlinde, Herman",
    title = "{Bekenstein-Hawking Entropy as Topological Entanglement Entropy}",
    eprint = "1308.2342",
    archivePrefix = "arXiv",
    primaryClass = "hep-th",
    doi = "10.1007/JHEP11(2013)208",
    journal = "JHEP",
    volume = "11",
    pages = "208",
    year = "2013"
}

@article{Shenker:2013pqa,
    author = "Shenker, Stephen H. and Stanford, Douglas",
    title = "{Black holes and the butterfly effect}",
    eprint = "1306.0622",
    archivePrefix = "arXiv",
    primaryClass = "hep-th",
    reportNumber = "SU-ITP-13-08",
    doi = "10.1007/JHEP03(2014)067",
    journal = "JHEP",
    volume = "03",
    pages = "067",
    year = "2014"
}

@article{Shenker:2013yza,
    author = "Shenker, Stephen H. and Stanford, Douglas",
    title = "{Multiple Shocks}",
    eprint = "1312.3296",
    archivePrefix = "arXiv",
    primaryClass = "hep-th",
    reportNumber = "SU-ITP-13-24",
    doi = "10.1007/JHEP12(2014)046",
    journal = "JHEP",
    volume = "12",
    pages = "046",
    year = "2014"
}

@article{Roberts:2014ifa,
    author = "Roberts, Daniel A. and Stanford, Douglas",
    title = "{Two-dimensional conformal field theory and the butterfly effect}",
    eprint = "1412.5123",
    archivePrefix = "arXiv",
    primaryClass = "hep-th",
    reportNumber = "MIT-CTP-4626",
    doi = "10.1103/PhysRevLett.115.131603",
    journal = "Phys. Rev. Lett.",
    volume = "115",
    number = "13",
    pages = "131603",
    year = "2015"
}

@article{dray1985gravitational,
  title={The gravitational shock wave of a massless particle},
  author={Dray, Tevian and Hooft, Gerard't},
  journal={Nuclear physics B},
  volume={253},
  pages={173--188},
  year={1985},
  publisher={Elsevier}
}

@article{vaidya1999external,
  title={The external field of a radiating star in general relativity},
  author={Vaidya, PC},
  journal={General Relativity and Gravitation},
  volume={31},
  number={1},
  pages={119--120},
  year={1999},
  publisher={Springer}
}

@article{Jiang:2018tlu,
    author = "Jiang, Jie",
    title = "{Holographic complexity in charged Vaidya black hole}",
    eprint = "1811.07347",
    archivePrefix = "arXiv",
    primaryClass = "hep-th",
    doi = "10.1140/epjc/s10052-019-6639-1",
    journal = "Eur. Phys. J. C",
    volume = "79",
    number = "2",
    pages = "130",
    year = "2019"
}

@article{Chapman:2018dem,
    author = "Chapman, Shira and Marrochio, Hugo and Myers, Robert C.",
    title = "{Holographic complexity in Vaidya spacetimes. Part I}",
    eprint = "1804.07410",
    archivePrefix = "arXiv",
    primaryClass = "hep-th",
    doi = "10.1007/JHEP06(2018)046",
    journal = "JHEP",
    volume = "06",
    pages = "046",
    year = "2018"
}

@article{Chapman:2018lsv,
    author = "Chapman, Shira and Marrochio, Hugo and Myers, Robert C.",
    title = "{Holographic complexity in Vaidya spacetimes. Part II}",
    eprint = "1805.07262",
    archivePrefix = "arXiv",
    primaryClass = "hep-th",
    doi = "10.1007/JHEP06(2018)114",
    journal = "JHEP",
    volume = "06",
    pages = "114",
    year = "2018"
}

@article{Castro:2016tlm,
    author = "Castro, Alejandra",
    title = "{Lectures on Higher Spin Black Holes in AdS$_3$ Gravity}",
    doi = "10.5506/APhysPolB.47.2479",
    journal = "Acta Phys. Polon. B",
    volume = "47",
    pages = "2479--2508",
    year = "2016"
}

@article{Nielsen:2005mkt,
    author = "Nielsen, Michael A.",
    title = "{A geometric approach to quantum circuit lower bounds}",
    eprint = "quant-ph/0502070",
    archivePrefix = "arXiv",
    doi = "10.26421/QIC6.3-2",
    journal = "Quant. Inf. Comput.",
    volume = "6",
    number = "3",
    pages = "213--262",
    year = "2006"
}

@article{Nielsen:2006cea,
    author = "Nielsen, Michael A. and Dowling, Mark R. and Gu, Mile and Doherty, Andrew C.",
    title = "{Quantum Computation as Geometry}",
    eprint = "quant-ph/0603161",
    archivePrefix = "arXiv",
    doi = "10.1126/science.1121541",
    journal = "Science",
    volume = "311",
    number = "5764",
    pages = "1133--1135",
    year = "2006"
}

@article{McGreevy:2022oyu,
    author = "McGreevy, John",
    title = "{Generalized Symmetries in Condensed Matter}",
    eprint = "2204.03045",
    archivePrefix = "arXiv",
    primaryClass = "cond-mat.str-el",
    doi = "10.1146/annurev-conmatphys-040721-021029",
    journal = "Ann. Rev. Condensed Matter Phys.",
    volume = "14",
    pages = "57--82",
    year = "2023"
}

@article{Schafer-Nameki:2023jdn,
    author = "Schafer-Nameki, Sakura",
    title = "{ICTP lectures on (non-)invertible generalized symmetries}",
    eprint = "2305.18296",
    archivePrefix = "arXiv",
    primaryClass = "hep-th",
    doi = "10.1016/j.physrep.2024.01.007",
    journal = "Phys. Rept.",
    volume = "1063",
    pages = "1--55",
    year = "2024"
}

@article{Bhardwaj:2023kri,
    author = "Bhardwaj, Lakshya and Bottini, Lea E. and Fraser-Taliente, Ludovic and Gladden, Liam and Gould, Dewi S. W. and Platschorre, Arthur and Tillim, Hannah",
    title = "{Lectures on generalized symmetries}",
    eprint = "2307.07547",
    archivePrefix = "arXiv",
    primaryClass = "hep-th",
    doi = "10.1016/j.physrep.2023.11.002",
    journal = "Phys. Rept.",
    volume = "1051",
    pages = "1--87",
    year = "2024"
}

@inproceedings{Shao:2023gho,
    author = "Shao, Shu-Heng",
    title = "{What's Done Cannot Be Undone: TASI Lectures on Non-Invertible Symmetries}",
    booktitle = "{Theoretical Advanced Study Institute in Elementary Particle Physics 2023}: {Aspects of Symmetry}",
    eprint = "2308.00747",
    archivePrefix = "arXiv",
    primaryClass = "hep-th",
    reportNumber = "YITP-SB-2023-19",
    month = "8",
    year = "2023"
}

@article{Petkova:2000ip,
    author = "Petkova, V. B. and Zuber, J. B.",
    title = "{Generalized twisted partition functions}",
    eprint = "hep-th/0011021",
    archivePrefix = "arXiv",
    reportNumber = "UNN-SCM-M-00-07, CERN-TH-2000-322",
    doi = "10.1016/S0370-2693(01)00276-3",
    journal = "Phys. Lett. B",
    volume = "504",
    pages = "157--164",
    year = "2001"
}

@article{Oshikawa:1996dj,
    author = "Oshikawa, Masaki and Affleck, Ian",
    title = "{Boundary conformal field theory approach to the critical two-dimensional Ising model with a defect line}",
    eprint = "cond-mat/9612187",
    archivePrefix = "arXiv",
    doi = "10.1016/S0550-3213(97)00219-8",
    journal = "Nucl. Phys. B",
    volume = "495",
    pages = "533--582",
    year = "1997"
}

@article{Frohlich:2006ch,
    author = "Frohlich, Jurg and Fuchs, Jurgen and Runkel, Ingo and Schweigert, Christoph",
    title = "{Duality and defects in rational conformal field theory}",
    eprint = "hep-th/0607247",
    archivePrefix = "arXiv",
    reportNumber = "KCL-MTH-06-08, ZMP-HH-06-11",
    doi = "10.1016/j.nuclphysb.2006.11.017",
    journal = "Nucl. Phys. B",
    volume = "763",
    pages = "354--430",
    year = "2007"
}

@article{Balasubramanian:2019wgd,
    author = "Balasubramanian, Vijay and Decross, Matthew and Kar, Arjun and Parrikar, Onkar",
    title = "{Quantum Complexity of Time Evolution with Chaotic Hamiltonians}",
    eprint = "1905.05765",
    archivePrefix = "arXiv",
    primaryClass = "hep-th",
    doi = "10.1007/JHEP01(2020)134",
    journal = "JHEP",
    volume = "01",
    pages = "134",
    year = "2020"
}

@article{Balasubramanian:2021mxo,
    author = "Balasubramanian, Vijay and DeCross, Matthew and Kar, Arjun and Li, Yue (Cathy) and Parrikar, Onkar",
    title = "{Complexity growth in integrable and chaotic models}",
    eprint = "2101.02209",
    archivePrefix = "arXiv",
    primaryClass = "hep-th",
    doi = "10.1007/JHEP07(2021)011",
    journal = "JHEP",
    volume = "07",
    pages = "011",
    year = "2021"
}

@article{Craps:2023rur,
    author = "Craps, Ben and De Clerck, Marine and Evnin, Oleg and Hacker, Philip",
    title = "{Integrability and complexity in quantum spin chains}",
    eprint = "2305.00037",
    archivePrefix = "arXiv",
    primaryClass = "quant-ph",
    doi = "10.21468/SciPostPhys.16.2.041",
    journal = "SciPost Phys.",
    volume = "16",
    number = "2",
    pages = "041",
    year = "2024"
}

@article{Baiguera:2025uss,
    author = "Baiguera, Stefano and Chagnet, Nicolas and Chapman, Shira and Shoval, Osher",
    title = "{CFT Complexity and Penalty Factors}",
    eprint = "2507.22118",
    archivePrefix = "arXiv",
    primaryClass = "hep-th",
    month = "7",
    year = "2025"
}

@article{Baiguera:2023bhm,
    author = "Baiguera, Stefano and Chapman, Shira and Policastro, Giuseppe and Schwartzman, Tal",
    title = "{The Complexity of Being Entangled}",
    eprint = "2311.04277",
    archivePrefix = "arXiv",
    primaryClass = "hep-th",
    doi = "10.22331/q-2024-09-12-1472",
    journal = "Quantum",
    volume = "8",
    pages = "1472",
    year = "2024"
}

@article{Erdmenger:2022lov,
    author = "Erdmenger, Johanna and Weigel, Anna-Lena and Gerbershagen, Marius and Heller, Michal P.",
    title = "{From complexity geometry to holographic spacetime}",
    eprint = "2212.00043",
    archivePrefix = "arXiv",
    primaryClass = "hep-th",
    doi = "10.1103/PhysRevD.108.106020",
    journal = "Phys. Rev. D",
    volume = "108",
    number = "10",
    pages = "106020",
    year = "2023"
}

@article{Chapman:2021jbh,
    author = "Chapman, Shira and Policastro, Giuseppe",
    title = "{Quantum computational complexity from quantum information to black holes and back}",
    eprint = "2110.14672",
    archivePrefix = "arXiv",
    primaryClass = "hep-th",
    doi = "10.1140/epjc/s10052-022-10037-1",
    journal = "Eur. Phys. J. C",
    volume = "82",
    number = "2",
    pages = "128",
    year = "2022"
}

@article{Baiguera:2025dkc,
    author = "Baiguera, Stefano and Balasubramanian, Vijay and Caputa, Pawel and Chapman, Shira and Haferkamp, Jonas and Heller, Michal P. and Halpern, Nicole Yunger",
    title = "{Quantum complexity in gravity, quantum field theory, and quantum information science}",
    eprint = "2503.10753",
    archivePrefix = "arXiv",
    primaryClass = "hep-th",
    reportNumber = "YITP-25-39",
    doi = "10.1016/j.physrep.2025.11.001",
    journal = "Phys. Rept.",
    volume = "1159",
    pages = "1--77",
    year = "2026"
}

@article{Ponsot:1999uf,
    author = "Ponsot, B. and Teschner, J.",
    title = "{Liouville bootstrap via harmonic analysis on a noncompact quantum group}",
    eprint = "hep-th/9911110",
    archivePrefix = "arXiv",
    reportNumber = "DIAS-STP-99-14, ESI-783, LPM-99-46",
    month = "11",
    year = "1999"
}

@article{Hofman:2008ar,
    author = "Hofman, Diego M. and Maldacena, Juan",
    title = "{Conformal collider physics: Energy and charge correlations}",
    eprint = "0803.1467",
    archivePrefix = "arXiv",
    primaryClass = "hep-th",
    doi = "10.1088/1126-6708/2008/05/012",
    journal = "JHEP",
    volume = "05",
    pages = "012",
    year = "2008"
}

@article{Hartman:2016lgu,
    author = "Hartman, Thomas and Kundu, Sandipan and Tajdini, Amirhossein",
    title = "{Averaged Null Energy Condition from Causality}",
    eprint = "1610.05308",
    archivePrefix = "arXiv",
    primaryClass = "hep-th",
    doi = "10.1007/JHEP07(2017)066",
    journal = "JHEP",
    volume = "07",
    pages = "066",
    year = "2017"
}

@article{Kundu:2021nwp,
    author = "Kundu, Arnab",
    title = "{Wormholes and holography: an introduction}",
    eprint = "2110.14958",
    archivePrefix = "arXiv",
    primaryClass = "hep-th",
    doi = "10.1140/epjc/s10052-022-10376-z",
    journal = "Eur. Phys. J. C",
    volume = "82",
    number = "5",
    pages = "447",
    year = "2022"
}

@article{Fuchs:2023ngi,
    author = {Fuchs, J{\"u}rgen and Schweigert, Christoph and Wood, Simon and Yang, Yang},
    title = "{Algebraic structures in two-dimensional conformal field theory}",
    eprint = "2305.02773",
    archivePrefix = "arXiv",
    primaryClass = "math.QA",
    reportNumber = "Hamburger Beitr. zur Mathematik Nr. 941; ZMP-HH/23-7",
    doi = "10.1016/B978-0-323-95703-8.00013-6",
    month = "5",
    year = "2023"
}

@article{Moore:1988qv,
    author = "Moore, Gregory W. and Seiberg, Nathan",
    title = "{Classical and Quantum Conformal Field Theory}",
    reportNumber = "IASSNS-HEP-88-39",
    doi = "10.1007/BF01238857",
    journal = "Commun. Math. Phys.",
    volume = "123",
    pages = "177",
    year = "1989"
}

@article{Verlinde:1988sn,
    author = "Verlinde, Erik P.",
    title = "{Fusion Rules and Modular Transformations in 2D Conformal Field Theory}",
    reportNumber = "THU-88/17",
    doi = "10.1016/0550-3213(88)90603-7",
    journal = "Nucl. Phys. B",
    volume = "300",
    pages = "360--376",
    year = "1988"
}

@article{Coussaert:1995zp,
    author = "Coussaert, Oliver and Henneaux, Marc and van Driel, Peter",
    title = "{The Asymptotic dynamics of three-dimensional Einstein gravity with a negative cosmological constant}",
    eprint = "gr-qc/9506019",
    archivePrefix = "arXiv",
    reportNumber = "ULB-TH-95-08",
    doi = "10.1088/0264-9381/12/12/012",
    journal = "Class. Quant. Grav.",
    volume = "12",
    pages = "2961--2966",
    year = "1995"
}

@article{Roberts:2014isa,
    author = "Roberts, Daniel A. and Stanford, Douglas and Susskind, Leonard",
    title = "{Localized shocks}",
    eprint = "1409.8180",
    archivePrefix = "arXiv",
    primaryClass = "hep-th",
    reportNumber = "MIT-CTP-4594, SU-ITP-14-20",
    doi = "10.1007/JHEP03(2015)051",
    journal = "JHEP",
    volume = "03",
    pages = "051",
    year = "2015"
}

@article{Brown:2015lvg,
    author = "Brown, Adam R. and Roberts, Daniel A. and Susskind, Leonard and Swingle, Brian and Zhao, Ying",
    title = "{Complexity, action, and black holes}",
    eprint = "1512.04993",
    archivePrefix = "arXiv",
    primaryClass = "hep-th",
    doi = "10.1103/PhysRevD.93.086006",
    journal = "Phys. Rev. D",
    volume = "93",
    number = "8",
    pages = "086006",
    year = "2016"
}

@article{Brown:2015bva,
    author = "Brown, Adam R. and Roberts, Daniel A. and Susskind, Leonard and Swingle, Brian and Zhao, Ying",
    title = "{Holographic Complexity Equals Bulk Action?}",
    eprint = "1509.07876",
    archivePrefix = "arXiv",
    primaryClass = "hep-th",
    doi = "10.1103/PhysRevLett.116.191301",
    journal = "Phys. Rev. Lett.",
    volume = "116",
    number = "19",
    pages = "191301",
    year = "2016"
}

@article{Stanford:2014jda,
    author = "Stanford, Douglas and Susskind, Leonard",
    title = "{Complexity and Shock Wave Geometries}",
    eprint = "1406.2678",
    archivePrefix = "arXiv",
    primaryClass = "hep-th",
    doi = "10.1103/PhysRevD.90.126007",
    journal = "Phys. Rev. D",
    volume = "90",
    number = "12",
    pages = "126007",
    year = "2014"
}

@article{Bhardwaj:2023idu,
    author = {Bhardwaj, Lakshya and Bottini, Lea E. and Pajer, Daniel and Sch{\"a}fer-Nameki, Sakura},
    title = "{Gapped phases with non-invertible symmetries: (1+1)d}",
    eprint = "2310.03784",
    archivePrefix = "arXiv",
    primaryClass = "hep-th",
    doi = "10.21468/SciPostPhys.18.1.032",
    journal = "SciPost Phys.",
    volume = "18",
    number = "1",
    pages = "032",
    year = "2025"
}

@article{Freed:2022qnc,
    author = "Freed, Daniel S. and Moore, Gregory W. and Teleman, Constantin",
    title = "{Topological symmetry in quantum field theory}",
    eprint = "2209.07471",
    archivePrefix = "arXiv",
    primaryClass = "hep-th",
    month = "9",
    year = "2022"
}

@article{muger2003subfactors,
  title={From subfactors to categories and topology II: The quantum double of tensor categories and subfactors},
  author={M{\"u}ger, Michael},
  journal={Journal of Pure and Applied Algebra},
  volume={180},
  number={1-2},
  pages={159--219},
  year={2003},
  publisher={Elsevier}
}

@article{Banados:1992wn,
    author = "Banados, Maximo and Teitelboim, Claudio and Zanelli, Jorge",
    title = "{The Black hole in three-dimensional space-time}",
    eprint = "hep-th/9204099",
    archivePrefix = "arXiv",
    reportNumber = "PRINT-92-0151 (CHILE), IASSNS-HEP-92-29",
    doi = "10.1103/PhysRevLett.69.1849",
    journal = "Phys. Rev. Lett.",
    volume = "69",
    pages = "1849--1851",
    year = "1992"
}

@article{Carlip:1995qv,
    author = "Carlip, Steven",
    title = "{The (2+1)-Dimensional black hole}",
    eprint = "gr-qc/9506079",
    archivePrefix = "arXiv",
    reportNumber = "UCD-95-15",
    doi = "10.1088/0264-9381/12/12/005",
    journal = "Class. Quant. Grav.",
    volume = "12",
    pages = "2853--2880",
    year = "1995"
}

@article{Banados:1998sm,
    author = "Banados, Maximo",
    editor = "D'Olivo, J. C. and Mondragon, M. and Lopez Castro, G.",
    title = "{Notes on black holes and three-dimensional gravity}",
    eprint = "hep-th/9903244",
    archivePrefix = "arXiv",
    doi = "10.1063/1.1301386",
    journal = "AIP Conf. Proc.",
    volume = "490",
    number = "1",
    pages = "198--216",
    year = "1999"
}

@article{Banados:1998gg,
    author = "Banados, Maximo",
    editor = "Falomir, H. and Gamboa Saravi, R. E. and Schaposnik, F. A.",
    title = "{Three-dimensional quantum geometry and black holes}",
    eprint = "hep-th/9901148",
    archivePrefix = "arXiv",
    doi = "10.1063/1.59661",
    journal = "AIP Conf. Proc.",
    volume = "484",
    number = "1",
    pages = "147--169",
    year = "1999"
}

@article{Brown:1986nw,
    author = "Brown, J. David and Henneaux, M.",
    title = "{Central Charges in the Canonical Realization of Asymptotic Symmetries: An Example from Three-Dimensional Gravity}",
    doi = "10.1007/BF01211590",
    journal = "Commun. Math. Phys.",
    volume = "104",
    pages = "207--226",
    year = "1986"
}

@article{Achucarro:1986uwr,
    author = "Achucarro, A. and Townsend, P. K.",
    editor = "Salam, A. and Sezgin, E.",
    title = "{A Chern-Simons Action for Three-Dimensional anti-De Sitter Supergravity Theories}",
    reportNumber = "Print-87-0078 (CAMBRIDGE)",
    doi = "10.1016/0370-2693(86)90140-1",
    journal = "Phys. Lett. B",
    volume = "180",
    pages = "89",
    year = "1986"
}

@article{Auzzi:2020idm,
    author = "Auzzi, Roberto and Baiguera, Stefano and De Luca, G. Bruno and Legramandi, Andrea and Nardelli, Giuseppe and Zenoni, Nicol{\`o}",
    title = "{Geometry of quantum complexity}",
    eprint = "2011.07601",
    archivePrefix = "arXiv",
    primaryClass = "hep-th",
    doi = "10.1103/PhysRevD.103.106021",
    journal = "Phys. Rev. D",
    volume = "103",
    number = "10",
    pages = "106021",
    year = "2021"
}

@article{Jefferson:2017sdb,
    author = "Jefferson, Ro and Myers, Robert C.",
    title = "{Circuit complexity in quantum field theory}",
    eprint = "1707.08570",
    archivePrefix = "arXiv",
    primaryClass = "hep-th",
    doi = "10.1007/JHEP10(2017)107",
    journal = "JHEP",
    volume = "10",
    pages = "107",
    year = "2017"
}

@inproceedings{Freed:2009qp,
    author = "Freed, Daniel S. and Hopkins, Michael J. and Lurie, Jacob and Teleman, Constantin",
    title = "{Topological Quantum Field Theories from Compact Lie Groups}",
    booktitle = "{A Celebration of Raoul Bott's Legacy in Mathematics}",
    eprint = "0905.0731",
    archivePrefix = "arXiv",
    primaryClass = "math.AT",
    reportNumber = "AIM-2009-21",
    month = "5",
    year = "2009"
}

@book{NielsenChuang,
  author = {Nielsen, Michael A. and Chuang, Isaac L.},
  title = {Quantum Computation and Quantum Information},
  publisher = {Cambridge University Press},
  year = {2000}
}

@misc{PreskillQI,
  author = {Preskill, John},
  title = {Lecture Notes for Physics 229: Quantum Information and Computation},
  note = {Available at \url{https://theory.caltech.edu/~preskill/ph229/}}
}

@article{Okada:2024qmk,
    author = "Okada, Masaki and Tachikawa, Yuji",
    title = "{Noninvertible Symmetries Act Locally by Quantum Operations}",
    eprint = "2403.20062",
    archivePrefix = "arXiv",
    primaryClass = "hep-th",
    doi = "10.1103/PhysRevLett.133.191602",
    journal = "Phys. Rev. Lett.",
    volume = "133",
    number = "19",
    pages = "191602",
    year = "2024"
}

@inproceedings{Susskind:2018pmk,
    author = "Susskind, Leonard",
    title = "{Three Lectures on Complexity and Black Holes}",
    eprint = "1810.11563",
    archivePrefix = "arXiv",
    primaryClass = "hep-th",
    doi = "10.1007/978-3-030-45109-7",
    publisher = "Springer",
    series = "SpringerBriefs in Physics",
    month = "10",
    year = "2018"
}
\end{document}